\documentclass[12pt,a4paper,openany]{book}
\usepackage{anysize}
\usepackage[utf8]{inputenc}
\usepackage[english]{babel}
\usepackage[ruled,chapter]{algorithm}
\usepackage{makeidx}
\usepackage{subfig}
\usepackage{amsmath}
\usepackage{amssymb}
\usepackage{amsthm}
\usepackage{bbm, dsfont}
\usepackage{bm}
\usepackage{amsfonts}
\usepackage{enumitem}
\usepackage{mathrsfs}
\usepackage{graphicx}
\usepackage{fancyhdr}
\usepackage{array}
\usepackage{cancel}
\usepackage{geometry}
\usepackage{lettrine}
\usepackage{simplewick}
\usepackage{latexsym}
\usepackage[all]{xy}
\usepackage{eufrak}
\usepackage{euscript}
\usepackage{enumerate}
\usepackage{dsfont}
\usepackage{slashed}
\usepackage{braket}
\usepackage{pdfpages}

\usepackage{xcolor} 

\usepackage[plainpages=false,pagebackref=true,pdftex]{hyperref}

\renewcommand*{\backref}[1]{}
\renewcommand*{\backrefalt}[4]{%
    \ifcase #1 %
    \or
        (Cited on page~\textcolor{blue}{#2})%
    \else
        (Cited on pages~\textcolor{blue}{#2})%
    \fi
}

\makeindex

\marginsize{25mm}{25mm}{25mm}{25mm}

\hypersetup{pdfpagelabels,hyperindex,colorlinks=true,breaklinks=true,bookmarks=true,linkcolor=blue,citecolor=blue,
urlcolor=black}

\begin{document}

\setlength{\abovecaptionskip}{0.0cm}
\setlength{\belowcaptionskip}{0.0cm}
\setlength{\baselineskip}{24pt}

\let\cleardoublepage\clearpage

\pagestyle{fancy}
\fancyhf{}
\rhead{\thepage}
\renewcommand{\headrulewidth}{0pt}

\thispagestyle{empty}
\vspace*{\fill}
\begin{figure}[H] \centering
	\includegraphics[scale=0.84]{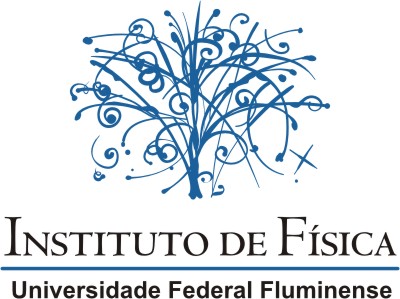}
\end{figure}

\vspace{15pt}

\begin{center}

\vspace{30pt}

{\Large \bf FORMULATIONS OF QUANTUM THERMODYNAMICS AND APPLICATIONS IN OPEN SYSTEMS}

\vspace{56pt}

{\large \bf John Milton Zapana Choquehuanca}

\vspace{50pt}

\vspace{90pt}

\textbf{Niterói - RJ}

\textbf{2025}

\end{center}
\vspace*{\fill}
\newpage   


\pagestyle{fancy}
\lhead{}
\chead{}
\rhead{\thepage}
\lfoot{}
\cfoot{}
\rfoot{}

\fancypagestyle{plain}
{
	\fancyhf{}
	\lhead{}
	\chead{}
	\rhead{\thepage}
	\lfoot{}
	\cfoot{}
	\rfoot{}
}

\renewcommand{\headrulewidth}{0pt}


\frontmatter 
\setcounter{page}{1} 
\thispagestyle{empty}

\begin{figure}[h]
	\includegraphics[scale=0.5]{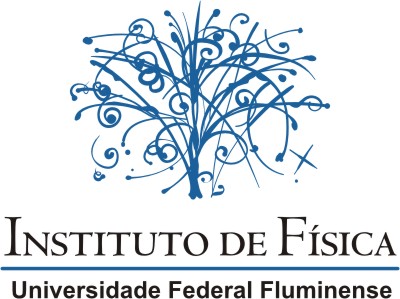}
\end{figure}

\vspace{15pt}

\begin{center}

\textbf{UNIVERSIDADE FEDERAL FLUMINENSE}

\textbf{INSTITUTO DE FÍSICA}

\vspace{30pt}

{\Large \bf FORMULATIONS OF QUANTUM THERMODYNAMICS AND APPLICATIONS IN OPEN SYSTEMS}

\vspace{20pt}

{\large \bf John Milton Zapana Choquehuanca}

\vspace{30pt}

\begin{flushright}
\parbox{10.3cm}{Tese de Doutorado apresentada ao Programa de Pós-Graduação em Física do Instituto de Física da Universidade Federal Fluminense - UFF, como parte dos requisitos necessários à obtenção do título de Doutor em Física, área de concentração, Óptica e Informação Quântica.

\vspace{18pt}

{\large \bf Orientador:~Dr.Marcelo Silva Sarandy}

\vspace{12pt}
{\large \bf Coorientador:~Dr.Fagner Muruci de Paula}}
\end{flushright}

\vspace{90pt}

\textbf{Niterói - RJ}

\textbf{Agosto de 2025}

\end{center}


\includepdf[pages=1, pagecommand={\thispagestyle{plain}}]{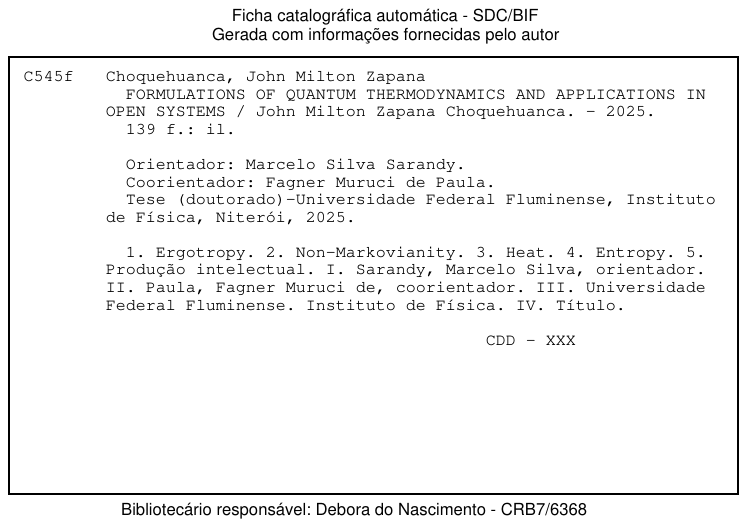}



\vspace*{\fill}
\begin{flushright}
\textit{A la vida,\\
por sus hermosas oportunidades.}
\end{flushright}
\vspace*{\fill}
\newpage

\newpage

\noindent

\vspace*{10pt}
\begin{center}
{\LARGE\bf Resumo}\\
\vspace{15pt}
{\Large\bf FORMULATIONS OF QUANTUM THERMODYNAMICS AND APPLICATIONS IN OPEN SYSTEMS}\\
\vspace{6pt}
{\bf John Milton Zapana Choquehuanca}\\
\vspace{12pt}
{\bf Orientador:~Dr.Marcelo Sarandy}\\
{\bf Coorientador:~Dr.Fagner Muruci de Paula}\\
\vspace{20pt}
\parbox{14cm}{Resumo da Tese de Doutorado apresentada ao Programa de Pós-Graduação em Física do Instituto de Física da Universidade Federal Fluminense - UFF, como parte dos requisitos necessários à obtenção do título de Doutor em Física, área de concentração, Óptica e Informação Quântica.}
\end{center}
\vspace*{25pt}

\noindent A termodinâmica quântica emergiu como um campo central para entender como são os processos de conversão de energia em sistemas microscópicos. Nesses sistemas, efeitos como coerência, entrelaçamento e não-markovianidade desempenham papéis fundamentais. Nesta tese, exploramos diferentes formas de descrever a termodinâmica quântica, usando duas abordagens principais: uma baseada em entropia e outra em ergotropia. Primeiro, introduzimos uma abordagem generalizada para quantificar a não-markovianidade por meio da quebra da monotonicidade em funções termodinâmicas. Nesse contexto, o fluxo de calor baseado em entropia surge como uma ferramenta prática para testemunhar e medir memória quântica em mapas unitais que não invertem o sinal da energia interna. Em seguida, analisamos a dinâmica da ergotropia em qubits abertos, tanto em evoluções markovianas quanto não-markovianas. Identificamos fenômenos como o congelamento e a morte súbita da ergotropia, além de estabelecer uma relação analítica entre a variação da ergotropia e o trabalho induzido pelo ambiente. Isso fornece uma interpretação física clara para o termo adicional da primeira lei na formulação baseada em entropia. Por fim, propomos uma formulação termodinâmica baseada em ergotropia, na qual o calor é reinterpretado em termos da variação do estado passivo associado ao operador densidade que governa a dinâmica quântica. Essa abordagem permite medir a não-markovianidade de mapas unitais de forma mais geral e precisa do que o fluxo de calor baseado em entropia. Essa vantagem surge da ligação direta entre calor e entropia de von Neumann, uma propriedade garantida pela invariância sob transformações passivas. Além disso, a temperatura fora do equilíbrio se mantém naturalmente não-negativa, de maneira análoga à termodinâmica no equilíbrio. Esses resultados consolidam a formulação entrópica e ergotrópica como ferramentas úteis para identificar a não-markovianidade. Na abordagem entrópica, o ambiente exerce uma influência sobre o sistema por meio de trabalho. Nós estabelecemos um limite para esse trabalho induzido pelo ambiente, baseado no custo energético necessário para transitar entre os estados passivos inicial e final da dinâmica quântica. A partir desse custo passivo, atribuímos a ele o papel de calor. Isso nos permite estabelecer a primeira lei da termodinâmica quântica, na qual parte do trabalho é representada pela variação da ergotropia.

\vspace{15pt}

\noindent \textbf{Palavras-chave:} termodinâmica quântica; ergotropia; não-markovianidade; coerência quântica; calor.



\newpage

\noindent

\vspace*{10pt}
\begin{center}
{\LARGE\bf Abstract}\\
\vspace{15pt}
{\Large\bf FORMULATIONS OF QUANTUM THERMODYNAMICS AND APPLICATIONS IN OPEN SYSTEMS}\\
\vspace{6pt}
{\bf John Milton Zapana Choquehuanca}\\
\vspace{12pt}
{\bf Orientador:~Dr.Marcelo Silva Sarandy}\\
{\bf Coorientador:~Dr.Fagner Muruci de Paula}\\
\vspace{20pt}
\parbox{14cm}{\emph{Abstract} da Tese de Doutorado apresentada ao Programa de Pós-Graduação em Física do Instituto de Física da Universidade Federal Fluminense - UFF, como parte dos requisitos necessários à obtenção do título de Doutor em Física, área de concentração, Óptica e Informação Quântica.} 
\end{center}
\vspace*{25pt}

\noindent Quantum thermodynamics has emerged as a central field for understanding how energy conversion processes occur in microscopic systems. In these systems, effects such as coherence, entanglement, and non-Markovianity play key roles. In this thesis, we explore different ways to describe quantum thermodynamics, using two main approaches: one based on entropy and the other on ergotropy. First, we introduce a generalized approach to quantify non-Markovianity through the breakdown of monotonicity in thermodynamic functions. In this context, the entropy-based heat flow serves as a practical tool to witness and measure quantum memory in unital maps that do not reverse the sign of the internal energy. Next, we analyze the dynamics of ergotropy in open qubits under both Markovian and non-Markovian evolutions. We identify phenomena such as freezing and sudden death of ergotropy, and we establish an analytical relation between the change in ergotropy and the environment-induced work . This provides a clear physical interpretation for the additional term in the first law in the entropy-based formulation. Finally, we propose an ergotropy-based thermodynamic formulation, in which heat is reinterpreted in terms of the change of the passive state associated with the density operator governing the quantum dynamics. This approach allows one to measure non-Markovianity of unital maps more generally and accurately than the entropy-based heat flow. This advantage comes from the direct link between heat and von Neumann entropy, a property ensured by invariance under passive transformations. Moreover, the out-of-equilibrium temperature naturally remains non-negative, similarly to equilibrium thermodynamics. These results establish the entropy- and ergotropy-based formulations as useful tools for identifying non-Markovianity. In the entropy-based approach, the environment influences the system through work. We set a limit for this environment-induced work based on the energy cost required to move between the initial and final passive states of the quantum dynamics. From this passive cost, we assign it the role of heat. This allows us to establish the first law of quantum  thermodynamics, in which part of the work is represented by the change in ergotropy.
\vspace{15pt}

\noindent \textbf{Keywords:} quantum thermodynamics; ergotropy; non-Markovianity; quantum coherence; heat.



\newpage

\noindent

\vspace*{15pt}

\begin{center}

{\LARGE\bf Agradecimientos}

\end{center}

\vspace*{20pt}

\noindent En primer lugar, quiero expresar mi más profundo agradecimiento a Dios, quien ha guiado cada paso de mi vida. A Él le debo todas las oportunidades que se me han presentado, las bendiciones recibidas y la fuerza para seguir adelante incluso en los momentos más difíciles.

\noindent En segundo lugar, quiero agradecer con todo mi corazón y amor a mi hermosa familia, que ha sido el pilar fundamental en mi camino personal y académico. A mi padre Hugo, a mi madre Gabina, a mi hermano Edwin y a mis abuelitos—que en paz descansen—, todos ellos estuvieron siempre a mi lado con amor incondicional, animándome a crecer y ayudándome a superar cada desafío. Nunca me dejaron solo y siempre me motivaron a dar lo mejor de mí. Los amo profundamente y es un orgullo para mí que sean mi familia.

\noindent También quiero agradecer profundamente a mi amada, la Dra. Lauziene, a quien tuve la fortuna de conocer en Brasil. Su presencia y su amor han sido de los mayores regalos que la vida me ha dado. Con paciencia y cariño, me ayudó a reconocer y superar mis debilidades, y su apoyo constante me dio la energía y la voluntad necesarias para continuar este camino. Te amo mucho, mi gringuinha.

\noindent Quiero expresar mi sincero agradecimiento a mi orientador, el Dr. Marcelo, y a mi coorientador, el Dr. Fagner, por su dedicación, paciencia y guía. Sus explicaciones y sugerencias ampliaron de forma significativa mi comprensión de la física y contribuyeron profundamente a mi crecimiento como investigador. Las publicaciones logradas durante este período no habrían sido posibles sin su apoyo continuo; ellos son, sin duda, protagonistas esenciales detrás de esos trabajos.
\vspace{1.4cm}\\
\noindent Agradezco también al CNPq, cuya beca me permitió dedicarme por completo a esta investigación. Gracias a su apoyo, pude participar en eventos académicos importantes, ampliar mi red científica y difundir los resultados de este trabajo a una comunidad más amplia.

\noindent Finalmente, quiero agradecer a todas las personas que conocí dentro y fuera de la universidad y que, con amabilidad y disposición, me ofrecieron su ayuda, su tiempo y sus consejos, contribuyendo de distintas maneras a mi vivencia aquí en Brasil.\\
\noindent A todos ustedes, mi gratitud más sincera.


\newpage

\noindent

\vspace*{15pt}

\begin{center}

{\LARGE\bf Acknowledgments}

\end{center}

\vspace*{20pt}

\noindent First of all, I want to express my deepest gratitude to God, who has guided every step of my life. To Him I owe all the opportunities that have appeared in my life, the blessings I have received, and the strength to move forward even in the most difficult moments.

\noindent Second, I want to thank with all my heart and love my beautiful family, who has been the fundamental pillar in my personal and academic path. To my father Hugo, my mother Gabina, my brother Edwin, and my grandparents—may they rest in peace—all of them were always by my side with unconditional love, encouraging me to grow and helping me to overcome each challenge. They never left me alone and always motivated me to give the best of myself. I love them deeply, and it is a great honor for me that they are my family.

\noindent I also want to deeply thank my beloved, Dr. Lauziene, whom I was fortunate to meet in Brazil. Her presence and her love have been some of the greatest gifts that life has given me. With patience and affection, she helped me recognize and overcome my weaknesses, and her constant support gave me the energy and will needed to continue on this path. I love you very much, my gringuinha.

\noindent I want to express my sincere gratitude to my advisor, Dr. Marcelo, and to my co-advisor, Dr. Fagner, for their dedication, patience, and guidance. Their explanations and suggestions significantly expanded my understanding of physics and deeply contributed to my growth as a researcher. The publications achieved during this period would not have been possible without their continuous support; they are, without a doubt, essential main figures behind those works.
\vspace{1.4cm}\\
\noindent I also thank CNPq, whose scholarship allowed me to fully dedicate myself to this research. Thanks to its support, I was able to participate in important academic events, expand my scientific network, and share the results of this work with a wider community.

\noindent Finally, I want to thank all the people I met inside and outside the university who, with kindness and willingness, offered me their help, their time, and their advice, contributing in different ways to my experience here in Brazil.\\
\noindent To all of you, my most sincere gratitude.

\cleardoublepage        
\vspace*{\fill}
\begin{flushright}
\textit{Vive con orgullo y la frente en alto,\\
no permitas que tus miedos y debilidades te alejen de tus objetivos,\\
mantén tu mente enfocada y tu corazón ardiente;\\
no importa que pase, sigue avanzando,\\
no te rindas a pesar de haberte caído,\\
recuerda que el tiempo no espera a nadie,\\
ni te da compañía ni comparte tus penas,\\
no te lamentes, eres un futuro pilar,\\
tu deber es florecer y crecer;\\
por favor; florece, crece y madura,\\
cuando sea tiempo, te convertirás en pilar,\\
tú puedes, confía plenamente en ello.} \\
\vspace{0.5cm}
--- John Milton
\end{flushright}
\vspace*{1cm}
\newpage


\newpage
\phantomsection
\addcontentsline{toc}{chapter}{Summary}
\tableofcontents

\newpage
\phantomsection
\addcontentsline{toc}{chapter}{List of Figures}
\listoffigures


\mainmatter
\begin{chapter}{INTRODUCTION}
\label{intro}

Modern technology increasingly depends on materials whose behaviors and properties are governed by quantum phenomena such as coherence and entanglement. This new kind of technology, known as second-generation quantum technology, includes applications ranging from quantum computers to quantum batteries and sensors. Its development has a strong impact on both the public and private sectors, supported by research programs with large financial investments~\cite{alexia2022,metzler2023}. In 2025, the United Nations declared the International Year of Quantum Science and Technology, highlighting its relevance in fields such as health, agriculture, communications, and security, and encouraging the creation of consortia and specialized startups~\cite{quantuminsider2023}. Efforts to advance these technologies have intensified through initiatives like the Quantum Energy Initiative (QEI), which since 2022 has promoted strategies to reduce the physical costs of emerging quantum devices~\cite{QEI2025}. These advances reflect the need for solid theoretical tools in quantum thermodynamics to understand and optimize energy processes in quantum systems.
\vspace{0.2cm}\\
\noindent Classical thermodynamics, developed in the 19th century in the context of heat engines~\cite{callen1985}, studies the relations between different forms of energy and processes of energy transfer. Although traditionally applied to macroscopic systems, in recent decades there has been growing interest in extending these concepts to nanoscale and atomic systems, giving rise to quantum thermodynamics~\cite{prasad2016}. This field explores the limits and potential of quantum devices operating out of equilibrium, recognizing the importance of thermodynamic laws in computational devices, where the intrinsic relation between information, heat, and work is already well established~\cite{gold20166,binder2019}. For this reason, quantum thermodynamics provides a strong foundation for the idea that information is physical, as illustrated by Landauer’s principle \cite{Esposito2011,Parrondo2015} or the von Neumann entropy, which is the quantum version of Shannon’s entropy and acts as a quantifier of information, linking thermodynamics with information theory. Moreover, information theory offers powerful tools to study and characterize quantum features such as entanglement and coherence \cite{deffner2019}.
\vspace{0.2cm}\\
\noindent In this regime, non-Markovian dynamics play a central role in the evolution of systems. One of the main challenges is to rigorously define thermodynamic quantities, such as work and heat, which are not standard Hermitian observables and therefore require alternative formulations~\cite{gemmer,peter,alicki}. A correct definition of these quantities is essential to analyze nonequilibrium processes and to understand how quantum resources like coherence and correlations affect the efficiency of quantum devices.
\vspace{0.2cm}\\
\noindent To address these issues, it is necessary to consider the interaction of the system with its environment, which can be weak or strong. Environments are often modeled as thermal baths of bosons (photons, phonons, quantum oscillators) or other large reservoirs with an extremely high number of degrees of freedom. Because of this complexity, their microscopic description is highly intricate, making the theory of open quantum systems indispensable. This theoretical framework is widely used in quantum optics, condensed matter physics, atomic physics, quantum information, and quantum thermodynamics, providing tools to understand how the environment influences the properties and dynamics of a system.
\vspace{0.2cm}\\
\noindent An important phenomenon in open systems is decoherence, which appears as the loss of quantum superpositions and the emergence of classical behavior. Decoherence not only limits the performance of quantum devices but also poses a central challenge for the implementation of technologies such as quantum computing, metrology, and cryptography, where fine control of quantum states is essential. To describe the evolution of these systems, different theoretical approaches are used: on one hand, quantum maps provide a discrete description applicable to both weak and strong interactions; on the other, master equations and stochastic differential equations allow a continuous description of the dynamics, although they require specific approximations.
\vspace{0.2cm}\\
\noindent Within this framework, a key development is the Gorini-Kossakowski-Lindblad-Sudarshan (GKLS) master equation, which describes the dynamics of systems weakly coupled to a very large environment. The dynamics captured by this equation are Markovian, characterized by an irreversible loss of information from the system to the environment. However, recent progress seeks to extend the theory to the strong coupling regime, where non-Markovian effects become prominent \cite{ Thomas2018,Huang2022,Tiwari2024}. In this case, the dynamics allow the environment to return information to the system, leading to complex and temporary behaviors that the GKLS equation cannot fully capture. This opens the way for the study of new quantum effects and highlights the need for new theoretical tools to correctly describe energy transfer and the evolution of quantum resources in open systems \cite{Gorini1976, Breuer:Book, rivas2014}.
\vspace{-0.7cm}\\
\noindent Several proposals have sought to reformulate the laws of quantum thermodynamics to address the challenges that arise in nonequilibrium systems. At the most fundamental level, the first law of quantum thermodynamics, formulated by Alicki~\cite{alicki}, describes internal energy as the expectation value of the system’s Hamiltonian, associating work with controlled changes in the energy levels via time-dependent Hamiltonians and identifying heat as the energy exchanged with the environment. Complementarily, the second law of quantum thermodynamics is expressed in terms of entropy production \cite{Landi2021}, which sets the direction of irreversible processes and serves as an essential tool to extend thermodynamic analysis to nonequilibrium and stochastic regimes. In this context, modern reformulations aim to consistently integrate quantum coherence, correlations, and non-Markovian effects into the definitions of work, heat, and entropy, in order to clarify the boundaries among these quantities and provide a unified framework for the study of open quantum systems~\cite{deffner2019, binder}. \vspace{0.1cm}
\vspace{-0.07cm}\\
\noindent Recent entropy-based formulations divide internal energy into a part associated with entropy changes (heat) and another independent of them (work). This naturally connects with quantum information theory and allows interpreting coherence and correlations as thermodynamic resources~\cite{alipour1, alipour2,Ahmadi2020,streltsov,cramer,Binder:18}. These reformulations have also shown that work may include components induced by the environment even when the Hamiltonian remains constant. This not only clarifies the limits between heat and work in nonequilibrium processes~\cite{Landi2021,Ahmadi2020, bernardo} but also emphasizes the need for a theoretical framework able to integrate coherence, correlations, and non-Markovian effects in order to fully understand the thermodynamics of open quantum systems. Moreover, such formulations have demonstrated that heat flows, coherence, and von Neumann entropy can act as indicators of non-Markovianity~\cite{titas, passo, haseli}, showing the close relationship between system dynamics and environmental memory.
\vspace{0.2cm}\\
\noindent Within this conceptual framework, the notion of ergotropy emerges, defined as the maximum extractable energy from a quantum state by means of unitary transformations~\cite{kosloff}. Ergotropy has become a central resource in quantum thermodynamics, with applications such as thermal engines, quantum batteries, and nonequilibrium energy control protocols~\cite{Allahverdyan:04,Rossnagel:16, Maslennikov:19,Joshi:22,Zhu:23}. Interaction with the environment adds further complexity, as phenomena such as decoherence and dissipation modify the balance between heat and work, even generating contributions like environment-induced work~\cite{vallejo}.
\vspace{0.2cm}\\
\noindent This thesis contributes to the study of quantum thermodynamics in open and non-Markovian systems through a set of theoretical developments and applications. Chapters 2 and 3 establish the theoretical foundations, addressing the dynamics of closed and open systems, quantum maps, master equations, and the thermodynamic concepts of work, heat, entropy, ergotropy, and coherence. Chapters 4, 5, and 6 present the results obtained during the PhD, published in Physical Review A: Chapter 4 applies an entropy-based formulation to heat flows in a non-Markovian qubit; Chapter 5 studies the relation between environment-induced work and ergotropy, analyzing ergotropy dynamics in open systems and describing freezing and sudden death phenomena; and Chapter 6 proposes an ergotropy-based formulation that connects heat with entropy variation and work with ergotropy variation, defining positive temperature out of equilibrium and an indicator of non-Markovianity in unitary maps. Finally, Chapter 7 summarizes the results, presents the conclusions, and offers perspectives for future research.

\end{chapter}

\begin{chapter}{QUANTUM DYNAMICS: FROM CLOSED TO OPEN SYSTEMS}
\label{cap2}

Quantum mechanics was originally developed to describe the theoretical behavior of closed quantum systems. However, the systems studied in laboratories are never perfectly isolated from their surroundings. Therefore, the study of open quantum systems becomes essential to understand more realistic scenarios, where the interaction with the environment must be considered.\\ This theory of open quantum systems is particularly relevant in the field of quantum thermodynamics, as it provides the necessary tools to explore how classical thermodynamic principles behave when quantum effects become significant. A fundamental tool offered by this framework is the Gorini–Kossakowski–Lindblad–Sudarshan (GKLS) master equation, which establishes a formal approach to describe the time evolution of a quantum system weakly coupled to a large external environment.\\
The theoretical foundations presented in this chapter are well-established and widely accepted, and they serve as a basis for the discussions in the following sections. 

\section{The concept of Density Operator}

\subsection{Pure state}

We know that a state vector $\ket{\psi_s(t)}$ contains information about the state of a physic system $S$. Based on this idea, the density operator $\hat{\rho}_s(t)$ for a pure state $\ket{\psi_s(t)}$ is given by, \begin{equation}\label{Eq:rhopuro}
    \hat{\rho}_s(t)=\ket{\psi_s(t)}\bra{\psi_s(t)}.
\end{equation} The matrix representation of $\hat{\rho}_s(t)$ in a specific basis is referred to as the density matrix. If the pure state is expressed as a superposition of basis states $\{\ket{k}\}$, we have $\ket{\psi_s(t)} = \sum_{k=0}^{d-1} c_k(t) \ket{k}$ where $d$ is the dimension of the system’s Hilbert space $\mathcal{H}_s$, and $c_k(t) \in \mathbb{C}$ is the amplitude coefficient associated with the basis state $\ket{k}$. In this basis, the density matrix becomes  \cite{maximilian2007},
\begin{equation}\label{matrixrho}
    \hat{\rho}_s(t)=\sum^{d-1}_{\{k,\,k^\prime\}=0} c_k(t) c^*_{k^\prime}(t)\ket{k}\bra{k'}.\\[5pt]
\end{equation}
The diagonal elements of the density matrix, $k=k^\prime$, represent the populations in a  basis, while the off-diagonal elements, $k\neq k^\prime$, indicate quantum coherence in that same basis. 

\noindent Coherence must always be interpreted with respect to a specific reference basis. It is worth emphasizing that a diagonal density matrix does not imply that the system lacks quantum properties; it simply means that, in that basis, there is no coherence between the basis states. Coherence may still be present in other representations \cite{maximilian2007}.

\subsection{Mixed state}

A pure state describes a fully known system, associated with an ensemble consisting of a single element, $\{1, \ket{\psi_s(t)}\}$. However, there are also situations in which the information about the system is incomplete. In such cases, we only know that the system is in a set of pure states, each associated with a classical probability $p_{k'}$ that satisfies the conditions $0 < p_{k'}(t) < 1$ and $\sum_{k'=0}^{d'-1} p_{k'}(t) = 1$. In other words, we are dealing with a statistical mixture of pure states, which are not necessarily orthogonal, and which are associated with an ensemble of $d'-$elements, $\{p_{k'}(t), \ket{\psi_{k'}(t)}\}$ \cite{lidar2019}. This type of state is known as a mixed state and is described by
\begin{equation}
\hat{\rho}_s(t)=\sum^{d'-1}_{k'=0} p_{k'}(t) \hat{\rho}_{k'}(t)=\sum^{d'-1}_{k'=0} p_{k'}(t) \ket{\psi_{k'}(t)}\bra{\psi_{k'}(t)}. \\[5pt]
\end{equation}
It is essential to distinguish between a statistical mixture and a quantum superposition. A mixed state reflects classical ignorance about which pure state the system actually occupies. In contrast, a superposition $\ket{\psi_s(t)} = \sum_{k=0}^{d-1} c_k(t) \ket{k}$ describes a single pure state that simultaneously involves all basis states. The density matrix of a pure state contains coherent terms of the form $c_k(t) c_{k'}^*(t)$, which indicate quantum superposition. On the other hand, a mixed state has only $d'$ diagonal elements associated with classical probabilities $p_{k'}(t)$, when expressed in the basis generated by the states $\ket{\psi_{k'}(t)}$.

\subsection{Properties  of the density operator}
Density matrix has the following properties \cite{lidar2019}:
\begin{itemize}
    \item \textbf{Unit trace} $\rightarrow$ This property can easily be seen  by following calculation
    \begin{equation*}
        \text{tr}[\hat{\rho}_s(t)]=\sum_{k'=0}^{d'-1} p_{k'}(t)\,\text{tr}[\ket{\psi_{k'}(t)}\bra{\psi_{k'}(t)}]= \sum_{k'=0}^{d'-1} p_{k'}(t)=1
    \end{equation*}
    \item \textbf{Hermiticity} $\rightarrow$  All density operator is Hermitian. This property can easily be seen by following calculation,
    \begin{equation*}
        \hat{\rho}_s^\dagger(t) = \sum_{k'=0}^{d'-1} p^\dagger_{k'}(t) \big(\ket{\psi_{k'}(t)}\bra{\psi_{k'}(t)}\big)^\dagger=\sum_{k'=0}^{d'-1} p_{k'}(t) \ket{\psi_{k'}(t)}\bra{\psi_{k'}(t)}=\hat{\rho}_s(t)
    \end{equation*}  
    \item \textbf{Positive definite} $\rightarrow$ For all vectors $\ket{v} \in \mathcal{H}$, the positivity of the density operator $\hat{\rho}_s(t)$ is given by 
    \begin{equation}
    \braket{v|\hat{\rho}_s|v}=\sum_{k'=0}^{d'-1} p_{k'}\braket{v\mid \psi_{k'}}\braket{\psi_{k'} \mid v}=\sum_{k'=0}^{d'-1} p_{k'}\left|\braket{ v \mid \psi_{k'}}\right|^{2} \geq 0 .
\end{equation}
The term $p_{k'}$ is non-negative as a result of its probabilistic nature.
\end{itemize}
For a pure state, the identity $\hat{\rho}_s^2(t) = \hat{\rho}_s(t)$ holds, which is not valid for mixed states. Another criterion for distinguishability is the called purity, defined as $P = \text{tr}[\hat{\rho}_s^2(t)]$, for which we have $P = 1$ in the case of a pure state and $P < 1$ for a mixed state. 
\vspace{0.2cm}\\
One of the main advantages of the density operator $\hat{\rho}_s(t)$ is that it allows us to compute all possible measurement outcomes on the system for any observable via the trace operation. Then, using the general quantum state that is the mixed state, the expectation value of any observable $\hat{O}$ is defined as \begin{equation}\label{expecvalue}
    \braket{\hat{O}}= \text{tr}\big[\hat{\rho}_s(t) \hat{O}\big] = \sum^{d'-1}_{k'=0} p_{k'}(t) \braket{\psi_{k'}(t) | \hat{O} | \psi_{k'}(t)},
\end{equation}
while in the case of a pure state, this reduces to $ \braket{\hat{O}}=\braket{\psi_s(t)|\hat{O}| \psi_s(t)}$.

\section{Closed System Dynamics}

In classical systems, time evolution is formulated through differential equations. In quantum theory, the first evolution equation for a quantum system is the well known Schrödinger equation, given by \cite{nakahara2008} \begin{equation}\label{schrodinger}
    \mathrm{i} \hbar \frac{d}{ dt}\ket{\psi_s(t)}=\hat{H}_s(t)\ket{\psi_s(t)}.
\end{equation}
This equation describes the behavior of a closed quantum system, where the pure state vector $\ket{\psi_s(t)} \in \mathcal{H}_s$. Here, $\mathcal{H}_s$ represents the Hilbert space of the system,  and $\hat{H}_s(t)$ is the Hamiltonian operator that defines the energy of the system $S$. By definition, a closed system does not exchange any information with another system, commonly referred to as the environment. This information may correspond to energy or matter. Theoretically, any closed system evolves under a unitary dynamics generated by  $\hat{H}_s(t) \in \mathcal{H}_s$.
From Eq. \eqref{schrodinger}, one can derive the evolution equation for the system's density operator $\hat{\rho}_s(t)$. This equation is known as the Liouville–von Neumann equation~\cite{manenti}, \begin{equation} \label{eqvonneuman}
    \frac{d}{d t} \hat{\rho}_s(t)=-i \hbar[\hat{H}_s(t), \hat{\rho}_s(t)].
\end{equation}
The von Neumann equation is valid for both pure and mixed states. It is important to note that the Hamiltonian used here is a Hermitian operator. Since it depends on time, this indicates that the system is under external control; in other words, it is controllable.
\vspace{0.1cm}
The formal solution of Eq.~\eqref{eqvonneuman} is given by \begin{equation}\label{solutionvonneuman}
    \hat{\rho}_s(t)=\hat{V}_s(t) \hat{\rho}_s(t_0=0)\hat{V}^\dagger_s(t),
\end{equation}

\noindent where $\hat{V}_s(t)$ is the time evolution operator. This operator satisfies the property  $\hat{V}^{-1}_s(t) =\hat{V}(t)$ where $\hat{V}^{-1}_s(t)$ represents the inverse operator of $\hat{V}_s(t)$. As a result, the evolution of a closed system is invertible, and the von Neumann equation describes a reversible process. The form of the operator $\hat{V}_s(t)$ depends on the properties of the Hamiltonian. For a time-independent Hamiltonian, such that the system is conservative, the unitary operator takes the form, \begin{equation} \label{unitarycte}
    \hat{V}_s(t)=\mathrm{e}^{-\mathrm{i} \hat{H}_s t / \hbar}.
\end{equation}
On the other hand, when the Hamiltonian depends on time, meaning the system is non-conservative, the evolution operator is written as \begin{equation}\label{unitarynocte}
    \hat{V}_s\left(t\right)=\hat{\mathcal{T}} \exp \left(\frac{-i}{\hbar} \int_{t_0=0}^{t} \hat{H}_s\left(t^{\prime}\right) d t^{\prime}\right),
\end{equation}
where $\hat{\mathcal{T}}$ is the time-ordering operator, which allows the general unitary operator in Eq. \eqref{unitarynocte} to be expressed as a series expansion known as the Dyson series (see Ref.\cite{manenti}). 

\section{Two-Level Systems}

Many physical systems can be modeled as a two-level system~\cite{Band2012}. A prime example is spin-1/2 particle systems. The polarization of photons, where right and left circularly polarized correspond to spin-up and spin-down states, provides another example. Additionally, an atom with two energy levels coupled by an electromagnetic field near resonance can also be treated as a two-level system. To achieve a better understanding of a two-level system, we begin from a classical context.
\vspace{0.2cm}\\
\noindent The magnetic dipole moment $\vec{\mu}$ of a charged particle is related to its angular moment $\vec{S}_\text{spin}$ and its charge $q$ and mass $m\,$ through,
\begin{equation}\label{eq:2-muclasico}
\vec{\mu}=\frac{q}{2m}\,\vec{S}_\text{spin}.
\end{equation}
In the quantum realm, the Eq.\eqref{eq:2-muclasico} is modified by a correction factor $g$ specific to each particle, 
\begin{equation}\label{eq:2-mucuantico}
\vec{\hat{\mu}}=\frac{g\,p}{2m}\,\vec{\hat{S}}_\text{spin}=\gamma_\mu\,\vec{\hat{S}}_\text{spin}.
\end{equation}  When a particle with a dipole moment is placed in a magnetic field $\vec{B}(t)$, the Hamiltonian of the system can be expressed as $\hat{H_s}(t)=-\vec{\hat{\mu}}.\vec{B}(t)= -\gamma_\mu \,\vec{\hat{S}}_\text{spin}.\vec{B}(t)\,$
where $\gamma_\mu\,$ is the gyromagnetic constant. If we include $\gamma_\mu$ within the components of $\vec{B}(t)$, the Hamiltonian can be written as $\hat{H_s}(t)=-\vec{h}(t)\,.\,\vec{\hat{S}}_\text{spin}$ where $\vec{h}(t)\,$ is the new magnetic field whose components $\big(h_x(t), h_y(t), h_z(t)\big)$ are real for $\hat{H_s}(t)$ to be Hermitian. This Hamiltonian can be written in terms of the Pauli matrices,
\begin{equation}\label{hamiltonianomagnetico}
    \hat{H}_s(t) = -\frac{\hbar}{2} \,\, \vec{h}(t)\cdot\vec{\hat{\sigma}}=-\frac{\hbar}{2} \,
    \begin{pmatrix}
h_z(t) & h_x(t) - i h_y(t) \\
h_x(t) + i h_y(t) & -h_z(t)
\end{pmatrix},
\end{equation}
where $\vec{\hat{S}}=-(\hbar/2)\,\vec{\hat{\sigma}}\,$ with el vector de operadores de Pauli $\vec{\hat{\sigma}}=( \hat{\sigma}_x, \hat{\sigma}_y, \hat{\sigma}_z),$\\
\begin{equation}\label{operadorespauli}
    \hat{\sigma}_x=\begin{pmatrix}
        0&1\\
        1&0
    \end{pmatrix},\quad \quad
    \hat{\sigma}_y=\begin{pmatrix}
        0&-i\\
        i&0
    \end{pmatrix},\quad \quad
    \hat{\sigma}_z=\begin{pmatrix}
        1&0\\
        0&-1
    \end{pmatrix},
\end{equation}
\noindent It is important to note that this Hamiltonian $\hat{H}_s(t)$ can also be obtained from an algebraic and quantum mechanical perspective, where it is shown that any Hermitian operator acting on a two-dimensional Hilbert space can be represented as $\hat{H}_s=a_0 \hat{\mathbb{I}}+a_1 \hat{\sigma}_x+a_2 \hat{\sigma}_y+a_3 \hat{\sigma}_z$ where  $a_0,a_1,a_2$ and $a_3$ are real functions. The term proportional to $\mathbb{I}$ adds only a global phase to the quantum state. Therefore, it is omitted in the Hamiltonian to focus on the terms that truly determine the system's evolution. 

\noindent For the case of a constant $\hat{H}_s$, the associated unitary evolution operator can be physically interpreted as inducing a rotation of the spin state around a fixed axis in three-dimensional space. The direction of this axis is determined by the form of the Hamiltonian, and the resulting motion is known as spin precession. In this regime, the spin vector keeps circling around $\vec{h}$. However, in a more realistic situation, the spin state gradually aligns in the same direction of the external magnetic field.
\vspace{0.2cm}\\
\noindent This damping process arises due to interactions with the environment, which are not captured by unitary Hamiltonian dynamics alone. To describe such non-unitary evolution, it is necessary to include the influence of the environment explicitly, typically via open quantum system. If the Hamiltonian becomes time-dependent $\hat{H}_s(t)$, the axis and rate of rotation also vary in time, leading to more complex but still unitary dynamics.

\subsection{Bloch Sphere}

The qubit is the simplest system in quantum theory. It is described by a two-dimensional complex Hilbert space $\mathcal{H}_s$. Within this space, one can choose a pair of orthonormal quantum states that form a basis~\cite{Nielsen-Book},
\begin{equation}\label{basecomputacional}
    \ket{0}=\begin{pmatrix}
1 \\
0
\end{pmatrix}, \quad \ket{1}=\begin{pmatrix}
0\\
1
\end{pmatrix}.
\end{equation} 
These two states are also known as the computational basis. Based on the principle of superposition, any qubit state can be written as
\begin{equation}\label{superpositionqubit}
    \ket{\psi_s}=\alpha_0 \ket{0}+\alpha_1\ket{1}
\end{equation} where $\alpha_0$  and $\alpha_1$ are complex numbers satisfying the  condition $|\alpha_0|^2+|\alpha_1|^2=1$. A generic pure state of a qubit can also be parametrized as
 \begin{equation}
    \ket{\psi_s}=\cos \frac{\theta}{2}\ket{0}+e^{i \phi} \sin \frac{\theta}{2}\ket{1},
\end{equation} where $\theta \in [0, \pi]$ and $\phi \in [0, 2\pi)$. In this representation, it is assumed that any global phase $e^{i\eta}\ket{\psi_s}$ is physically irrelevant, and therefore $\ket{\psi_s}$ and $e^{i\eta}\ket{\psi_s}$ describe the same physical state. If we adopt spherical coordinates, we can identify $\theta$ as the polar angle and $\phi$ as the azimuthal angle. The parameter $\phi$ is often referred to as the relative phase of the qubit.

\begin{figure}[htbp]
    \centering
\includegraphics[width=0.3\textwidth]{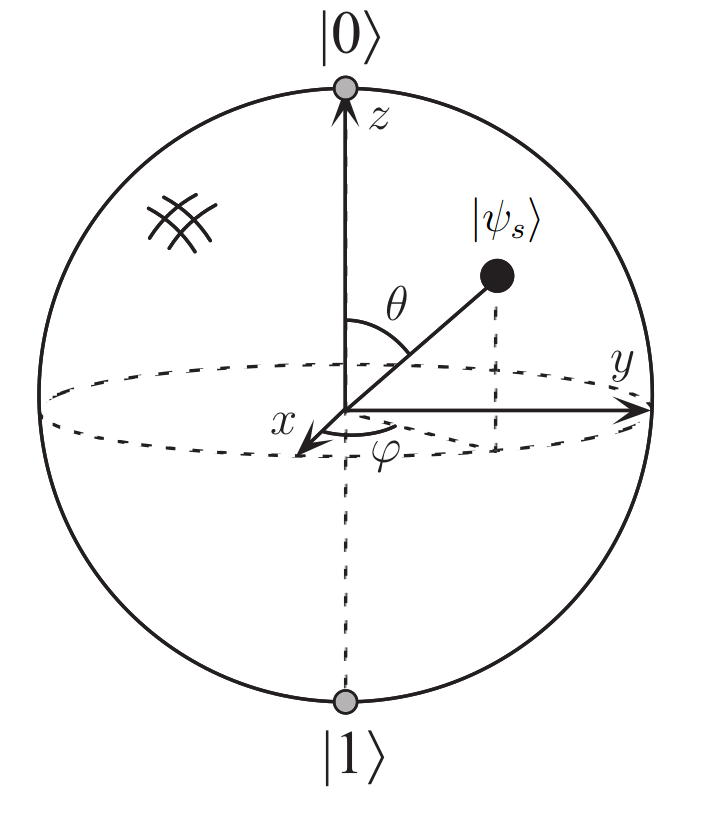}
    \caption{Representation of a qubit on the Bloch sphere. Reproduced from Nielsen and Chuang ~\cite{Nielsen-Book}.}
    \label{fig:mi_figura}
\end{figure}
\noindent Starting from this generic state expression, one can construct a geometric representation of the qubit in the form of a sphere, known as the Bloch sphere. This visualization provides a powerful tool for analyzing how quantum maps~(will be discussed later) act on the qubit state. Mathematically, the Bloch sphere representation is given by 
\begin{equation}\label{densityqubit}
    \hat{\rho}_s(t)=\frac{1}{2} \left(\hat{\mathbb{I}}+\vec{r}(t)\,.\, \vec{\hat{\sigma}}\right)=\frac{1}{2}\begin{pmatrix}
1+r_z(t) & r_x(t)-i r_y(t) \\
r_x(t)+i r_y(t) & 1-r_z(t)
\end{pmatrix},
\end{equation}
where $\vec{\hat{\sigma}}$ denotes the vector of Pauli matrices and $\vec{r}(t) = (r_x(t), r_y(t), r_z(t))$ is known as the Bloch vector. The foundation of this expression lies in the fact that any single qubit density operator $\hat{\rho}_s(t)$ is Hermitian and provides a successful representation of a two-level quantum system. It can be expressed as a linear combination of the identity and the three Pauli matrices $\hat{\rho}_s(t) = b_0(t) \hat{\mathbb{I}} + b_1(t) \hat{\sigma}_x + b_2(t) \hat{\sigma}_y + b_3(t) \hat{\sigma}_z$. The coefficients $b_i(t)$ can be determined using the properties of the density operator.\\The Bloch representation becomes particularly convenient when expressed in spherical coordinates, since the purity of the state determines the magnitude of the Bloch vector. Pure states lie on the surface of the Bloch sphere ($|\vec{r}|=r=1$), while mixed states lie inside it ($r<1$). In fact, from the purity relation $\text{tr}[\hat{\rho}_s^2]=\frac{1}{2}(1+r^2)$, we obtains the equation of a sphere in three dimensions $r^2=r_x^2+r^2_y+r_z^2$, which naturally justifies the parametrization of the qubit state in terms of $(r,\theta,\phi)$.

\noindent The components of the Bloch vector in spherical coordinates are given by,
  \begin{equation}\label{vectorblochesfrico}
    r_x(t)=r(t)\cos \phi(t)  \sin \theta(t),\quad r_y(t)=r(t)\sin \phi(t)  \sin \theta(t), \quad r_z(t)=r(t)\cos \theta(t),
\end{equation} where $r(t)$ is the magnitude of the Bloch vector. The positivity condition of $\hat{\rho}_s$ implies that $0 \leq r(t) \leq 1$. Furthermore, $r(t)$ provides insight into the eigenvalues $r_n(t)$ with $n=\{0,1\}$ of the single qubit density operator $\hat{\rho}_s(t)$, which are necessary for its spectral decomposition, \begin{equation} \label{eigenvaluesqubit}
    r_n(t)=\frac{1\pm r(t)}{2}.
\end{equation} An additional advantage of the parameter $r(t)$ is that it quantifies the purity of the qubit state, where a pure state corresponds to $r(t) = 1$, which lies on the surface of the Bloch sphere, while a maximally mixed state corresponds to $r(t) = 0$, located at the center of the sphere.

\section{Open system Dynamics}

\subsection{Quantum map}

We know that the evolution of a quantum system as a whole is unitary. However, the situation changes when we focus on subsystems. In practice, these subsystems unavoidably interact with external degrees of freedom, leading to dynamics that are generally non-unitary. To adequately describe such open quantum systems, we resort to the formalism of quantum maps, which provides an effective framework for modeling their evolution. Beyond its foundational role in quantum information theory, this formalism is also essential in quantum thermodynamics, where it allows for the consistent description and analysis of noisy processes from an energetic perspective. In this context, noise is not merely a source of decoherence, but also a map through which energy, entropy, and information flow between the system and its environment.
\begin{figure}[H] 
\centering
\includegraphics[width=0.47\columnwidth]{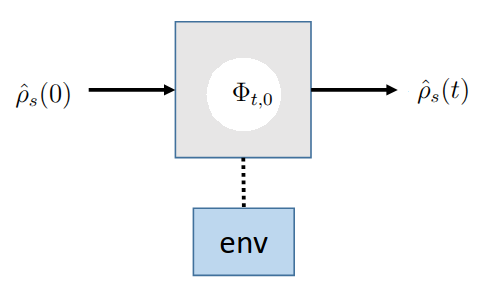}
\caption{(Color online). Representation of the evolution of an open system through a map $\Phi_{t,0}$, which transforms the state $\hat{\rho}_{s}(0)$ into $\hat{\rho}_{s}(t)$. The map includes environmental influences such as decoherence, dissipation, or noise. Adapted from Ref.
 \cite{Preskill2020}.}
\label{fig:mapaCPTP}
\end{figure}

\noindent A quantum map $\Phi_{t,0}$ is a linear transformation that maps one quantum state into another quantum state and captures any dynamical change in the state of a quantum system (see Fig. \ref{fig:mapaCPTP}). Formally,
\begin{equation}\label{eq2:map}
\Phi_{t,0} :\, \hat{\rho}_s(0) \mapsto \hat{\rho}_s(t)  \quad \leftrightarrow \quad \hat{\rho}_s(t)= \Phi_{t,0}\big[\hat{\rho}_s(0)\big].
\end{equation} If one wants to give a physical motivation to the definition of quantum maps, it is necessary to impose a set of constraints on Eq.~\eqref{eq2:map}, such that $\Phi_{t,0}$ describes a physical process leading from a well-defined physical state to other equally valid states. 

\noindent The following constraints must be satisfied~\cite{lidar2019}:
\begin{itemize}
    \item[\textbf{(1)}] \textbf{Trace preservation} - necessary to preserve the physical validity of the quantum state,  
    \begin{equation}\label{eq2:traza}
    \text{tr}\big[\,\Phi_{t,0}[\hat{\rho}_s(0)]\,\big]=\text{tr}\big[\hat{\rho}_s(t)\big]=1.
    \end{equation}
    \item[\textbf{(2)}] \textbf{Convexity Preservation} - necessary to ensure that the evolution of a quantum system remains consistent with classical notions of uncertainty. If a system is initially in a statistical mixture of states, represented as $ \hat{\rho}_s(0) = \sum_{k'} p_{k'} \hat{\rho}_{k'}$, with  $ p_{k'} \geq 0 $ and $ \sum_{k'} p_{k'}(t) = 1 $, then the quantum map $\Phi_{t,0}$ must evolve the mixture as,
    \begin{equation}\label{eq2:convexity}
        \Phi_{t,0}[\hat{\rho}(0)] = \Phi_{t,0}[\,\sum_{k'} p_{k'} \hat\rho_{k'}\,]=\sum_{k'} p_{k'} \Phi_{t,0}[\hat\rho_{k'}].
    \end{equation}
\item[(3)] \textbf{Complete positivity-}  First, a quantum map $\Phi$ is said to be positive if it maps positive operators to positive operators:
\begin{equation}
    \hat{\rho}_s(0) \geq 0 \quad \Rightarrow \quad \Phi_{t,0}\big(\hat{\rho}_s(0)\big) \geq 0.
\end{equation}
To ensure that the map remains physically valid even if the system is entangled with
an ancillary system of arbitrary dimension, complete positivity will be necessary
\begin{equation}
    (\Phi_{t,0} \otimes \mathbb{I}_e)\big(\hat{\rho}_{se}(0)\big) \geq 0, \quad \forall \hat{\rho}_{se}(0) \geq 0.
\end{equation}
\end{itemize}

\noindent Therefore, all quantum map that satisfies these three conditions will be called a completely positive trace-preserving (CPTP) map.

\section{Properties of CPTP Maps}

CPTP maps present certain properties that make them essential tools in both quantum information and quantum thermodynamics~\cite{manzano2018}. These are the following:

\begin{itemize}
    \item [\textbf{(a)}] CPTP maps induce contractivity with respect to certain metrics in the space of density operators. If we consider the trace distance metric $D\big(\hat{\rho}_{s_0}, \hat{\sigma}_{s_0}\big)$, contractivity is expressed as
    \begin{equation}
     D(\hat{\rho}_{s_0}, \hat{\sigma}_{s_0}) \geq D(\Phi_{t,0}(\hat{\rho}_{s_0}), \Phi_{t,0}(\hat{\sigma}_{s_0})),
    \end{equation}
    where 
    \begin{equation}
     D(\hat{\rho}_{s_0}, \hat{\sigma}_{s_0}) = \frac{1}{2} \| \hat{\rho}_{s_0} - \hat{\sigma}_{s_0} \|_1,   
    \end{equation}
    with $\| \hat{A} \|_1 := \text{tr} \left[ \sqrt{\hat{A}^\dagger \hat{A}} \right]$. This inequality indicates that, if we subject two quantum states, $\hat{\rho}_{s_0}=\hat{\rho}_s(0)$ and $\hat{\sigma}_{s_0}=\hat{\sigma}_s(0)$, to the same physical evolution allowed by $\Phi_{t,0}$, the ability to distinguish them does not increase, reducing the accessible information. 
    \vspace{0.2cm}\\
    \\
    \\
    Concerning the metric based on quantum fidelity $F(\hat{\rho}_{s_0}, \hat{\sigma}_{s_0})$, which quantifies how similar two states are in terms of their superposition, contractivity implies that the inequality
    \begin{equation}
        F(\hat{\rho}_{s_0}, \hat{\sigma}_{s_0}) \leq F\big(\Phi_{t,0}(\hat{\rho}_{s_0}), \Phi_{t,0}(\hat{\sigma}_{s_0})\big)
    \end{equation}
    holds with $F(\hat{\rho}_{s_0}, \hat{\sigma}_{s_0}) = \left( \text{tr} \left[ \sqrt{ \sqrt{\hat{\rho}_{s_0}} \, \hat{\sigma}_{s_0} \, \sqrt{\hat{\rho}_{s_0}} } \, \right] \right)^2$. This means that fidelity can increase under the action of a CPTP map, making the states more similar to each other as they evolve.  Both metrics are consistent with each other.

\item [\textbf{(b)}] CPTP maps induce monotonicity. This property can be understood as a generalization of contractivity, since it is not limited to symmetric metrics between states, but also applies to more general functionals that are asymmetric, such as relative entropy $S(\hat{\rho}_{s_0}||\hat{\sigma}_{s_0})$ or mutual information $I_{\hat{\rho}_{AB}}(A:B)$. If we consider the quantum relative entropy, an asymmetric distance between $\hat{\rho}_{s_0}$ and $\hat{\sigma}_{s_0}$, defined as:
\begin{equation}
S(\hat{\rho}_{s_0} \| \hat{\sigma}_{s_0}) = \text{tr} \left[ \hat{\rho}_{s_0} \left( \ln{\hat{\rho}_{s_0}} - \ln{\hat{\sigma}_{s_0}} \right) \right],
\end{equation}
the property of monotonicity is reflected in the following inequality, known as the Uhlmann inequality:
\begin{equation}
S\big(\Phi_{t,0}(\hat{\rho}_{s_0}) \big\| \Phi_{t,0}(\hat{\sigma}_{s_0}) \big) \leq S(\hat{\rho}_{s_0} \| \hat{\sigma}_{s_0}).
\end{equation}
This inequality acquires a significant meaning in the context of the second law of quantum thermodynamics. In this setting, if we interpret $\hat{\sigma}_{s_0}$ as the thermal equilibrium state $\hat{\rho}_s^{\, \mathrm{th}} = \frac{1}{Z} e^{-\beta \hat{H}_s}$ with $\beta = \frac{1}{k_B T_e}$, then the relative entropy quantifies how far the system is from equilibrium. Therefore, the inequality implies that
\begin{equation}
    \frac{d}{dt} S\big( \hat{\rho}_s(t) \, \| \, \hat{\rho}_s^{\,\text{th}} \big) \leq 0.
\end{equation}
If the system's evolution is governed by a CPTP map, it cannot spontaneously move away from thermal equilibrium. To violate this inequality, one must either inject an external source of work or energy, or enforce a non-Markovian dynamics.

    \item [\textbf{(c)}] CPTP maps that preserve the identity operator, 
    \begin{equation}
     \Phi_{t,0}(\hat{\mathbb{I}}) = \hat{\mathbb{I}}   
    \end{equation}
    are called unital maps. Physically, this means that the maximally mixed state remains invariant under the evolution, that is, $\Phi_{t,0}(\hat{\mathbb{I}}/d) = \hat{\mathbb{I}}/d$, where $d$ is the dimension of the system. A key feature of unital maps is that they admit the maximally mixed state as an invariant state, which corresponds to a state of maximal entropy. Consequently, a unital map $\Phi_{t,0}$ does not introduce additional information into the system nor drives it toward a preferred state, in contrast with thermal relaxation processes. For this reason, unital maps play an important role in thermodynamically non-dissipative processes.
    
    In this context, unital maps never decrease the von Neumann entropy. For any state $\hat{\rho}_s$, a unital map satisfies
\begin{equation}
S(\Phi_{t,0}(\hat{\rho}_{s_0})) \geq S(\hat{\rho}_{s_0}),
\end{equation}
where $S(\hat{\rho}_{s_0}) = \text{tr}[\hat{\rho}_{s_0} \ln{\hat{\rho}_{s_0}}]$ is the von Neumann entropy. Therefore, a unital map $\Phi_{t,0}$ cannot make a quantum state more ordered. This inequality holds only for unital maps or for closed system dynamics. Examples include unitary evolution, ideal projective measurements, the depolarizing map.

    \item [\textbf{(d)}] Every CPTP map always admit at least one invariant state (a fixed point) $\hat{\rho}_s^{\,\text{fixed}}$. If a CPTP map is strictly contractive with respect to a complete metric (for example, the trace distance $D$ or the Bures distance), meaning that for every pair of states $\hat{\rho}_{s_0}$, $\hat{\sigma}_{s_0}$  we have
    \begin{equation}
     D(\hat{\rho}_{s_0}, \hat{\sigma}_{s_0}) > D(\Phi(\hat{\rho}_{s_0}), \Phi(\hat{\sigma}_{s_0})),   
    \end{equation}
    then the fixed point is unique. This unique fixed point, called the stationary state, is the state toward which the system inevitably evolves, independent of the initial condition.
    
    An example of a map with multiple fixed points is the phase damping map, which suppresses coherences in the eigenbasis of $\hat{\sigma}_z$ while leaving populations unchanged. In contrast, the depolarizing map is unital and admits a single fixed point, the maximally mixed state $\hat{\rho}_s^{\,\text{fixed}} = \hat{\mathbb{I}}/d$; this map erases all information, both classical and quantum.
    
    Another widely studied example in quantum thermodynamics is the thermal map, which models the interaction with a thermal bath at temperature $T_e$. Its fixed point is the thermal equilibrium (Gibbs) state $\hat{\rho}^{\,\text{fixed}}_s = \hat{\rho}_s^{\,\text{th}}$. This map is strongly dissipative, ensuring that any information about the initial state is irreversibly lost over long timescales.

    \item[\textbf{(e)}] CPTP maps acting on the space state of a quantum system form a semigroup $(\mathcal{S}, \circ)$ under the operation of functional composition $\circ$. For any two CPTP maps $\Phi_{t,0}$ and $\Lambda_{t,0}$ , their composition is defined as 
\begin{equation}\label{composition}
\Phi_{t_2,t_1}  \circ \Lambda_{t_1,0} (\hat{\rho}_{s_0}) = \Phi_{t_2,t_1} \big[\Lambda_{t_1,0}(\hat{\rho}_{s_0})\big],
\end{equation}
and is itself a CPTP map, reflecting the closure of $\mathcal{S}$.
This composition is generally non-commutative, i.e.,
\begin{equation}
\Phi_{t_2,t_1} \circ \Lambda_{t_1,0}(\hat{\rho}_{s_0}) \neq \Lambda_{t_1,0} \circ \Phi_{t_2,t_1}(\hat{\rho}_{s_0}).
\end{equation}
The semigroup $(\mathcal{S}, \circ)$ satisfies the properties of closure, associativity, and the existence of an identity map $\mathcal{I}(\hat{\rho}_{s_0}) = \hat{\rho}_{s_0}$. However, not all elements have an inverse within the set, meaning the property of invertibility does not hold in general. Therefore, $\mathcal{S}$ is not a group. In particular, a CPTP map $\Phi_{t,0}$ is invertible by another CPTP map $\Phi_{t,0}^{-1}$, such that $\Phi_{t,0}^{-1} \,\circ\, \Phi_{t,0} = \mathcal{I}
$; only if $\Phi_{t,0}$ corresponds to a unitary transformation $\Phi_{t,0}[\hat{\rho}_{s_0}]=\hat{V}_s(t,0)\,\hat{\rho}_{s_0}\,\hat{V}_s^\dagger(t,0)$, where $\hat{V}_s(t,0)$ is a unitary operator. Therefore, within the set of CPTP maps, only those representing unitary evolutions are strictly invertible.

\end{itemize}

\section{Amplitude Damping: Standard and Generalized}

Amplitude damping standard (AD) is an important CPTP map that describes energy dissipation effects in a quantum system, modeling the thermodynamic loss of energy from a quantum state \cite{manenti, manzano2018}. This AD map represents the physical phenomenon of spontaneous emission, which is a process where an atom, nucleus, or other quantum systems undergo a transition from the highest energy state to the lowest energy state, releasing energy into the bath (see Fig. \ref{fig:emission}). This energy relaxation can occur through the emission of a photon or a phonon. In order to describe this emission process without the possibility of excitation, the bath temperature must be considered as $T_e = 0$, meaning that the bath contains no thermal energy and is in its vacuum or ground state. 
\begin{figure}[H] 
\centering
\includegraphics[width=0.7\columnwidth]{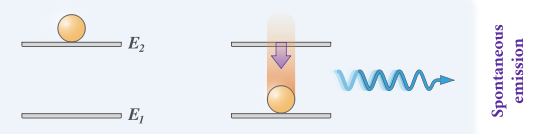}
\caption{(Color online)  Spontaneous emission process, where the excited state with energy $E_2$ decays to the ground state with energy $E_1$, releasing its energy into the environment. Adapted from Ref.\cite{Kaya2025}}
\label{fig:emission}
\end{figure}

\noindent For the characterization of the AD map, we consider the decay of the excited state of a two-level atom, represented by a single qubit. The ground state is denoted by $\ket{0}$, whose Bloch vector is $(0, 0, 1)$, while the excited state is denoted by $\ket{1}$, whose Bloch vector is $(0, 0, -1)$. This quantum map is defined using the Kraus representation as
\begin{equation}
    \Phi_{t,0}^{\text{AD}}(\hat{\rho}_{s_0})=\hat{K}^{\text{AD}}_0\,\hat{\rho}_{s_0}\,\hat{K}^{\text{AD}^\dagger}_0+\hat{K}^{\text{AD}}_1\,\hat{\rho}_{s_0}\,\hat{K}^{\text{AD}^\dagger}_1\,,
\end{equation}
where the operators $\hat{K}^{\text{AD}}_0, \hat{K}^{\text{AD}}_1$ are given by

\begin{equation}
\begin{array}{ll}
\hat{K}^{\text{AD}}_0 = \sqrt{p} \ket{0}\bra{1} \qquad&\quad ,  \qquad 
\hat{K}^{\text{AD}}_1 = \ket{0}\bra{0} + \sqrt{1 - p} \ket{1}\bra{1} ,\\
\\
\hat{K}^{\text{AD}}_0 = \begin{pmatrix}
0 & \sqrt{p} \\
0 & 0
\end{pmatrix} ,
&  \qquad \qquad
\hat{K}^{\text{AD}}_1 = \begin{pmatrix}
1 & 0 \\
0 & \sqrt{1 - p}
\end{pmatrix}.
\end{array}
\end{equation}

\noindent The parameter $0 \leq p \leq 1$ represents the transition probability during a short time interval $\Delta t$. Since the time dependence is encoded in $p(t)$, the kraus operators inherit this dependence. Consequently, the operator $\hat{K}^{\text{AD}}_0$ represents the transition from the excited state to the ground state, while $\hat{K}^{\text{AD}}_1$ preserves the ground state \cite{manenti}.

\begin{figure}[H] 
\centering
\includegraphics[width=0.49\columnwidth]{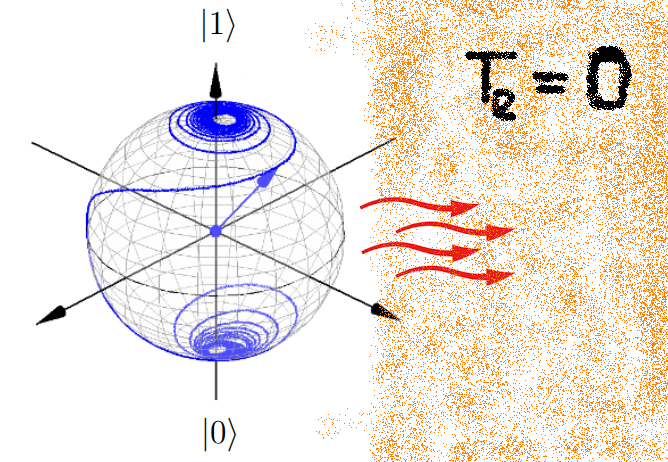}
\caption{(Color online) Schematic illustration of the AD for a single qubit, where the reservoir is at zero temperature. The blue curve shows the relaxation of the Bloch vector from $|1\rangle$ toward $|0\rangle$, while the red curve indicates the irreversible loss of coherence and energy into the environment, driving the system to its unique fixed point $|0\rangle$. Adapted from Ref.\cite{Arie2008}}
\label{fig:ADbloch}
\end{figure}

\noindent Recall that the general state of a qubit is represented by the density matrix in Eq.~\eqref{densityqubit}, and when applying the AD map, the output state becomes
\begin{equation}
\Phi_{t,0}^{\text{AD}}(\hat{\rho}_{s_0}) = \frac{1}{2} \begin{pmatrix}
1 + p + (1 - p) r_z & \quad \sqrt{1 - p}(r_x - i r_y) \\[1.7ex]
\sqrt{1 - p}(r_x + i r_y) & \quad 1 - p - (1 - p) r_z
\end{pmatrix}. \\[4pt]
\end{equation}
\noindent Therefore, the initial Bloch vector $(\,r_x\,,\,r_y\,,r_z\,)$ under the AD map transforms as
\begin{equation}
    \left(r_{x}\,,\, r_{y}\,,\, r_{z}\right) \quad \rightarrow \quad\left(\sqrt{1-p} \,r_{x}\,, \,\sqrt{1-p}\, r_{y}\,,\, p+(1-p) r_{z}\right) .
\end{equation}
The evolution of the Bloch vector toward the ground state is illustrated in Fig.~\ref{fig:ADbloch}.
\noindent Regarding this AD map, it has the ability to increase the purity of the input state. For example, if the initial state is the maximally mixed state $\hat{\mathbb{I}}/2$, the action of the AD map will drive it towards the ground state $\ket{0}\bra{0}$ when $p \rightarrow 1$. Consequently, the AD map is CPTP but non-unital, since it does not preserve the identity operator. Additionally, it is a strictly contractive map whose unique fixed point is the pure ground state $\ket{0}\bra{0}$.

\noindent Now, by applying the AD map discretely and consecutively over an arbitrary qubit state (a two-level system) $\hat{\rho}_{s_0}=\begin{pmatrix}
\rho_{00} & \rho_{01} \\
\rho_{10} & \rho_{11}
\end{pmatrix}$, we can describe the physical process of relaxation towards the ground state (spontaneous emission). Each discrete step can be interpreted as the evolution during a small time step $\Delta t \ll 1$, with a transition or decay probability $p$ per step, approximating continuous evolution. The relaxation rate is defined as $\gamma = p/\Delta t$. After $n$ steps, with total time $t = n\Delta t$, the qubit state evolves to

\begin{equation}
    \big[\Phi^{\rm AD}_{\Delta t,0}\big]^{\rm n-times}(\hat{\rho}_{s_0}) =
\begin{pmatrix}
\rho_{00} + \left[1 - (1 - p)^n \right] \rho_{11} & (1 - p)^{n/2} \rho_{01} \\
(1 - p)^{n/2} \rho_{10} & (1 - p)^n \rho_{11}
\end{pmatrix}.\\[10pt]
\end{equation}
where $\big[\Phi^{\rm AD}_{\Delta t,0}\big]^{\rm n-times}=\hat{\rho}_{s_0} \xrightarrow{\Phi_{\Delta t,0}^{\text{AD}}} \hat{\rho}_s(\Delta t) \xrightarrow{\Phi_{2\Delta t,\Delta t}^{\text{AD}}} \hat{\rho}_s(2 \Delta t) \xrightarrow{\Phi_{3\Delta t,2\Delta t}^{\text{AD}}} \cdots \xrightarrow{\Phi_{n\Delta t,(n-1)\Delta t}^{\text{AD}}} \hat{\rho}_s(n \Delta t).$

\noindent The AD map describes how, in this case, a qubit loses energy as it couples to an environment at temperature $T_e = 0$. During this process, the system tends to decay into the ground state through a form of quantum thermal dissipation. After $n$ applications of the AD map, the initial probability $\rho_{11}$ of finding the qubit in the excited state is reduced by a factor of $(1 - p)^n\,\rho_{11}$. In the limit $n \to \infty$, this reduction follows an exponential decay

\begin{equation}
\lim_{n \to \infty} \left(1 - \frac{\gamma t}{n} \right)^n \rho_{11} = e^{-\gamma t} \rho_{11} = e^{-t/T_\gamma} \rho_{11},\\[8pt]
\end{equation}

\noindent where $T_\gamma = 1/\gamma$ is defined as the relaxation time, meaning the characteristic time at which the qubit loses its energy to the environment and reaches the ground state. This indicates that the population decays exponentially as $e^{-t/T_\gamma}$. On the other hand, the coherence terms also decay but with a slower rate, following $e^{-t/2T_\gamma}$, showing that the AD map affects populations more strongly than coherence. 

\begin{equation}
\Phi_{t,0}^{\rm AD}\big[\hat{\rho}_{s_0}\big]=\hat{\rho}_s(t) =
\begin{pmatrix}
1 - \rho_{11} e^{-t/T_\gamma} & \rho_{01} e^{-t/2T_\gamma} \\
\rho_{10} e^{-t/2T_\gamma} & \rho_{11} e^{-t/T_\gamma}
\end{pmatrix},\\[5pt]
\end{equation}

\noindent In the limit $t \to \infty$, the state evolves towards the ground state $
\lim_{t \to \infty} \Phi_{t,0}^{\text{AD}}(\hat{\rho}_{s_0}) =
\begin{pmatrix}
1 & 0 \\
0 & 0
\end{pmatrix}
$. 

\noindent This behavior reflects the complete loss of energy and coherence, projecting the system onto a classical pure state based solely on populations. In quantum thermodynamics, this map models an irreversible process, where the decay of populations represents the loss of internal energy into the environment in the form of thermodynamic heat. Simultaneously, the decay of coherences represents the degradation of purely quantum properties into the environment, which can be interpreted as informational heat \cite{manenti, manzano2018}.

\noindent The AD map we applied to a qubit describes thermodynamic dissipation assuming the environment is at zero temperature. The generalized amplitude damping (GAD) map, on the other hand, describes the same type of dissipation but for a thermal environment at finite temperature. In this case, there is not only energy loss but also the possibility of thermal excitation, meaning the environment can occasionally provide energy to the qubit. The key difference is that while the AD channel always drives the system toward the ground state, the GAD channel drives it toward thermal equilibrium \cite{manenti, manzano2018} (see Fig.~\ref{fig:GADbloch} ). The action of the GAD map on a qubit state is given by
\begin{equation}
    \Phi_{t,0}^{GAD}(\hat{\rho}_{s_0})=\hat{K}_0^{GAD}\,\hat{\rho}_{s_0}\,\hat{K}^{GAD^{\dagger}}_0+\hat{K}_1^{GAD}\,\hat{\rho}_{s_0}\,\hat{K}^{GAD^{\dagger}}_1+\hat{K}_2^{GAD}\,\hat{\rho}_{s_0}\,\hat{K}^{GAD^{\dagger}}_2+\hat{K}_3^{GAD}\,\hat{\rho}_{s_0}\,\hat{K}^{GAD^{\dagger}}_3,
\end{equation}
where
\begin{equation}
\begin{aligned}
\hat{K}_0^{GAD} &= \sqrt{p}
\begin{pmatrix}
1 & 0 \\
0 & \sqrt{1 - a}
\end{pmatrix}, \qquad \qquad
\hat{K}_1^{GAD}= \sqrt{p}
\begin{pmatrix}
0 & \sqrt{a} \\
0 & 0
\end{pmatrix}, \\[1em]
\hat{K}_2^{GAD} &= \sqrt{1 - p}
\begin{pmatrix}
\sqrt{1 - a} & 0 \\
0 & 1
\end{pmatrix}, \qquad
\hat{K}_3^{GAD} = \sqrt{1 - p}
\begin{pmatrix}
0 & 0 \\
\sqrt{a} & 0
\end{pmatrix}.
\end{aligned}
\end{equation}
\noindent For a process involving a two-level system with energy separation $E=\hbar\,\omega_0$ weakly coupled to a thermal bosonic bath, the parameters $p$ and $a$ have the following interpretations
\begin{itemize}
    \item $a = 1 - e^{-t/\tau_R}$ quantifies the degree of completion of the dissipative process toward thermal equilibrium over time $t$. Here, $\tau_R$ is the characteristic relaxation time of the process and is inversely proportional to the thermal occupation of the bath, given by $(2n_{\text{th}}-1)^{-1}$ \cite{manzano2018}.
    \item $p = \frac{n_{\text{th}} + 1}{2n_{\text{th}} + 1}$, con $n_{\text{th}} = \left(e^{\beta E} - 1\right)^{-1}$ y $\beta=1/k_BT_e$,  is the thermal probability of finding the qubit system in the ground state. The term $n_{\text{th}}$, coming from the Bose-Einstein distribution, represents the average number of thermal excitations of the environment at the frequency $\omega _0$ and energy equal to the qubit’s transition energy $E=\hbar\,\omega_0$. The parameter $p$ also dictates the balance between relaxation and thermal excitation in the GAD map. At high temperatures $T_e\rightarrow\infty$, $n_{\text{th}}\rightarrow\infty$ and therefore the probability $p \rightarrow1/2$  indicating that both levels are equally populated. At low temperatures  $T_e\rightarrow0$,  $n_{\text{th}}\rightarrow0$, which means the system is found in its ground state $p\rightarrow1$, as thermodynamically expected for two-level systems \cite{manzano2018}.
\end{itemize}

\noindent The GAD map is a CPTP but non-unital map,  $\Phi_{\text{GAD}}(\hat{\mathbb{I}})\neq \hat{\mathbb{I}}$. This map is also characterized by being strictly contractive, with its unique fixed point given by the thermal Gibbs state. The matrix representation of this fixed point, written in the eigenbasis of the qubit Hamiltonian $\hat{H}_s=-\hbar \omega_0 \hat{\sigma} _{z} /2$ is:
\begin{equation}
\hat{\rho}_s^{\text{th}} = \frac{1}{2 \cosh\left( \frac{\hbar \omega_0}{2 k_B T_e} \right)} 
\begin{pmatrix}
e^{\frac{\hbar \omega_0}{2 k_B T_e}} & 0 \\
0 & e^{-\frac{\hbar \omega_0}{2 k_B T_e}}
\end{pmatrix}.
\end{equation}

\begin{figure}[H] 
\centering
\includegraphics[width=0.52\columnwidth]{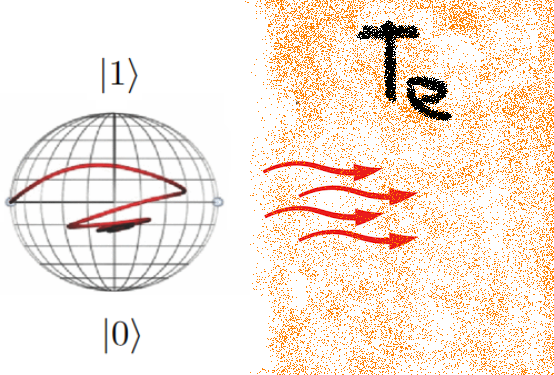}
\caption{(Color online) Schematic illustration of the GAD for a single qubit interacting with a finite temperature environment $T_e$. The red arrows represent irreversible loss of coherence and energy to the reservoir, while the trajectory inside the Bloch sphere describes the relaxation of the qubit state. Unlike the zero-temperature AD case (which relaxes to $|0\rangle$), here the relaxation leads the qubit to a thermal fixed point determined by $T_e$, $\hat{\rho}_s^{\, \rm th}$. Adapted from Ref.\cite{Renaud2011}}
\label{fig:GADbloch}
\end{figure}

\noindent If we apply the GAD map many times to any initial state $\hat{\rho}_s$, in the limit $t \rightarrow \infty$ the system evolves to the thermal state  $\hat{\rho}_s^{\,\text{th}}=\begin{pmatrix}
p & 0 \\
0 & 1-p
\end{pmatrix}$ where the probability of occupying the ground state at thermal equilibrium is $p = e^{\frac{\hbar \omega_0}{2 k_B T_e}}/2 \cosh\left( \frac{\hbar \omega_0}{2 k_B T_e} \right)$. This demonstrates that the GAD map guarantees the complete thermalization of the system~\cite{DEOLIVEIRA2020}. Moreover, there are extensions of the GADC that allow modeling more complex environments, such as squeezed thermal reservoirs~\cite{Srikanth2008}.

\section{Phase damping}

Phase Damping (PD), also known as Dephasing, is a type of interaction occurs particularly in quantum computing platforms such as superconducting qubits, due to the interaction with the environment, which generates random fluctuations in the qubit's frequency \cite{krantz2019}. Unlike the AD and GAD maps, the PD map, denoted as $\Phi^{\text{PD}}_{t,0}$, represents a process that occurs solely in quantum mechanics, describing the loss of quantum information without energy loss.

\noindent From a mathematical point of view, $\Phi^{\text{PD}}_{t,0}$  is a non-unitary map that acts on the system’s density operator $\hat{\rho}_{s_0}$ according to
\begin{equation}
    \Phi^{\text{PD}}_{t,0}(\hat{\rho}_{s_0})= (1-p)\,\hat{\rho}_{s_0}+ p\,\hat{\sigma}_z\,\hat{\rho}_{s_0}\,\hat{\sigma}_z,
\end{equation}

\noindent where $0\leq p\leq 1$  represents the probability that a phase-flip error (a phase inversion via $\hat{\sigma}_z$) occurs within a short time interval $\Delta t$. Since the time dependence is encoded in $p(t)$, the kraus operators inherit this dependence. Physically, the populations of the system state remain unchanged, but the relative coherence between states $\ket{0}$ and $\ket{1}$ is lost. 

\noindent Formally, the action of the dephasing map on a quantum state can be modeled using the following Kraus representation
\begin{equation}
    \Phi^{\text{PD}}_{t,0}(\hat{\rho}_{s_0})=\hat{K}^{\text{PD}}_0\,\hat{\rho}_{s_0}\,\hat{K}^{\text{PD}^\dagger}_0+\hat{K}^{\text{PD}}_1\,\hat{\rho}_{s_0}\,\hat{K}^{\text{PD}^\dagger}_1,
\end{equation}
where the operators $\hat{K}^{\text{PD}}_0, \hat{K}^{\text{PD}}_1$ are given by
\begin{equation}
\begin{array}{ll}
\hat{K}^{\text{PD}}_0 = \sqrt{1-p}\,(\, \ket{0}\bra{0}+\ket{1}\bra{1}\,) \qquad&\quad ,  \qquad 
\hat{K}^{\text{PD}}_1 = \sqrt{p}\,(\,\ket{0}\bra{0}- \ket{1}\bra{1}\,) ,\\
\\
\hat{K}^{\text{PD}}_0 = \begin{pmatrix}
\sqrt{1-p} & 0 \\
0 & \sqrt{1-p}
\end{pmatrix} ,
&  \qquad \qquad
\hat{K}^{\text{PD}}_1 = \begin{pmatrix}
\sqrt{p} & 0 \\
0 & -\sqrt{p}
\end{pmatrix},
\end{array}
\end{equation}

\noindent Considering again the general qubit state, Eq.~\eqref{densityqubit}, the action of the dephasing map is
\begin{equation}
\Phi^{\text{PD}}_{t,0}(\hat{\rho}_{s_0}) = \frac{1}{2} \begin{pmatrix}
1 + r_z & \quad (1 - 2p)(r_x - i r_y) \\[1.7ex]
(1 - 2p)(r_x + i r_y) & \quad 1 - r_z
\end{pmatrix}. 
\end{equation}

\noindent If the initial qubit state is represented by the Bloch vector $(r_x\,, \,r_y\,,\, r_z)$, the final output after applying the dephasing map becomes
\begin{equation}
    \left(r_{x}\,,\, r_{y}\,,\, r_{z}\right) \quad \rightarrow \quad\left((1-2p) r_{x}\,, \,(1-2p) r_{y}\,,\,r_{z}\right) .
\end{equation}
This means that the coherent components $r_x$ and $r_y$ are attenuated by a factor of $(1-2p)$, while the $r_z$ component (associated with the population) remains unchanged. In other words, the PD map does not induce energy exchange but only leads to coherence loss. Illustrated in Fig.~\ref{fig:PDBloch}. Another important characteristic of this map is that it never increases the state’s purity during Markovian evolution, since $\Phi_{t,0}^{\text{PD}}(\hat{\mathbb{I}}_s)=\hat{\mathbb{I}}_s$; therefore, it is a unital map. However, this unitality reflects that the PD map preserves the maximally mixed state but does not imply convergence towards it. Consequently, the PD map is not strictly contractive and admits a whole family of fixed points. Specifically, any state diagonal in the computational basis remains invariant under this map \cite{manenti}.

\noindent As in the AD case, the PD map can be seen as a discrete step of the state over a short interval $\Delta t \ll 1$, approximating continuous evolution. To better understand how this PD map works, let us start from a two-level system (qubit) described by the density operator $\hat{\rho}_{s_0}=\begin{pmatrix}
\rho_{00} & \rho_{01} \\
\rho_{10} & \rho_{11}
\end{pmatrix}$ and apply the PD map continuously over a total time $t=n \Delta t$
\begin{equation}\label{eq:nPD}
    \big[\Phi^{\text{PD}}_{\Delta t,0}\big]^{\rm n-times}(\hat{\rho}_{s_0}) =
\begin{pmatrix}
\rho_{00} & (1 - 2p)^{n} \rho_{01} \\[1.7ex]
(1 - 2p)^{n} \rho_{10} & \rho_{11}
\end{pmatrix}.
\end{equation}
where $\big[\Phi^{\text{PD}}_{\Delta t,0}\big]^{\rm n-times}=\hat{\rho}_{s_0} \xrightarrow{\Phi_{\Delta t,0}^{\text{PD}}} \hat{\rho}_s(\Delta t) \xrightarrow{\Phi_{2\Delta t,\Delta t}^{\text{PD}}} \hat{\rho}_s(2 \Delta t) \xrightarrow{\Phi_{3\Delta t,2\Delta t}^{\text{PD}}} \cdots \xrightarrow{\Phi_{n \Delta t,(n-1)\Delta t}^{\text{PD}}} \hat{\rho}_s(n \Delta t)$.

\noindent If we define the dephasing rate as the probability of a phase flip per unit time, $\gamma_\phi = 2p/\Delta t$, we can observe an exponential decay in the off-diagonal elements $(1 - 2p)^{n} \rho_{01}$ and $(1 - 2p)^{n} \rho_{10}$, which in the limit $n \rightarrow \infty$ approaches
\begin{equation}
    \lim _{n \rightarrow \infty} \rho_{10}\left(1-\frac{\gamma_{\phi} t}{n}\right)^{n}=\rho_{10} e^{-\gamma_{\phi} t}=\rho_{10} e^{-t / T_{\phi}},
\end{equation}
Therefore, Eq. \eqref{eq:nPD} becomes 
\begin{equation}
    \Phi_{t,0}^{\rm PD}\big[\hat{\rho}_{s_0}\big]=\hat{\rho}_s(t)=\left(\begin{array}{cc}
\rho_{00} & \rho_{01} e^{-t / T_{\phi}} \\
\rho_{10} e^{-t / T_{\phi}} & \rho_{11}
\end{array}\right),
\end{equation}
where the dephasing time $T_{\phi}=1/\gamma_\phi$ was introduced. This time characterizes how quickly the relative phase information of the qubit state is lost. From a thermodynamic perspective, dephasing is a purely entropy-generating process that increases the von Neumann entropy $\Delta S\geq 0$ without involving energy exchange. This behavior arises from the irreversibility caused by the collapse of coherences. Therefore, dephasing generates entropy without being associated with thermodynamic heat flow. This property is fundamental to understanding quantum irreversibility mechanisms that not tied to energy exchange~\cite{kosloff}.
\begin{figure}[H] 
\centering
\includegraphics[width=0.42\columnwidth]{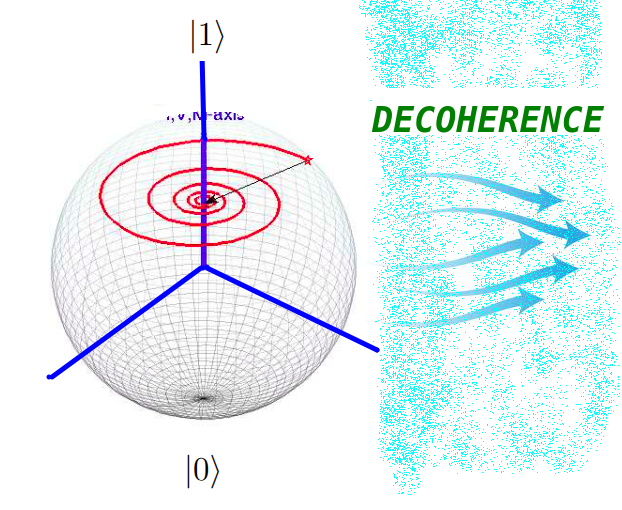}
\caption{(Color online) Schematic illustration of the dephasing for a single qubit. In this map, irreversibility comes only from the loss of coherence (decoherence), while the energy of the qubit remains constant, meaning that the population of the energy levels does not change during the evolution. The light blue arrows that connect the system to the environment represent the flow of coherence into the reservoir. The red trajectory inside the Bloch sphere shows the evolution of the qubit state on a plane with fixed energy. At long times, the state approaches a fixed point determined by its initial energy coordinate. Adapted from Ref.\cite{Gong2015}.}
\label{fig:PDBloch}
\end{figure}

\section{Kraus-operator formalism}

A highly general formalism that provides a way to represent specific properties of the environment and its influence on the reduced density matrix of the system is the Kraus representation. This approach has the advantage of not requiring any explicit reference to the detailed form of the total Hamiltonian of the composite system (system + environment).

\noindent Consider that the composite system $S-e$ behaves as a closed system whose evolution is governed by a unitary transformation of the form $\hat{V}_{se}(t)=\mathrm{e}^{-\mathrm{i} \hat{H}(t)\, t}$, with the total Hamiltonian 
\begin{equation}
    \hat{H}(t)=\hat{H}_{s} \otimes \mathbb{I}_{e}+\mathbb{I}_{s} \otimes \hat{H}_{e}+g \,\hat{H}_{i n t},
\end{equation}
where $\mathbb{I}_{s(e)}$ denotes the identity operator on the Hilbert space of the system (environment), $\hat{H}_{s(e)}$ is the free Hamiltonian of each subsystem, and $g$ is a real parameter that determines the strength of the interaction between the system and the environment, described by $\hat{H}_{\text{int}}$. When $g = 0$, the system and environment evolve independently and remain uncorrelated throughout the evolution. In contrast, when  $g \neq 0$,  the interaction generates correlations between the system and the environment during the dynamics.

\noindent Let us now consider the case $g=0$, such that the initial density matrix of the composite system is a product state: 
\begin{equation}
    \hat{\rho}(0)=\hat{\rho}_{s}(0) \otimes \hat{\rho}_{e}(0)
\end{equation}
where the initial state of the environment is expressed in its spectral decomposition as $\hat{\rho}_{e}(0)=\sum_{i} p_{i}\left|e_{i}\right\rangle\left\langle e_{i}\right|$.
Assuming further that $\hat{H}$ is time-independent, the evolution of the reduced state of the system $S$ is given by

\begin{equation}
\hat{\rho}_{s}(t) = \text{tr}_{e} \left\{ \hat{V}_{se}(t) \left[ \hat{\rho}_{s}(0) \otimes \left( \sum_i p_i \ket{e_i} \bra{e_i} \right) \right] \hat{V}_{se}^\dagger(t) \right\} .
\end{equation}

\noindent By explicitly performing the partial trace over the environmental basis $\ket{e_i}$, we obtain
\begin{equation}
        \hat{\rho}_{s}(t)=\sum_{i j} \hat{K}_{i j} \, \hat{\rho}_{s}(0) \, \hat{K}_{i j}^{\dagger}
\end{equation}

\noindent where the Kraus operators are defined as 
\begin{equation}
  \hat{K}_{ij} \equiv \sqrt{p_i} \, \braket{e_j | \hat{V}_{se}(t) | e_i}.
\end{equation}

\noindent All the information about the influence of the environment on the system is fully captured within the Kraus operators. It is important to note that any completely positive quantum map acting on a system of dimension $N$ can be represented by at most $N^2$ Kraus operators. This upper bound arises from the dimension of the space of linear operators acting on the system. However, depending on the internal structure of the map, some processes may require fewer operators. Furthermore, the Kraus representation is not unique. A given quantum map can be described by different sets of Kraus operators, which are related to each other through unitary transformations acting on the Kraus index space.

\noindent Since the evolution of the composite system is unitary, the effective map describing the dynamics of the system $S$ is CPTP. This is reflected in the fact that the Kraus operators satisfy the completeness condition, also referred to as the trace-preserving condition
\begin{equation}
    \sum_{i j} \hat{K}^{\dagger}_{i j} \hat{K}_{i j} = \hat{\mathbb{I}}_{s}.
\end{equation}

 \noindent  If this condition is not fulfilled, it can be understood that the evolution of the composite system is no longer strictly unitary, suggesting the existence of an additional external environment interacting with the combined system $S-e$. In such cases, the completeness condition is relaxed to the following inequality $\sum_{i j} \hat{K}^{\dagger}_{i j} \hat{K}_{i j} \leq \hat{\mathbb{I}}_{s}.$ which characterizes maps associated with conditional processes or selective measurements.

\section{Master Equation}

\subsection{From CPTP Maps to Markovian Master Equations}

In the previous sections, we modeled the dynamics of open quantum systems using CPTP maps. These maps represent transformations between the states of a system at two distinct times,$t_0$ and $t_2$,  and therefore provide a discrete time description of the system's evolution. However, if we aim to develop a continuous time description of the system’s dynamics, it becomes necessary to derive differential equations for the density operator $\hat{\rho}_s(t)$, a common strategy in classical mechanics. For open systems, such differential equations are typically referred to as master equations \cite{manzano2018}.

\noindent Extending the theory of CPTP maps to continuous time requires careful treatment. As a starting point, we consider the composition property of CPTP maps (see Eq.\eqref{composition}). Briefly, the concatenation of two CPTP maps also yields a CPTP map. However, the converse is not generally true. To be more specific, let us consider a general CPTP map  $\Phi_{t_2,t_0}$ describing the system's evolution between times $t_0$ and $t_2$.  In general, this map cannot be decomposed into the composition of two intermediate CPTP maps,
\begin{equation}\label{nodivisibilitymapascptp}
 \Phi_{t_2,t_0}\neq\Phi_{t_2,t_1} \circ \Phi_{t_1,t_0} ,
\end{equation}

\noindent where each map corresponds to a subinterval of time. The failure of this decomposition is due to the emergence of system-environment correlations after the first time interval governed by the map $\Phi_{t_1,t_0}$ \cite{manzano2018,Rivas2010}.

\noindent In other words, although the overall evolution described by $\Phi_{t_2,t_0}$ assumes that the system and environment are initially uncorrelated, this is no longer true at intermediate times. The global CPTP map takes the form
\begin{equation}\label{mapacptp20}
    \Phi_{t_2,t_0}\big(\hat{\rho}_s(t_0)\big) = \text{tr}_{e} \left[ \hat{V}_{\rm se}(t_2, t_0) \left( \hat{\rho}_s(t_0) \otimes \hat{\rho}_{e}(t_0) \right) \hat{V}_{\rm se}^\dagger(t_2, t_0) \right],
\end{equation}
where  $\hat{\rho}_s(t_0)$ and $\hat{\rho}_e(t_0)$ are initially uncorrelated. 

\noindent In contrast, the intermediate evolution governed by $\Phi_{t_2,t_1}$ acts not on a product state but on a correlated total state $\hat{\rho}(t_1)$,
\begin{equation}\label{mapacptp12}
    \Phi_{t_2,t_1}\big(\hat{\rho}_s(t_1)\big) = \text{tr}_{e} \Big[ \hat{V}_{se}(t_2, t_1)  \underbrace{\left(\hat{\rho}_s(t_1) \otimes \hat{\rho}_{e}(t_1) +\delta \rho_{corre}(t_1)\right)}_{\displaystyle \hat{\rho}(t_1)} \hat{V}_{se}^\dagger(t_2, t_1) \Big].
\end{equation}
The correlations $\delta \rho_{corre}(t_1)$ are generated by the global unitary evolution $\hat{V}_{ se}(t_1,t_0)$ due to the system-environment interaction encoded in the total Hamiltonian. Explicitly, these correlations arise from the transformation \begin{equation}\label{mapacptp01correlacionado}
    \hat{\rho}(t_1)=  \hat{V}_{se}(t_1, t_0) \left( \hat{\rho}_s(t_0) \otimes \hat{\rho}_{e}(t_0) \right) \hat{V}_{se}^\dagger(t_1, t_0).
\end{equation} As a consequence, these correlations prevent a consistent definition of continuous dynamics based on infinitesimal subdivisions of CPTP maps~\cite{manzano2018,rivas2011}.
\noindent The possibility (or impossibility) of decomposing a global CPTP map into a sequence of intermediate CPTP maps is commonly referred to as the divisibility criterion of quantum maps. If a system's evolution satisfies this criterion, then the dynamics is said to be Markovian. In this context, Markovianity reflects the physical idea that the system’s future evolution depends only on its current state and not on its history, which means the absence of memory effects. Conversely, if system-environment correlations persist and influence the dynamics, the evolution becomes non-Markovian, indicating the presence of memory and a backflow of information from the environment to the system~\cite{rivas2014, Rivas2010}.

\noindent In order to derive a differential equation that describes the time evolution of an open quantum system using CPTP maps, certain assumptions must be introduced. Formally, given an arbitrary state at time $t$ denoted by $\hat{\rho}_s(t)$, its evolution over a short time interval $\tau$ can be expressed as \cite{manzano2018, rivas2011}, \begin{equation}\label{mapacptpcontau}
    \hat{\rho}_s(t+\tau)= \Phi_{t+\tau,t}\big(\hat{\rho}_s(t)\big),
\end{equation}
where $\Phi_{t+\tau,t}$ is a CPTP map that depends on both  $t$ and $\tau$.

\noindent Using the definition of the derivative, we obtain
\begin{equation}\label{definicionderivadacptp}
    \frac{d}{dt} \hat{\rho}_s(t) = \lim_{\tau \to 0} \frac{\hat{\rho}_s(t+\tau)+\hat{\rho}_s(t)}{\tau} = \lim_{\tau \to 0} \frac{\Phi_{t+\tau,t} - \hat{\mathbb{I}}}{\tau} \left( \hat{\rho}_s(t) \right) \equiv \mathcal{L}_t \left( \hat{\rho}_s(t) \right).
\end{equation}

\noindent The term $\mathcal{L}_t$ is known as the Lindbladian. The equation~\eqref{definicionderivadacptp} is valid whenever the limit exists and the evolution is sufficiently smooth, meaning there are no discontinuities in infinitesimal time intervals. This ensures that $\mathcal{L}_t$ is well defined and that the system dynamics can be described by a time-local differential equation. In addition, since $\Phi_{t+\tau,t}$ must be a CPTP map, it is therefore necessary that the total state $\hat{\rho}(t)$ can be approximated by a product state such as \cite{manzano2018}, \begin{equation}\label{condicionproductoestado}
\hat{\rho}(t) = \hat{\rho}_s(t) \otimes \hat{\rho}_e(t) + \delta\hat{\rho}_{\text{corr}}(t) \approx \hat{\rho}_s(t) \otimes \hat{\rho}_e(t).
\end{equation}
This condition \eqref{condicionproductoestado} is typically satisfied under weak coupling and when the environment is large enough to remain close to thermal equilibrium at all times. Physically, this means that the environment rapidly absorbs and dissipates the information it receives from the system, distributing it across its many degrees of freedom. As a result, the information is effectively lost from the perspective of the system and does not influence its future evolution. This process involves two key time scales, namely the environment correlation time $\tau_c$, which characterizes how quickly the environment forgets the influence of the system, and the system evolution time $\tau_r$, which is the time scale over which the system states change as a result of their interaction with the environment~\cite{Breuer:Book, manzano2018}.

\noindent If we choose a time interval $\tau$ such that $\tau_c \ll \tau \ll \tau_r$, it becomes possible to apply a Markovian coarse-graining description. This procedure indicates that the correlations inevitably generated by the interaction between the system and the environment will have a negligible effect on the future evolution of the system, see Eq.\eqref{condicionproductoestado}. This Markovian coarse-graining procedure supports the validity of the Markov approximation (which will be discussed later), under which the system's evolution is assumed to be independent of its past history. Consequently, this allows the open system dynamics to be described by a time-local master equation, specifically the Lindblad equation~\cite{Breuer:Book, manzano2018}.

\subsection{Markovian Master equation}

We know that the reduced density matrix of the system $\hat{\rho}_s(t)$ is obtained from the total state of the system plus environment $\hat{\rho}_{s+e}(t)$, through
\begin{equation*}
    \hat{\rho}_s(t) = \text{tr}_e \big[ \hat{\rho}_{s+e}(t) \big],
\end{equation*}
\begin{equation} \label{eqreduzida}
\hat{\rho}_s(t) = \mathrm{tr}_e \big[ \hat{V}_{se}(t) \hat{\rho}_{s+e}(0) \hat{V}_{se}^\dagger(t) \big],
\end{equation}
where $\hat{V}_{se}(t)$ is the evolution operator of the total system. However, this approach requires full knowledge of the global dynamics, which is usually unfeasible for complex systems. On the other hand, the derivation of a master equation for $\hat{\rho}_{s}(t)$ starts by introducing a quantum map $\Lambda_{t,0}$ acting on the initial state $\hat{\rho}_s(0)$ \cite{maximilian2007},
\begin{equation}\label{mastereqgeneral}
\hat{\rho}_s(t) = \Lambda_{t,0} \big( \hat{\rho}_s(0)\big).
\end{equation}
This map $\Lambda_{t,0}$ generates the system’s evolution. The Eq. \eqref{mastereqgeneral} is the most general form a master equation can take. If Eq. \eqref{mastereqgeneral} is exact (meaning that no approximations are needed), then it is equivalent to Eq. \eqref{eqreduzida}. The real advantage of general master equations emerges when suitable approximations are applied. These allow one to describe the reduced dynamics locally in time using a first-order differential equation, without needing access to the full system-environment evolution \cite{maximilian2007}. 

\noindent In the following, we introduce the approximations required to derive this differential master equation,
\begin{itemize}
    \item \textbf{Born Approximation} - This approximation relies on the assumption that the system and the environment are weakly coupled. The environment is considered large enough that any influence the system may exert on it is negligible. As a result, the state of the environment remains approximately constant over time, and the total state of the system and environment can be approximated as a product state $\hat{\rho}_{s+e}(t) \approx\hat{\rho}_s(t) \otimes\hat{\rho}_e$ \cite{maximilian2007,manenti,benenti2004}.
    \item \textbf{Markov Approximation} - This approximation assumes that the state of the system at a given time depends only on its present state and not on its past history. In other words, the time derivative of $\hat{\rho}_s(t)$ is determined solely by $\hat{\rho}_s(t)$, without needing to consider how the system evolved to that state.   This idea is based on the assumption that the environment is memoryless, meaning that any influence the system exerts on it (such as autocorrelations and internal correlations) is rapidly dissipated or forgotten. As a result, the environment does not retain information about past interactions with the system. Consequently, there is no feedback of information from the environment to the system, making the flow of information essentially unidirectional, from the system to the environment. Within this context, it is reasonable to treat quantum noise as a purely dissipative process~\cite{maximilian2007,manenti,benenti2004}.
\end{itemize}
Let $\Lambda_{t,0}=\Lambda_t$ be a quantum map satisfying the approximations previously mentioned. This allows us to apply Stone's theorem, expressing it in exponential form as \cite{manenti}
\begin{equation*}
    \hat{\rho}_s(t) = \Lambda_t\big(\hat{\rho}_s(0)\big) 
\end{equation*}
\begin{equation}\label{Lcte}
    \hat{\rho}_s(t) = e^{\mathcal{L}t}\big(\hat{\rho}_s(0)\big),
\end{equation}
where $\mathcal{L}$  is referred to as a superoperator since it acts on the density operator $\hat{\rho}_s(0)$.  Assuming that $\mathcal{L}$ is time-independent, calculating the time derivative of Eq.\eqref{Lcte} becomes straightforward. The equation of motion for the system is then given by
\begin{equation}\label{eqmovLcte}
    \frac{d \hat{\rho}_s(t)}{dt} = \mathcal{L}(\hat{\rho}_s(t)).
\end{equation}
Recall that, in closed systems, the Hamiltonian $\hat{H}_s$ generates the dynamics. Similarly, the Lindbladian $\mathcal{L}$ serves as the generator for the evolution of open quantum systems. The form of this generator $\mathcal{L}$ must ensure that the map $\Lambda_t=e^{\mathcal{L} t}$ is completely positive and trace-preserving for all $t\geq0$. The family of superoperators $\{\Lambda_t\}_{t\geq 0}$ forms a one-parameter semigroup, where the only evolution parameter is time $t$, due to the time-independence of $\mathcal{L}$. 

\noindent This semigroup obeys three fundamental properties, which are \cite{lidar2019}
\begin{itemize}
    \item \textbf{Identity element} - $\Lambda_t=\hat{\mathbb{I}}$
    \item \textbf{Closure under temporal composition} - $\Lambda_t \circ \Lambda_s = e^{\mathcal{L} t} e^{\mathcal{L} s} = e^{\mathcal{L} (t + s)} = \Lambda_{t+s}$
   \item \textbf{Associativity} - $(\Lambda_t \circ\Lambda_s)\circ \Lambda_r = \Lambda_t \circ(\Lambda_s \circ \Lambda_r)$
\end{itemize}

\noindent However, not every element in this family has an inverse satisfying $\Lambda_t \Lambda_t^{-1}=\hat{\mathbb{I}}$, which prevents it from forming a full group structure. This typical behavior of quantum maps lacking inverse elements is responsible for the decoherence of quantum states, as occurs in physical processes such as amplitude damping and dephasing \cite{manenti}. Additionally, as the map $\Lambda_t$ has an exponential form and generates a semigroup, it automatically satisfies the divisibility criterion for CPTP maps, as justified by the semigroup composition property. For this reason, $\Lambda_t$ is called a Markovian evolution operator \cite{Breuer:Book,lidar2019}.

\subsection{Lindblad Equation}

\noindent Under the approximations introduced in the previous section, the evolution of $\hat{\rho}_s(t)$ takes the form of a first-order differential equation that is local in time
\begin{equation}\label{partesmasterequation}
\frac{d\hat{\rho}_s(t)}{dt} = \mathcal{L}[\hat{\rho}_s(t)] = -\frac{i}{\hbar}[\hat{H}_s, \hat{\rho}_s(t)] + \mathcal{D}[\hat{\rho}_s(t)].
\end{equation}
Here, the superoperator $\mathcal{L}$ has two main contributions. The first term is unitary and comes from the commutator with the system Hamiltonian $\hat{H}_s$ (which may include energy corrections due to the environment, such as the Lamb shift). The second term $\mathcal{D}$, is non-unitary and is usually called the dissipator. It accounts for decoherence and dissipation effects caused by the environment \cite{maximilian2007}.

\noindent This structure leads to the most well-known master equation, developed through the work of Gorini, Kossakowski, Sudarshan, and Lindblad. They derived the unique form of $\mathcal{L}$ that ensures the map $\Lambda_t$ defines a CPTP semigroup, which is divisible at all times \cite{Gorini1976, Lindblad1976}. Therefore, Eq. \eqref{partesmasterequation} takes the following form 
\begin{equation}\label{lindbladhomogeneo}
\frac{d \hat{\rho}_s(t)}{dt} = -\frac{i}{\hbar}[\hat{H}_s, \hat{\rho}_s(t)] + \sum_{k=1}^{K} \gamma_k \left( \hat{L}_k \hat{\rho}_s(t) \hat{L}_k^\dagger - \frac{1}{2} \left\{ \hat{L}_k^\dagger \hat{L}_k, \hat{\rho}_s(t) \right\} \right) ,
\end{equation} where $\hat{H}_s$ is the system Hamiltonian, $\hat{L}_k$ $\hat{L}_k$ are the so-called Lindblad operators, and $\gamma_k$ are decay rates with units of [1/time]. Both $\hat{L}_k$ and $\gamma_k$ are determined by the coupling between the system and its environment \cite{manzano2018,manenti}.

\noindent Equation \eqref{lindbladhomogeneo} is commonly referred to as the Lindblad equation and describes purely Markovian dynamics. This Markovian master equation is time-homogeneous because $\mathcal{L}$ does not depend on time. That is, $\hat{H}_s$, $\hat{L}_k$, and $\gamma_k \geq 0$ are all time-independent \cite{manenti}. The fact that the decay rates $\gamma_k$ are positive follows from the requirement that $\mathcal{L}$ generates a CPTP divisible map at all times~\cite{Gorini1976, lidar2019, Lindblad1976}.

\noindent There are different approaches to derive Eq.\eqref{lindbladhomogeneo}. From the semigroup perspective, one obtains the general structure of $\mathcal{L}$ that guarantees CPTP dynamics with semigroup properties. Another method starts from a microscopic model of the system and its environment, under standard approximations. A third approach uses the formalism of quantum operations, where the short time expansion of CPTP maps leads to an approximate master equation, while the coarse graining method, also based on the Kraus representation, introduces an intermediate time scale that allows for a more general effective description of the reduced dynamics.

\noindent We now show how to find such a generator $\mathcal{L}$ for very short time evolutions using only a short time expansion within the Kraus representation formalism . To this end, we start by performing a Taylor expansion of the system's density operator around $t = 0$ \cite{lidar2019}, \begin{equation}
\hat{\rho}_s(dt) = \hat{\rho}_s(0) + \frac{d\hat{\rho}_s(0)}{dt}\, dt + O(dt^2).
\label{eq:taylor}
\end{equation}
On the other hand, the Kraus operator formalism allows us to describe the quantum evolution over an infinitesimal time interval $dt$ as \begin{equation}
\hat{\rho}(dt) = \sum_{\alpha=\{i,j\}} \hat{K}_{ij}(dt)\, \hat{\rho}(0)\, \hat{K}_{ij}^\dagger(dt).
\label{eq:kraus}
\end{equation}
In order for expressions \eqref{eq:taylor} and \eqref{eq:kraus} to agree up to order $O(dt)$, we look for appropriate operators $\hat{K}_{\alpha=\{i,j\}}$. Since the first term in the Taylor expansion \eqref{eq:taylor} is simply  $\hat{\rho}_s(0)$,  at least one of the Kraus operators must contain the identity operator. This motivates the choice
\begin{equation}
\hat{K}_0 = \hat{\mathbb{I}} + \hat{L}_0\, dt.
\label{eq:K0}
\end{equation}
Expanding the first term in the Kraus representation \eqref{eq:kraus} yields
\begin{equation}\label{fisrtkraus}
\hat{K}_0 \hat{\rho}_s(0) \hat{K}_0^\dagger = \hat{\rho}_s(0) + \left( \hat{L}_0 \hat{\rho}_s(0) + \hat{\rho}_s(0) \hat{L}_0^\dagger \right) dt + O(dt^2).
\end{equation}
Although this expression is of the correct order, it is not sufficient. An evolution governed by a single Kraus operator describes a unitary evolution, and therefore we must include additional operators to account for the non-unitary part of the evolution characteristic of an open quantum system. We thus consider
\begin{equation}
\hat{K}_\alpha = \sqrt{dt}\, \hat{L}_\alpha, \quad \alpha \geq 1.
\label{eq:Kalpha}
\end{equation}
These operators \eqref{eq:Kalpha} contribute 
\begin{equation}\label{secondkraus}
\hat{K}_\alpha \hat{\rho}_s(0) \hat{K}_\alpha^\dagger = \hat{L}_\alpha \hat{\rho}_s(0) \hat{L}_\alpha^\dagger \, dt.
\end{equation}
Imposing the trace preservation condition $\,\sum_{\alpha=0} \hat{K}_\alpha^\dagger \hat{K}_\alpha = \hat{\mathbb{I}}$, together with Eqs.\eqref{eq:K0} and \eqref{eq:Kalpha}, we obtain
\begin{equation*}
    \hat{K}_0^\dagger \hat{K}_0 + \sum_{\alpha=1} \hat{K}_\alpha^\dagger \hat{K}_\alpha = \hat{\mathbb{I}},
\end{equation*}
\begin{equation}
    \hat{\mathbb{I}} + dt \left( \hat{L}_0 + \hat{L}_0^\dagger + \sum_{\alpha \geq 1} \hat{L}_\alpha^\dagger \hat{L}_\alpha \right) + \cancel{O(dt^2)}^{ \,\approx0} = \hat{\mathbb{I}},
\end{equation}
which leads to the condition 
\begin{equation}
\hat{L}_0 + \hat{L}_0^\dagger + \sum_{\alpha \geq 1} \hat{L}_\alpha^\dagger \hat{L}_\alpha = 0.
\end{equation}
We now decompose the linear operator  $\hat{L}_0$ into its Hermitian and anti-Hermitian parts as $\hat{L}_0 = \hat{A} - i \hat{H}$, where $\hat{A} = \hat{A}^\dagger,\ \hat{H} = \hat{H}^\dagger$ , yielding
\begin{equation}\label{operadorA}
\hat{A} = -\frac{1}{2} \sum_{\alpha \geq 1} \hat{L}_\alpha^\dagger \hat{L}_\alpha.
\end{equation}
The above decomposition of $\hat{L}_0$ is intended to associate the term $i\hat{H}$ with the unitary evolution of the system, in such a way that one recovers the form of the von Neumann equation. Consequently, the remaining term $\hat{A}$, is associated with the non-unitary effects in the system's dynamics. By combining Eqs. \eqref{fisrtkraus}, \eqref{secondkraus} and \eqref{operadorA}, we obtain
\begin{equation*}
\begin{aligned}
\hat{\rho}_s(dt) &= \hat{K}_0 \hat{\rho}_s(0) \hat{K}_0^\dagger + \sum_{\alpha \geq 1} \hat{K}_\alpha \hat{\rho}_s(0) \hat{K}_\alpha^\dagger ,\\
\hat{\rho}_s(dt) &= \hat{\rho}_s(0) + (\hat{A} - i \hat{H}) dt \, \hat{\rho}_s(0) + \hat{\rho}_s(0) (\hat{A} + i \hat{H}) dt + \sum_{\alpha \geq 1} \hat{L}_\alpha \hat{\rho}_s(0) \hat{L}_\alpha^\dagger dt +O(dt^2), \\
\hat{\rho}_s(dt) &= \hat{\rho}_s(0) - i [\hat{H}, \hat{\rho}_s(0)] dt + \{ \hat{A}, \hat{\rho}_s(0) \} dt + \sum_{\alpha \geq 1} \hat{L}_\alpha \hat{\rho}_s(0) \hat{L}_\alpha^\dagger dt + O(dt^2), \\
\hat{\rho}_s(dt) &= \hat{\rho}_s(0) - i [\hat{H}, \hat{\rho}_s(0)] dt + \sum_{\alpha \geq 1} \left( \hat{L}_\alpha \hat{\rho}_s(0) \hat{L}_\alpha^\dagger - \frac{1}{2} \left\{ \hat{L}_\alpha^\dagger \hat{L}_\alpha, \hat{\rho}_s(0) \right\} \right) dt + O(dt^2).
\end{aligned}
\end{equation*}
Hence,
\begin{equation}\label{diferenciarho}
    \hat{\rho}_s(dt) - \hat{\rho}_s(0) = - i [\hat{H}, \hat{\rho}_s(0)] dt + \sum_{\alpha \geq 1} \left( \hat{L}_\alpha \hat{\rho}_s(0) \hat{L}_\alpha^\dagger - \frac{1}{2} \left\{ \hat{L}_\alpha^\dagger \hat{L}_\alpha, \hat{\rho}_s(0) \right\} \right) dt + O(dt^2).
\end{equation}
Now, taking the definition of the time derivative of $\hat{\rho}_s(t)$ evaluated at $t=0$
\begin{equation}\label{derivadadiferenciarho}
\frac{d\hat{\rho}_s(t)}{dt}\Big|_{t=0} = \lim_{dt \to 0} \frac{\hat{\rho}_s(t+dt) - \hat{\rho}_s(t)}{dt}\Big|_{t=0}=\lim_{dt \to 0} \frac{\hat{\rho}_s(dt) - \hat{\rho}_s(0)}{dt}.
\end{equation}
Substituting Eq.\eqref{diferenciarho} into Eq.\eqref{derivadadiferenciarho} and taking the limit $dt \rightarrow 0$, which makes the $O(dt^2)$ term vanish, we obtain
\begin{equation}
    \frac{d\hat{\rho}_s(t)}{dt}\Big|_{t=0} = -i[\hat{H}, \hat{\rho}(0)] + \sum_{\alpha \geq 1} \left( \hat{L}_\alpha \hat{\rho}(0) \hat{L}_\alpha^\dagger - \frac{1}{2} \left\{ \hat{L}_\alpha^\dagger \hat{L}_\alpha, \hat{\rho}(0) \right\} \right).
\end{equation}
Initially, we take the operators $\hat{L}_\alpha$ with dimensions $[1/\sqrt{\text{time}}]$, so that if we wish to make them dimensionless, we perform the following rescaling $\hat{L}_\alpha \rightarrow \sqrt{\gamma_\alpha}, \hat{L}_\alpha$, where $\gamma_\alpha \geq 0$ has units of $[1/\text{time}]$
\begin{equation}\label{prelindblad}
\frac{d\hat{\rho}_s(t)}{dt}\Big|_{t=0}  = -i[\hat{H}, \hat{\rho}_s(0)] + \sum_{\alpha \geq 1} \gamma_\alpha \left( \hat{L}_\alpha \hat{\rho}_s(0) \hat{L}_\alpha^\dagger - \frac{1}{2} \left\{ \hat{L}_\alpha^\dagger \hat{L}_\alpha, \hat{\rho}_s(0) \right\} \right).
\end{equation}
Finally, under the Markov approximation, one assumes that the system does not retain memory of its past, allowing the extension of Eq. \eqref{prelindblad} to all times $t \geq 0$. Consequently, Eq. \eqref{prelindblad} becomes time-local, leading to the general form of the Lindblad master equation
\begin{equation}\label{lindbladequationkraus}
\frac{d\hat{\rho}_s(t)}{dt} = -i[\hat{H}, \hat{\rho}_s(t)] + \sum_\alpha \gamma_\alpha \left( \hat{L}_\alpha \hat{\rho}_S(t) \hat{L}_\alpha^\dagger - \frac{1}{2} \left\{ \hat{L}_\alpha^\dagger \hat{L}_\alpha, \hat{\rho}_s(t) \right\} \right).
\end{equation}

\subsection{Non-Markovian Master equation}

In the previous sections, we established that the Markovian master equation (in the Lindblad form) is valid only under specific conditions, namely when the system–environment coupling is weak and memory effects from the environment can be neglected. However, in physically relevant scenarios, such as those involving low environmental temperatures or strong system–environment interactions, these assumptions break down. A concrete example is that of a superconducting qubit strongly coupled to an environment composed of other two-level systems at low temperatures, where both the Born and Markov approximations fail to hold~\cite{maximilian2007}.

\noindent When the environment exhibits significant memory effects, the evolution of the reduced density operator $\hat{\rho}_s(t)$ is expected to depend on its entire history. As a consequence, a description based on time-local differential equations becomes inadequate. In such cases, one must resort to integro-differential equations with memory kernels, such as those derived from the projection operator formalism developed by Nakajima and Zwanzig  \cite{maximilian2007}. 

\noindent In essence, the starting point is the von Neumann equation governing the evolution of the total system-environment state $\hat{\rho}(t)$, which evolves unitarily under the action of the total Liouvillian $\mathbb{L}$,
\begin{equation}\label{vonNeumancompuesto}
    \frac{d}{dt} \hat{\rho}(t) = \mathbb{L} \hat{\rho}(t) = -\frac{i}{\hbar} [\hat{H}_{\text{tot}}(t), \hat{\rho}(t)].
\end{equation}
To derive an effective dynamics for the system $S$, a projection operator $\hat{P}$ is introduced, which extracts the relevant information about the system from the total state. A common choice for $\hat{P}$ assumes that the initial state is factorized,
\begin{equation}\label{proyectorP}
    P \hat{\rho}(t) = \text{tr}_e[\hat{\rho}(t)] \otimes \hat{\rho}_e = \hat{\rho}_s(t) \otimes \hat{\rho}_e,
\end{equation}
where $\hat{\rho}_e$ is taken to be a fixed reference state, which from a thermodynamic perspective corresponds to an equilibrium state. The complementary projection operator is defined as $\hat{Q} = \mathbb{I} - \hat{P}$, and it acts on the irrelevant part of the dynamics, which may include thermal fluctuations of the environment and system–environment correlations. 

\noindent Following a rigorous derivation, as detailed in Ref.\cite{Breuer:Book}, one arrives at the following exact evolution equation for the projected part, \begin{equation}\label{ecuacionproyectada}
    \frac{d}{dt} \hat{P}\hat{\rho}(t) = -\frac{i}{\hbar}\hat{P}\mathcal{L} \hat{P}\hat{\rho}(t) -\frac{1}{\hbar^2} \int_{t_0}^t d\tau\, \hat{P}\mathbb{L}\hat{Q} \,e^{-\frac{i}{\hbar}\hat{Q} \mathbb{L} \hat{Q} (t - \tau)} \hat{Q} \mathbb{L} \hat{P}\hat{\rho}(\tau),
\end{equation}
Taking the partial trace over the environmental degrees of freedom leads to the non local in time equation for the reduced density operator
\begin{equation}\label{ecuacionnakajimareduzido}
    \frac{d}{dt} \hat{\rho}_s(t) = -\frac{i}{\hbar}[\hat{H}_s(t), \hat{\rho}_s(t)] -\frac{1}{\hbar^2} \int_{t_0}^{t} \mathcal{K}_{\tau}[\hat{\rho}_s(t-\tau)] \, d\tau
\end{equation}
where $t \geq \tau \geq t_0$. Equation \eqref{ecuacionnakajimareduzido} is known as the generalized Nakajima-Zwanzig master equation. The superoperator $\mathcal{K}\tau$, often referred to as the memory kernel, governs the influence of the system's past state $\hat{\rho}_s(t-\tau)$ on its present evolution $\frac{d}{dt} \hat{\rho}_s(t)$. This memory kernel is precisely what gives rise to non-Markovian effects. In the regime where the environment relaxes much faster than the characteristic time-scale of the system, the information transferred from the system to the bath is quickly dissipated. This rapid dissipation significantly suppresses memory effects, causing $\mathcal{K}(\tau) \rightarrow \delta(\tau)$, which transforms the non local in time equation into a time local form, and subsequently allows it to be reduced to a Markovian description~\cite{Breuer:Book}.

\noindent Nevertheless, obtaining a closed analytical expression for the memory kernel is extremely challenging due to the exponential operator $e^{-\frac{i}{\hbar}\hat{Q} \mathbb{L} \hat{Q} (t - \tau)}$. This complexity represents one of the main obstacles to the explicit solution of the Nakajima-Zwanzig equation. Remarkably, there are many physical processes that, although exhibiting non-Markovian dynamics, can still be described by time-local master equations, as presented in Eq. \eqref{definicionderivadacptp}
\begin{equation}\label{Ldependientedetiempo}
    \frac{d}{dt}\hat{\rho}_s(t)=\mathcal{L}_t\big(\hat{\rho}_s(t)\big),
\end{equation}
where the superoperator $\mathcal{L}_t$ depends explicitly on time, but not on past times. One way to arrive in Eq. \eqref{Ldependientedetiempo} is by applying the time-convolutionless approximation to Eq. \eqref{ecuacionnakajimareduzido} (see Ref. \cite{Breuer:Book}). 

\noindent As a result, although the dynamics remains non-Markovian in general, it is now governed by a time-local equation for $\hat{\rho}_s(t)$. The form of $\mathcal{L}_t$ must represent a physically valid process, meaning that the associated quantum map must be CPTP. In the previous sections, we presented a way to derive a time-independent version of $\mathcal{L}$ from the Kraus formalism. In that case, we assumed constant positive decoherence rates $\gamma_k$, which corresponds to fully Markovian dynamics. However, in realistic physical scenarios, these rates are typically time dependent, and thus approximations such as Born and Markov are no longer hold. 

\noindent In Ref.\cite{rivas2011}, the original result by Gorini–Kossakowski–Lindblad–Sudarshan was extended to include time-dependent generators $\mathcal{L}_t$,  ensuring that the dynamics remains CPTP and Markovian. The general form of $\mathcal{L}_t$ is given by
\begin{equation} \label{lindbladnohomogeneo}
\frac{d \hat{\rho}_s(t)}{dt} = -\frac{i}{\hbar}[\hat{H}_s(t), \hat{\rho}_s(t)] + \sum_{k=1}^{K} \gamma_k(t) \left( \hat{L}_k(t) \hat{\rho}_s(t) \hat{L}_k^\dagger(t) - \frac{1}{2} \left\{ \hat{L}_k^\dagger(t) \hat{L}_k(t), \hat{\rho}_s(t) \right\} \right) .
\end{equation}
Here, the operators $\hat{H}_s(t)$ and $\hat{L}_k(t)$, as well as the decay rates $\gamma_k(t)$, are explicitly time dependent. In particular, when $\gamma_k(t) \geq 0$ , the dynamics can be considered fully Markovian. In this case, the evolution is referred to as a time inhomogeneous Markovian master equation. On the other hand, if at least one decay rate  $\gamma_k(t) \leq 0$ becomes negative over a finite time interval, the process is generally classified as non-Markovian \cite{Breuer:Book}. 

\noindent The distinction between the time homogeneous and time inhomogeneous Markovian master equations,  \eqref{lindbladhomogeneo} and \eqref{lindbladnohomogeneo},  lies in the properties of their associated quantum maps. First, the map corresponding to a time homogeneous Markovian equation takes the form $\Lambda_{t - t_0} = e^{\mathcal{L}(t - t_0)}$ being $t \geq t_0$, and the set of such maps forms a one-parameter quantum semigroup. As previously discussed, in this semigroup we have $\Lambda_{t, t_0} = \Lambda_{t - t_0}$~\cite{Breuer2009}. 

\noindent Similarly, the time inhomogeneous Markovian master equation generates a family of quantum maps CPTP depending on two time parameters, expressed as
\begin{equation}\label{mapanohomgeneo}
    \Lambda_{t, t'} = \hat{\mathcal{T}} \exp \left\{ \int_{t'}^{t} \mathcal{L}
    _\tau d\tau \right\},
\end{equation}
where $\hat{\mathcal{T}}$ denotes the time-ordering operator, and $t \geq t' $.

\noindent It is important to emphasize that both equations \eqref{lindbladhomogeneo} and \eqref{lindbladnohomogeneo} must satisfy the divisibility condition for CPTP maps. In the homogeneous case, the divisibility criterion is derived from the closure property of semigroups \begin{equation} \label{divisibilidadhomogenea}
\Lambda_{t - t_0} = \Lambda_{t - t'} \circ \Lambda_{t' - t_0}.
\end{equation}
In contrast, in the time-inhomogeneous case, the divisibility condition takes the form \begin{equation} \label{divisibilidadnohomogenea}
\Lambda_{t, t_0} = \Lambda_{t, t'} \circ \Lambda_{t', t_0}.
\end{equation}
In both expressions, \eqref{divisibilidadhomogenea} and \eqref{divisibilidadnohomogenea}, each map must be CPTP for $t \geq  t' \geq t_0$. However, these two criteria are not equivalent. In the homogeneous case, the maps depend only on the time interval, $\Lambda_{t,t_0}=\Lambda_{t-t_0}$, and the divisibility follows from the semigroup composition property. In contrast, in the non-homogeneous case, the quantum map depends on the initial and final times independently(see Eq.~\eqref{mapanohomgeneo}), which implies $\Lambda_{t,t_0} \neq \Lambda_{t-t_0}$~\cite{Breuer2009}.

\section{Trace Distance Measure}

The trace distance between two quantum states reflects the probability of distinguishing them. Based on this idea, Markovian and non-Markovian processes can be differentiated. In a Markovian process, the distinguishability between states decreases, while in a non-Markovian process it increases. This growth in distinguishability can be interpreted as a backflow of information from the environment to the system \cite{Breuer2009}.

\noindent The construction of this measure based on the trace distance $D(t)$ was developed in the work of Ref. \cite{Breuer2009}, where the measure is also referred to as the BLP measure. To define this non-Markovianity measure, we first need the definition of the trace distance between two quantum states of the same system, $\hat{\rho}_s(t)$ and $\hat{\sigma}_s(t)$, which is given by
\begin{equation}\label{distanciatraza}
D\big(\hat{\rho}_s(t),\hat{\sigma}_s(t)\big) = \frac{1}{2} \text{tr} \| \hat{\rho}_s(t) - \hat{\sigma}_s(t) \|,
\end{equation}
where $\|\hat{A}\| = \sqrt{\hat{A}^\dagger \hat{A}}$. 

\noindent The trace distance $D$ defines a metric in the space of physical states of the system and is widely used in quantum information theory. The value of $D$ represents a quantitative measure of distinguishability between two quantum states, and it satisfies
\begin{equation}\label{limitedistancia}
0 \leq D\big(\hat{\rho}_s(t),\hat{\sigma}_s(t)\big) \leq 1.
\end{equation}
A key feature of this metric is its behavior under the action of CPTP quantum maps, denoted by $\Phi_{t,t_0}$, for which the following inequality must hold
\begin{equation}\label{decrecimientoCPTPdeD}
D\Big(\Phi_{t,t_0} \big(\hat{\rho}_s(t_0)\big)\,, \,\Phi_{t,t_0}\big( \hat{\sigma}_s(t_0)\big)\Big) \leq D\big(\hat{\rho}_s(t), \hat{\sigma}_s(t)\big),
\end{equation}
Maps of the form $\Phi_{t,t_0}$ contract the trace distance and describe a Markovian evolution if Eq. \eqref{decrecimientoCPTPdeD} holds at all times. This implies that, under a Markovian dynamic, the distinguishability between states decreases monotonically over time, reflecting a continuous loss of information from the system to the environment. Therefore, when $D(t)$ is decreasing, one has $\frac{d}{dt}D(t) \leq 0$ for all Markovian quantum processes, whether they follow a dynamical semigroup or are time-dependent \cite{Breuer2009}. 

\noindent Physical processes for which $\frac{d}{dt}D(t) \geq 0$ during certain time intervals represent non-Markovian dynamics. Physically, this means that in a non-Markovian evolution, the distinguishability between the pair of states increases at certain times due to a temporary information backflow. In this context, non-Markovian character implies that the associated quantum map does not satisfy the divisibility criterion for quantum maps, given by Eq. \eqref{divisibilidadnohomogenea}, as evidenced by the increase in $D(t)$. It is important to emphasize that such information feedback is a natural feature of many physical systems and does not necessarily imply the absence of thermalization at long times, as expected from a thermodynamic perspective \cite{Breuer2009}.

\noindent Consequently, the measure based on $D(t)$ should quantify the total increase in distinguishability throughout the entire time evolution, that is, the total amount of information recovered from the environment by the system. 

\noindent Therefore, the measure is defined as
\begin{equation}\label{medidorblp}
\mathcal{N}_{\text{D}} = \max_{\hat{\rho}_s(t_0), \hat{\sigma}_s(t_0)} \int_{\frac{d}{dt} D(t) > 0} \frac{d}{dt} D(t) \, dt,
\end{equation}
where the integral is taken over the time intervals in which the distinguishability between the quantum states increases. A pair of initial states evolving under the same quantum map that leads to the maximum possible recovery of information will maximize the value of the integral.

\noindent Since the function $D(t)$ is continuous (due to the continuity of the trace norm) and piecewise differentiable (as a result of well-defined master equations generating smooth evolutions), it is possible to identify the intervals $(a_i, b_i)$ in which $\frac{d \,D(t)}{dt}>0$. In each of these intervals $i$, the fundamental theorem of calculus can be applied to rewrite the integral as a sum of finite differences
\begin{equation}\label{teoremafundamentalcalculo}
\int_{a_i}^{b_i} \frac{d}{dt} D(t) \, dt = D(b_i) - D(a_i).
\end{equation}
Thus, the non-Markovianity measure can be expressed as the sum of the net increases in trace distance over all intervals where it grows
\begin{equation}\label{medidorblpdiscreto}
\mathcal{N}_{\text{D}} = \max_{\hat{\rho}_s(t_0), \hat{\sigma}_s(t_0)} \sum_i \left[ D\big(\hat{\rho}_s(b_i), \hat{\rho}_s(b_i)\big) - D\big(\hat{\rho}_s(a_i), \hat{\rho}_s(a_i)\big) \right].
\end{equation}
An important point about this measure arises from Eq. \eqref{decrecimientoCPTPdeD}, which is not only valid for CPTP maps but also for a broader class of maps that are positive and trace-preserving, even when their corresponding Markovian master equations are not in Lindblad form. Therefore, the measure defined in Eq. \eqref{medidorblp} also applies to Markovian dynamics which, although not in Lindblad form, still preserve positivity and trace \cite{Breuer2009}.

\end{chapter}
\begin{chapter}{QUANTUM THERMODYNAMICS REVIEW}
\label{cap3}
\vspace{-0.1cm}
\noindent When extending the concepts of classical thermodynamics to the quantum domain, a fundamental question arises: how should one define the internal energy, heat, and work of a system interacting with its environment? For a closed system composed of a subsystem and its surroundings, the total energy is described by the expectation value of the total Hamiltonian. However, in the theory of open quantum systems, the focus is on characterizing only the reduced system, which makes it challenging to determine which part of the total energy should be considered as internal to the system and which belongs to the environment.
\vspace{0.2cm}\\
\noindent In weak-coupling regimes, it is generally safe to associate the internal energy of the system and the environment with the expectation values of their respective Hamiltonians. Yet, this simplification fails in cases of strong coupling, structured environments, non-Markovian effects, or finite baths, where no consensus has been reached on how to properly define internal energy, heat, work, or entropy production.
\vspace{0.2cm}\\
\noindent Several approaches have been proposed to address these challenges, based on different criteria such as entropy, coherence, ergotropy, or stochastic methods, with the most commonly used being the one proposed by \cite{alicki}. Nevertheless, none of these approaches can be considered universal, as they exhibit inconsistencies in certain thermodynamic processes.

\section{Classical Thermodynamics}

The thermodynamic study of a natural physical process begins by dividing it into two parts, the system under study and its environments. The system can be defined as a thermodynamic system because of the existence of boundaries. These boundaries separate the system from its environments and regulate how they interact with each other, which can occur through the exchange of energy or matter~\cite{Greiner1997}.

\noindent Classically, thermodynamic systems are categorized into three types. Isolated systems do not exchange energy or matter with their surroundings. Closed systems allow energy exchange, but do not matter. And open systems allow the exchange of both energy and matter with their environments, resulting in irreversible processes.

\noindent The state of a thermodynamic system can be described in terms of volume $V$, pressure $P$, temperature $T$, and the number of moles $N$. These parameters are defined as macroscopic state variables. Moreover, thermodynamics introduces the concepts of energy, $U=U(T,V,N)$, and entropy, $S=S(T,V,N)$, which are functions of state variables, often called as state functions. The concepts $U$ and $S$ form the foundation of the first and second laws of thermodynamics~\cite{Greiner1997}.

\subsection{First Law}

In every real physical process, the principle of energy conservation is fundamental. Empirical evidence indicates that, to date, this principle holds true at both macroscopic and microscopic scales.
For a closed system, the change in internal energy during an arbitrary change of state, whether in a reversible or irreversible process, can be expressed as the sum of the work $\delta W$ and heat $\delta Q$ exchanged with its surroundings~\cite{Greiner1997},
\begin{equation}\label{eq:1Law-classical}
    dU=\delta W + \delta Q
\end{equation}
Note that $\delta W$ and $\delta Q$ associated with a small change of state may depend on the thermodynamic process, so they are not exact differentials. The symbol $\delta$ is used to indicate that they are inexact differentials. In contrast, the change $dU$ depends only on the initial and final states of the system, regardless of the process. For this reason, internal energy has an exact differential. It is important to emphasize that the first law remains valid irrespective of whether the state change is reversible or irreversible~\cite{Greiner1997},
\begin{equation}\label{eq:1LawRevIrr}
    dU=\delta W_{rev}+\delta Q_{rev}=\delta W_{irr}+\delta Q_{irr}
\end{equation}

\noindent The case  of isothermal expansion and compression of an ideal gas serves as a starting point to illustrate a broader principle. In general, the absolute value of the work performed is greater in a reversible process than in an irreversible one. Similarly, the work required for an irreversible compression always exceeds that of a reversible process. More generally , considering the sign, we arrive at the conclusion that~\cite{Greiner1997},
\begin{equation}\label{eq:WrevWirr}
    \delta W_{rev} \leq \delta W_{irr} 
\end{equation}
This mean that a reversible change is more efficient than any irreversible process between the same two states.
From Eq.\eqref{eq:WrevWirr}, it can be concluded that in an irreversible process, the magnitude of the heat dissipated to the surroundings always exceeds than in a reversible process and less heat is needed \cite{Greiner1997},
\begin{equation}
    \delta Q_{irr}\leq \delta Q_{rev} 
\end{equation}

\subsection{Second Law}

The second law of thermodynamics originates from the pioneering studies of Carnot on the efficiency of heat engines during the Industrial Revolution \cite{manzano2018}. Carnot identified heat flow as essential for work generation and established that the maximum efficiency of an ideal engine operating between a hot reservoir $T_{\rm hot}$ and a cold reservoir $T_{\rm cold}$ is achieved only in reversible processes \cite{manzano2018},
\begin{equation}
\eta_C = \frac{W}{Q_{\rm hot}} = 1 - \frac{T_{\rm cold}}{T_{\rm hot}}.\\[5pt]
\end{equation}
In a thermodynamic process connecting two equilibrium states $A$ and $B$, multiple paths may be followed, each associated with different amounts of work and heat exchanged. Among these, reversible paths are special, as the system remains infinitesimally close to equilibrium at every step. The existence of such reversible trajectories led Clausius to introduce the concept of thermodynamic entropy, defined through its change in a reversible process \cite{manzano2018}
\begin{equation}
\Delta S^{\rm \,thermo} = S^{\rm \,thermo}_{(B)} - S^{\rm \,thermo}_{(A)} = \int_A^B \frac{\delta Q_{\rm rev}}{T},\\[5pt]
\end{equation}
where $\delta Q_{\rm rev}$ is the heat absorbed by the system in a reversible step. 
\vspace{0.2cm}\\
\noindent Since thermodynamic entropy depends only on the initial and final states of the process, the differential becomes exact $dS^{\rm \,thermo} = \delta Q_{\rm rev}/T$. For irreversible processes, entropy satisfies the Clausius inequality
\begin{equation}
\Delta S^{\rm \,thermo} \ge \int_A^B \frac{\delta Q_{\rm irr}}{T}.
\end{equation}
The entropy change $\Delta S^{\rm \,thermo}$ can be split as
\begin{equation}
\Delta S^{\rm \,thermo} = \Delta_i S^{\rm \,thermo} + \Delta_e S^{\rm \,thermo},
\end{equation}
which provides a modern formulation of the second law. Here, the entropy flow $\Delta_e S^{\rm \,thermo}$ accounts for exchanges with the environment
\begin{equation}
\Delta_e S^{\rm \,thermo} = \int_A^B \frac{\delta Q}{T},
\end{equation}
where $\delta Q$ is the heat exchanged between the system and the environment. The remaining term, called the entropy production $\Delta_i S^{\rm \,thermo}$, quantifies the entropy generated within the system due to irreversible processes and satisfies
\begin{equation}
\Delta_i S^{\rm \,thermo} = \Delta S^{\rm \,thermo} - \int_A^B \frac{\delta Q}{T} \ge 0.
\end{equation}
If the environment acts as a thermal reservoir, maintaining its state even while exchanging energy with the system, the entropy flow satisfies
\begin{equation}
\Delta_e S^{\rm \,thermo} = - \Delta S_{\rm env}^{\rm \,thermo}.
\end{equation}
Consequently, the total entropy production can be expressed as the sum of changes in the system and environment
\begin{equation}
\Delta_i S^{\rm \,thermo} = \Delta S^{\rm \,thermo}_{\rm sys} + \Delta S_{\rm env}^{\rm \,thermo}
\end{equation}
This formulation summarizes the second law of thermodynamics, stating that the sum of entropy changes in a system and its environment never decreases, reflecting the fundamental irreversibility of natural processes \cite{manzano2018}.

\section{First Law of Quantum Thermodynamics}

\subsection{First Law under Weak Coupling}

As we have seen in previous sections, the first law of quantum thermodynamics is based on the principle of energy conservation. In macroscopic classical thermodynamics, defining energy does not present conceptual difficulties, since there is a clear separation between system and environment and no complex quantum effects come into play. However, in open quantum systems, the internal energy is usually identified as $U(t)=\braket{\hat{H}_s(t)}$, as established in quantum statistical mechanics. It is important to highlight that this expression is only valid in the weak coupling regime, without significant correlations. When the system interacts strongly with its environment, or when quantum coherence and entanglement are present, part of the energy can be stored in the interaction term or in correlations with the environment. This makes the definition of internal energy ambiguous and, as a consequence, the specific definitions of heat and work in the quantum regime become, in general, nontrivial~\cite{Picatoste2024, rivas2019}.
\vspace{0.2cm}\\
\noindent Let us first identify the problem regarding the internal energy of the system by starting from the total Hamiltonian of the open system, given by \begin{equation}\label{Htotal}
    \hat{H}_\text{tot}(t)= \hat{H}_s(t)+\hat{H}_e(t)+g\,\hat{H}_\text{int}(t)
\end{equation} Taking the expectation value, we obtain
\begin{equation*}
    \braket{\hat{H}_\text{tot}(t)} = \braket{\hat{H}_s(t)} + \braket{\hat{H}_e(t)} + g\,\braket{\hat{H}_\text{int}(t)}
\end{equation*}
\begin{equation}\label{Utotal}
    U_\text{tot}(t)=\text{tr}\big[\hat{\rho}_s(t) \hat{H}_s(t)\big] + \text{tr}\big[\hat{\rho}_e(t) \hat{H}_e(t)\big] + g\,\, \text{tr}\big[\hat{\rho}(t) \hat{H}_\text{int}(t)\big]\\[7pt]
\end{equation}
From the total energy $U_\text{tot}(t)= \braket{\hat{H}_\text{tot}(t)}$, it is natural to identify $\braket{\hat{H}s(t)}$ as part of the internal energy of the system. However, it is not clear which fraction of the interaction term $\braket{\hat{H}_\text{int}(t)}$ should also be assigned to it. In the weak coupling limit, determined by $g$, Eq. \eqref{Utotal} reduces to \begin{equation}
    U_\text{tot}(t)=\braket{\hat{H}_\text{tot}(t)}\approx \text{tr}\big[\hat{\rho}_s(t) \hat{H}_s(t)\big] + \text{tr}\big[\hat{\rho}_e(t) \hat{H}_e(t)\big]
\end{equation}
Within this assumption, the total internal energy of the composite is approximately equal to the sum of the system and the environment, $U_\text{tot}(t) \approx U_s(t)+U_e(t)$. Consequently, the internal energy of the system can be consistently defined as the expectation value of its Hamiltonian, representing the average energy during the evolution
\begin{equation}\label{Usweakcoupling}
    U_s(t)= \text{tr}\big[\hat{\rho}_s(t)\, \hat{H}_s(t)\big]
\end{equation}
Following the work of R. Alicki \cite{alicki}, this provides the standard formalism of quantum thermodynamics in the weak coupling regime. Accordingly, the first law of quantum thermodynamics states that any change in the internal energy of the system can only result from an energy exchange with the environment, in the form of either heat or work. Taking the time derivative of Eq. \eqref{Usweakcoupling}, we obtain 

\begin{equation}\label{dUsweak}
\frac{dU_s(t)}{dt}=\frac{d}{dt} \, \text{tr} \left[ \hat{H}_{s}(t) \, \hat{\rho}_{s}(t) \right]
= \underbrace{\text{tr} \left[ \frac{d \hat{H}_{s}(t)}{dt} \, \hat{\rho}_{s}(t) \right]}_{\text{WORK}}
+ \underbrace{\text{tr} \left[ \hat{H}_{s}(t) \, \frac{d\hat{\rho}_{s}(t)}{dt} \right]}_{\text{HEAT}}\\[7pt]
\end{equation}
The first term in \eqref{dUsweak} is identified as the instantaneous work flux $\delta W_\text{stand}(t)$, which originates from changes in the energy levels or eigenstates of the system Hamiltonian. The second term in \eqref{dUsweak} corresponds to the heat flux $\delta Q_\text{stand}(t)$, which is produced by changes in the state of the system, either in the populations or in the eigenbasis of the density operator.
\vspace{0.2cm}\\
\noindent Thus, the instantaneous work flux and the total average work exchanged in a time interval are given by
\begin{align}
    \delta W_\text{stand}(t)&=\text{tr} \left[ \frac{d \hat{H}_{s}(t)}{dt} \, \hat{\rho}_{s}(t) \right]\\[5pt]
    W_\text{stand}(t)&=\int_{t_0}^{t} \text{tr}\left[\frac{d \hat{H}_{s}(\tau)}{d\tau} \, \hat{\rho}_{s}(\tau)\right]\, d \tau
\end{align}
Similarly, the instantaneous heat flux and the average heat exchanged are given by
\begin{align}
    \delta Q_\text{stand}(t)&=\text{tr} \left[ \hat{H}_{s}(t) \, \frac{d\hat{\rho}_{s}(t)}{dt} \right]\\[5pt]
    Q_\text{stand}(t)&=\int_{t_0}^{t} \text{tr} \left[ \hat{H}_{s}(\tau) \, \frac{d\hat{\rho}_{s}(\tau)}{d\tau} \right]\, d \tau
\end{align}
It is worth noting that only in the case of a time-dependent Hamiltonian is work performed on the system. Therefore, the source of work lies in the external driving.

\subsection{Semiclassical First Law}

If a physical system can exchange energy with its environment in such a way that the system changes from state A to state B, the First Law of Thermodynamics from the classical perspective is defined as the sum of work and heat being equal to the change in the state function U \cite{Balian},
\begin{equation}\label{eq:3-1ley}
U_{A}-U_{B}= w+q
\end{equation}
From the point of view of statistical mechanics, the internal energy $U$ is defined as the mean value of the Hamiltonian $\hat{H}$ for the system in a  quantum state, which is  represented by the density operator $\hat{\rho_s}$,
\begin{equation}\label{eq:3.1-internal energy}
U=\text{tr}[\hat{\rho_s} \ \hat{H}_s]    
\end{equation}
If, $\left[\hat{\rho_s},\hat{H}_s\right]=0$, the state of the system can be expressed in the Hamiltonian eigenbasis as $\hat{\rho}_s = \sum_n P_n |E_n\rangle \langle E_n|$, with probabilities $P_n$ for each energy eigenstate. Let us consider
a Hamiltonian $\hat{H}_s= \sum^{ }_{n}E_n \ket{E_n}\bra{E_n}$, where $E_n$ and $\ket{E_n}$ are the $n$-th energy eigenvalues and eigenstates, respectively. In this case, the internal energy becomes \cite{alipour1, bernardo, Balian}
\begin{equation}\label{eq:3.1-calorclasico}
    U=\sum^{ }_{n=0}P_n E_n,
\end{equation}
where $P_n = \bra{E_n} \hat{\rho} \ket{E_n}$ is the probability of the system being in the $n$-th state.

\noindent When the system is open, energy exchange with the environment can occur. Microscopically, heat is associated with changes in the populations of energy levels
\begin{equation}\label{eq:3.1-deltacalor}
\delta q = \mathrm{Tr}[d\hat{\rho}_s \, \hat{H}_s] = \sum_n E_n \, dP_n,
\end{equation}
while work arises from changes in the energy levels themselves, caused by external manipulations
\begin{equation}
\delta w = \mathrm{Tr}[\hat{\rho}_s \, d\hat{H}_s] = \sum_n P_n \, dE_n.
\end{equation}
Thus, the semiclassical version of the first law reads,
\begin{equation}
dU = \delta q + \delta w = \sum_n E_n \, dP_n + \sum_n P_n \, dE_n.
\end{equation}
Physically, this means that changes in internal energy occur through two mechanisms \cite{Balian}:
\begin{itemize}
    \item Heat-energy transfer due to redistribution probabilistic of populations, increasing when higher-energy states become more probable.
    \item Work-energy transfer caused by deterministic shifts in energy levels without necessarily altering the populations.
\end{itemize}
\noindent However, this semiclassical decomposition into $q$ and $w$ is no longer valid when 
$[\hat{\rho}_s, \hat{H}_s] \neq 0$. In this situation, the state shows coherences in the energy eigenbasis, and these contribute energy in a way that cannot be classified as heat or work. Therefore, a generalized thermodynamic framework is required to properly describe the energy flow in quantum systems~\cite{bernardo}.

\section{Quantum Coherence}

It is known that quantum coherence is a resource that depends on the choice of basis. Each preferred basis is dictated by the physics of the problem; for example, the energy eigenbasis is commonly used in transport and thermodynamic studies. Considering this, let us take a reference orthonormal basis ${\ket{i}}_{i=0,1,...,d-1}$, where $d$ is the Hilbert space dimension. States that are strictly diagonal in this basis are called incoherent states, as they do not contain quantum coherence. These incoherent states have the following form \cite{cramer}
\begin{equation}\label{estadoincoherente}
\hat{\rho}_s^{\,\text{incoh}} = \sum_{j=0}^{d-1} p_j \, \ket{j}\bra{j}, 
\quad p_j \geq 0, \quad \sum_{j=0}^{d-1} p_j = 1.
\end{equation}
Here, the coefficients $p_j$ represent probabilities. The set composed of all incoherent states is denoted by $\mathcal{I}$. Consequently, this type of state does not include any coherent elements, and one way to generate such states is through the full dephasing or classicalization map $\Delta[\cdot]$. This map projects any state onto the reference basis by eliminating all coherences, thus making the state classical in the chosen basis \cite{streltsov}
\begin{equation}\label{fulldephasingmap}
    \Delta(\hat{\rho}_s) = \sum_{j=0}^{d-1} \braket{j|\hat{\rho}_s|j} \, \ket{j}\bra{j}.
\end{equation}
Having defined an incoherent state, we now discuss the existence and definition of incoherent maps $\Phi^{\text{icptp}}$. These maps are completely positive and trace-preserving (CPTP) maps whose Kraus operators additionally satisfy the following condition~\cite{vershynina}
\begin{equation}\label{mapaicptp}
   \hat{K}_j \, \hat{\rho}_s^{\,\text{incoh}} \, \hat{K}_j^{\dagger} \in \mathcal{I}, \quad \forall \, \hat{\rho}_s^{\,\text{incoh}} \in \mathcal{I}.
\end{equation}
An abbreviated way to refer to this type of map is by using the label ICPTP maps. Additionally, a subclass of ICPTP maps is called genuinely ICPTP maps, which are quantum maps that preserve any incoherent state \cite{vershynina}
\begin{equation}
    \Phi^{\,\text{g-icptp}}\, \big(\hat{\rho}_s^{\,\text{incoh}}\big)=\hat{\rho}_s^{\,\text{incoh}}.
\end{equation}
This type of map is valid only if all Kraus representations have diagonal operators in the pre-fixed basis. 

\noindent Another class of ICPTP maps is called strictly ICPTP, for which the Kraus operators commute with the full dephasing map $\Delta$ \cite{vershynina},
\begin{equation}
    \hat{K}_j \, \Delta(\hat{\rho}_s) \, \hat{K}_j^{\dagger} \;=\; \Delta\!\left( \hat{K}_j \, \hat{\rho}_s \, \hat{K}_j^{\dagger} \right).
\end{equation}
With the existence of incoherent maps established, the coherent terms of a state are linked to the definition of quantum coherence, extensively discussed in the literature. Quantum coherence, denoted $C(\hat{\rho}_s)$, acts as a functional that quantifies the coherence of an arbitrary quantum state $\hat{\rho}_s$. A widely used approach, proposed in Ref. \cite{cramer}, consists of measuring the distance between a state $\hat{\rho}_s$ and the set of incoherent states,
\begin{equation}\label{funcionalC}
    C(\hat{\rho}_s) = \min_{\hat{\rho}_s^{\, \text{incoh}}\,\in\, \mathcal{I}} \left\{\, D(\hat{\rho}_s, \hat{\rho}_s^{\, \text{incoh}})\right\},
\end{equation}
where $D(\hat{\rho}_s, \hat{\rho}_s^{\, \text{incoh}})$ represents a suitable distance measure, such as the trace norm or the relative entropy. Based on this definition, different coherence measures can be constructed, and the functional $C$ is considered a valid coherence measure if it satisfies the following conditions~\cite{streltsov, vershynina}
\begin{itemize}
    \item $C(\hat{\rho}_s) \geq 0$, and $C(\hat{\rho}_s) = 0$ if and only if $\rho \in \mathcal{I}$.
    \item Non-selective monotonicity under incoherent map - \begin{equation}
        C(\hat{\rho}_s) \geq C(\Phi^\text{icptp}(\hat{\rho}_s)).
    \end{equation}
    \item Selective monotonicity under incoherent maps -
    \begin{equation}
         C(\hat{\rho}_s) \ge \sum_n p_n C(\hat{\rho}_{s_n}),
    \end{equation}
    with outcomes $p_j = \mathrm{Tr}(\hat{K}_j \hat{\rho} K_j^\dagger)$, post-measurement states $\hat{\rho}_{s_j} = \hat{K}_j \hat{\rho}_s \hat{K}_j^\dagger / p_j$, and “incoherent Kraus operators $\hat{K}_j$.
     \item Convexity - For any set of states $\hat{\rho}_{s_j}$ and any probability distribution $p_j$
    \begin{equation}
    \sum_j p_j \, C(\hat{\rho}_{s_j}) \;\geq\; 
    C\!\left( \sum_j p_j \, \hat{\rho}_{s_j} \right), 
    \end{equation}
\end{itemize}
Among the proposed coherence measures, the most commonly used are those based on norms, in particular the $l_p$-norm applied to the off-diagonal elements of the system’s density matrix \cite{cramer}
\begin{equation}\label{normalp}
C_{l_p}(\hat{\rho}_s) = \left( \sum_{i \neq j} |\hat{\rho}_{s_{ij}}|^p \right)^{1/p}, \quad p \ge 1,\\[5pt]
\end{equation}
where $\hat{\rho}_{s_{ij}} = \bra{i} \hat{\rho}_s \ket{j}$ are the off-diagonal elements in the reference basis ${\ket{i}}$, and $|\cdot|$ denotes the complex modulus. The parameter $p$ allows flexibility in choosing the norm, being $l_1$ the most commonly used due to its simplicity and its compliance with the required properties discussed above \cite{cramer}
\begin{equation}\label{normal1}
    C_{l_1}(\hat{\rho}_s) = \sum_{i \neq j} |\hat{\rho}_{s,ij}|.
\end{equation}

\subsection{Coherence as a Measure of Non-Markovianity}

Here, we adopt a definition of $l_1$-norm coherence that is highly sensitive to the influence of the environment. 
\noindent The non-Markovianity measure constructed from  $C_{l_1}$, relies on the fact that, under ICPTP dynamics, the coherence must decrease monotonically over time. Therefore, if a temporary increase in coherence is observed, this is interpreted as a clear signature of non-Markovianity \cite{titas}
\begin{equation}\label{nomonotonocl1}
    \frac{d}{dt}C_{l_1}\big(\hat{\rho}_s(t)\big)>0.
\end{equation}
Equation \eqref{nomonotonocl1}  indicates a backflow of information from the environment to the system. Based on this idea, a coherence based non-Markovianity measure is defined as 
\begin{equation}\label{medidorcoherencia}
    \mathcal{N}_C(\Lambda) = \max_{\hat{\rho}_s(0) \in \mathcal{I}^c} \int_{\frac{d}{dt} C_{l_1}(\hat{\rho}_s(t)) > 0} \frac{d}{dt} C_{l_1}\big(\hat{\rho}_s(t)\big) \, dt, \\[7pt]
\end{equation}

\noindent where the maximization is taken over all initial states $\hat{\rho}_s(0)$ belonging to the set of coherent states $\mathcal{I}^c$. An important point to highlight is that this measure is valid for dynamics that are incoherent with respect to a fixed basis. The sufficient conditions for such incoherent dynamics are discussed in Ref.\cite{titas}.

\section{Quantum Thermal Behavior}

\subsection{Thermal States}
The density operator formalism provides a clear description of the state of a system in contact with a thermal bath at constant temperature $T_e$. Consider a system governed by a Hamiltonian $\hat{H}_s$ interacting with a thermal bath. From the perspective of statistical mechanics, the thermal equilibrium state is the one that maximizes the entropy. Classically, this corresponds to the maximization of the Shannon entropy, while in the quantum mechanics the von Neumann entropy is used. In both cases, the maximization is performed under the constraints of fixed average energy, $U = \text{tr}\left[\hat{\rho}_s \hat{H}_s\right]$, and normalization. In the quantum context, normalization is imposed on the density operator, $\text{tr}[\hat{\rho}_s] = 1$, whereas in the classical case, it applies to the probability distribution over microstates \cite{manenti}.
\vspace{0.2cm}\\
\noindent The maximization of entropy under these constraints is formalized through the following Lagrange functional
\begin{equation}
    \mathscr{L}(\hat{\rho}_s)=S(\hat{\rho}_s)-\beta_e \, \text{tr}\left[\hat{\rho}_s \hat{H}_{s}\right],\\[5pt]
\end{equation}
\noindent  where $\beta_e$ is the Lagrange multiplier associated with the average energy constraint and physically corresponds to the inverse temperature of the environment $\beta_e= \frac{1}{k_B T_e}$. Solving this optimization problem yields the density operator known as the thermal or Gibbs state, given by \cite{manenti}
\begin{equation}\label{gibbsestate}
    \hat{\rho}_s^{\,\rm th}=\frac{e^{-\beta_e \hat{H}_{s}}}{Z}     \\[5pt]
\end{equation}
with
\begin{equation}
    Z=\text{tr}\left[e^{-\beta_e \hat{H}_{s}}\right],
\end{equation}

\noindent where $Z$ is the partition function, which ensures the normalization of the thermal state and encodes all thermodynamic information about the system. 
\subsubsection{Quantum Oscillator in Equilibrium}
\noindent Let us now consider the case of a quantum harmonic oscillator in equilibrium with a thermal bath at constant temperature. This setup is widely implemented in optical cavities. The oscillator is described by the Hamiltonian $ \hat{H}_{s}=\hbar \omega_{r} \hat{a}^{\dagger} \hat{a}$, where $\omega_r$ is the oscillator frequency and $\hat{a}^\dagger, \hat{a}$ are the creation and annihilation operators, respectively. The eigenstates of this Hamiltonian are known as number states $\ket{n}$, associated with discrete energy levels for $n=0,1,2,3,\ldots$~\cite{manenti}. Using the definition of the Gibbs state Eq.~\eqref{gibbsestate}, we first compute the partition function
\begin{equation}
    Z=\text{tr}\left[e^{-\beta_e \hat{H}_{s}}\right]=\sum_{n=0}^{\infty} \bra{n} \sum_{m=0}^{\infty} e^{-\beta_e \hbar \omega_{r}\,m }\ket{m}\bra{m}n\rangle=\sum_{n=0}^{\infty} e^{-\beta_e \hbar \omega_{r} n}.\\[7pt]
\end{equation}

\noindent  Applying the identity $\sum_{n=0}^{\infty} x^{n}=1 /(1-x), x \in [0,1\rangle$, we obtain:
\begin{equation}
    Z=\frac{1}{1-e^{-\beta_e \hbar \omega_{r}}}.
\end{equation}

\noindent The thermal state of the oscillator can then be written as
\begin{equation}
   \hat{\rho}_s^{\,\text{th}}=\sum_{n=0}^{\infty} \frac{e^{-\beta_e \hbar \omega_{r} n}}{Z}\ket{n}\bra{n}=\sum_{n=0}^{\infty} p(n)\ket{n}\bra{n}.\\[5pt]
\end{equation}
\noindent The term $p(n)$ represents the probability distribution for the occupation of the $n-$th excitation level. These excitations are physically interpreted as photons in the case of electromagnetic resonators or phonons in mechanical systems. Based on the definition of $p(n)$, the average number of excitations in the thermal state is given by \cite{manenti}

\begin{equation*}
    \langle\hat{n}\rangle=\bar{N}=\text{tr}\left[\hat{\rho}_s^{\rm th} \,\hat{n}\right]=\sum_{n=0}^{+\infty}\bra{n} \hat{n} \sum_{m=0}^{+\infty}p(m)\ket{m}\bra{m}n\rangle=\sum_{n=0}^{+\infty}p(n)\bra{n} \hat{n}\ket{n},
\end{equation*}

\begin{equation}
    \bar{N}=\frac{1}{Z} \sum_{n=0}^{\infty} n e^{-\beta \hbar \omega_{r} n}\,.\\[5pt]
\end{equation}
Applying the identity $\sum_{n=0}^{\infty} n x^{n}=\frac{x}{(1-x)^{2}}, x \in [0,1]$, to the above expression, we obtain the well-known Bose–Einstein distribution:
\begin{equation}
    \bar{N}=\frac{1}{e^{\hbar \,\omega_{r} / k_{B} \, T_e}-1}\, .
\end{equation}
\subsubsection{Qubit with a Thermal Bath}

Now let us consider a qubit as the system in contact with a thermal bath at temperature $T_e$ \cite{manenti}. The Hamiltonian governing the qubit's evolution is
\begin{equation}
        \hat{H}_{s}=-\frac{\hbar \omega_{q} \hat{\sigma}_{z}}{2}
\end{equation}

\noindent where $\omega_q$ is the qubit transition frequency, and the computational basis coincides with the energy eigenbasis, $\ket{E_0}=\ket{0},\ket{E_1}=\ket{1}$, with eigenvalues $E_{0}=-\hbar \omega_{\mathrm{q}} / 2$ and $E_{1}=\hbar \omega_{\mathrm{q}} / 2$. 
\vspace{0.2cm}
\noindent Analogously to the harmonic oscillator case, we determine the partition function
\begin{equation*}
    Z=\text{tr}\left[e^{-\beta_e \hat{H}_{s}}\right] =\sum_{i=0}^{1}\langle i|\left( e^{-\beta_e E_0 }\ket{0}\bra{0}+ e^{-\beta_e E_1 }\ket{1}\bra{1}\right)|i\rangle=e^{-\beta_e E_{0}}+e^{-\beta_e E_{1}},
\end{equation*}
\begin{equation}
    Z=e^{\beta_e \hbar \omega_{q} / 2}+e^{-\beta_e \hbar \omega_{q} / 2},
\end{equation}
\begin{equation}
    Z=2 \cosh \left(\frac{\hbar \omega_{q}}{2 k_{B} T_e}\right).\\[7pt]
\end{equation}

\noindent  Since the Gibbs state for the qubit is given by $\hat{\rho}_s^{\rm \,th}=e^{\beta_e \frac{\hbar \omega{q} \hat{\sigma}_{z}}{2}}/Z$ and taking into account that $\hat{\sigma_z}=\ket{0}\bra{0}-\ket{1}\bra{1}$, the thermal state can be expressed as
\begin{equation*}
    \hat{\rho}_s^{\rm \,th}=\frac{e^{\beta_e \hbar \omega_{q} / 2}}{Z}\ket{0}\bra{0}+\frac{e^{-\beta_e \hbar \omega_{q} / 2}}{Z}\ket{1}\bra{1}\\[7pt]
\end{equation*}
\begin{equation}
\hat{\rho}_s^{\rm \,th}=\frac{1}{2 \cosh \left(\frac{\hbar \omega_{q}}{2 k_{B} T_e}\right)} \begin{pmatrix}
e^{\frac{\hbar \omega_{q}}{2k_{B} T_e}} & 0 \\[0.5em]
0 & e^{-\frac{\hbar \omega_{q}}{2k_{B} T_e}}
\end{pmatrix}\\[10pt]
\end{equation}

\noindent In the low-temperature limit, $T_e \rightarrow 0$, the qubit approaches its ground state, meaning that the population $\hat\rho_{s_{00}}^{\rm \, th}$ reflects the probability of finding the qubit in the ground state, while $\hat\rho_{s_{11}}^{\rm \, th}$ corresponds to the population in the excited state. At finite temperatures, the qubit state becomes mixed. In the high-temperature regime, where $\hbar \omega_q \ll k_B T_e$, the qubit tends to a maximally mixed state \cite{manenti}.

\subsection{Thermal Maps}

Having defined what a thermal state is, we now move on to the definition of a thermal map $\Phi^\text{th}$, which is a subset of CPTP maps. This type of map constitutes a fundamental framework for quantum thermodynamics and resource theory. Every thermal map involves the concepts of a thermal state and a global fixed point. 
\vspace{0.2cm}\\
\noindent We know that a fixed point of a quantum map $\Phi$ is a state $\hat{\rho}^{\,\text{fixed}}$ that satisfies \cite{Landi2021}
\begin{equation}\label{puntofijolocal}
    \Phi \big(\hat{\rho}_s^{\,\text{fixed}}\big)=\hat{\rho}_s^{\,\text{fixed}}.
\end{equation}
On the other hand, when we say that a map has a global fixed point, it means that this state $\hat{\rho}_s^{\,\text{fixed}}$ satisfies \cite{Landi2021}
\begin{equation}\label{puntofijoglobal}
    \hat{V}_{se} \left( \hat{\rho}^{\,\text{fixed}}_s \otimes \hat{\rho}_e \right) \hat{V}_{se}^\dagger = \hat{\rho}^{\,\text{fixed}}_s \otimes \hat{\rho}_e
\end{equation}
This condition is much stronger than \eqref{puntofijolocal}, since it not only requires that $\hat{\rho}^{\,\text{fixed}}_s$ be invariant under the quantum map on the system, but also imposes a condition on the joint state with the environment. This restriction is a particular feature of thermal maps.
\vspace{0.2cm}
\noindent The formal definition of thermal maps $\Phi^{\,\text{th}}$ is given by
\begin{equation}\label{mapatermico}
    \hat{\rho}_s(t) \;=\; \Phi^\text{th}_t\big(\hat{\rho}_s(t_0)\big) 
= \text{tr}_e \!\left[ \hat{V}_{se}(t,t_0) \, \bigl(\hat{\rho}_s(t_0) \otimes \hat{\rho}^{\text{th}}_e \bigr) \, \hat{V}_{se}^\dagger(t,t_0) \right].
\end{equation}
The environment must be in a fixed thermal state $\hat{\rho}_e^{\,\text{th}}$, and the unitary operator of the composite system $\hat{V}_{se}(t,t_0)$ must satisfy the energy conservation condition. In other words, the energy exchanged between the system and the environment is conserved and not trapped in the interaction~\cite{Landi2021, salimi2020}
\begin{equation}\label{strictconservacionenergia}
    \bigl[\hat{V}_{se}(t,t_0), \, \hat{H}_s + \hat{H}_e \bigr] = 0.
\end{equation}

\section{Quantum Entropy}

\subsection{von Neumann Entropy}

A quantity that is highly useful in both quantum statistical mechanics and quantum information theory is the von Neumann entropy, defined as \begin{equation}
    S(\hat{\rho}_s)=-\text{tr} (\hat{\rho}_s \ln \hat{\rho}_s).
\end{equation} This definition becomes particularly important in quantum thermodynamics. The von Neumann entropy acts as a functional of a quantum state $f(\hat{\rho}_s)$. If we consider the spectral decomposition of the state $\hat{\rho}_s= \sum_k r_k \ket{r_k}\bra{r_k}$, the entropy takes the form \begin{equation}
    S(\hat{\rho}_s)=-\sum_k (r_k \ln r_k).
\end{equation}  The expression of $S(\hat{\rho}_s)$ in terms of the eigenvalues of the system's state is formally identical to the Shannon entropy from classical information theory~\cite{nakahara2012,sagawa2022}. Therefore, the von Neumann entropy can be understood as a quantifier of the statistical uncertainty (or classical randomness) associated with a quantum state.

\subsection{Properties}

\begin{itemize}[itemsep=-5pt, topsep=0pt]
    \item A fundamental property of $S(\hat{\rho}_s)$  is its invariance under unitary transformations of the quantum state $\hat{\rho}_s \rightarrow \hat{V}_s \hat{\rho}_s \hat{V}_s^\dagger$~\cite{LandiNotes2018}, where $\hat{V}_s$ is a unitary operator. To prove this invariance, we use \begin{equation}
        f(\hat{V}_s \hat{\rho}_s \,\hat{V}_s^\dagger) = \hat{V}_s f(\hat{\rho}_s) \hat{V}_s^\dagger.
    \end{equation}
   known as the infiltration property of unitary operators. This identity holds for any functional that can be written as a power series in $\hat{\rho}_s$. Then, using the cyclic property of  the trace, we obtain \begin{equation*}
S(\hat{V}_s \hat{\rho}_s \hat{V}_s^\dagger)=-\text{tr} \left[ \hat{V}_s \hat{\rho}_s \hat{V}_s^\dagger \ln(\hat{V}_s \hat{\rho}_s \hat{V}_s^\dagger) \right] 
= \text{tr} \left[ \hat{V}_s \hat{\rho}_s \hat{V}_s^\dagger \hat{V}_s (\ln \hat{\rho}_s) \hat{V}_s^\dagger \right]
= \text{tr} (\hat{\rho}_s \ln \hat{\rho}_s),
\end{equation*}
\begin{equation}
S(\hat{V}_s \hat{\rho}_s \hat{V}_s^\dagger) = S(\hat{\rho}_s).
\end{equation}

This property implies that in closed systems governed by unitary evolution, $S(\hat{\rho}_s)$ remains constant. Physically, this invariance reflects the preservation of global quantum information under unitary dynamics. Consequently, any effective increase in entropy must originate from non-unitary processes such as decoherence or interactions with an external environment.

\item  The von Neumann entropy is always non-negative \cite{LandiNotes2018,Bertlmann:Book},
\begin{equation}
    S(\hat{\rho}_s)\geq 0.
\end{equation}
This follows from the fact that the entropy can be written as a sum of terms of the form  $-r_k \ln r_k$, where $r_k \in [0,1]$. Each term is non-negative, so the total entropy is also non-negative. The minimum value, $S_{\min}(\hat{\rho}_s) = 0$, occurs when one $r_k = 1$ and all others are zero, that is, when the system is in a pure state. Conversely, the maximum entropy occurs when all eigenvalues are equal, $r_k = 1/D$, with $D$ the dimension of the system, corresponding to a maximally mixed state. In this case, the entropy takes the value $S_{\max}(\hat{\rho}_s) = \ln D$. Therefore, the von Neumann entropy, which quantifies the mixedness of a state, is bounded as \begin{equation}
    0 \leq S(\hat{\rho}_s) \leq \ln D.
\end{equation}

\item When a unital CPTP map $\Phi$ acts on the system, the von Neumann entropy is monotonic under this class of quantum map \cite{sagawa2022}. This fundamental property is referred to as monotonicity, \begin{equation}
    S(\hat{\rho}_s)\leq S(\Phi(\hat{\rho}_s)).
\end{equation} 

\item For a bipartite quantum state 
$\hat{\rho}_{AB}$ 
with reduced states 
$\hat{\rho}_{A}$ 
and 
$\hat{\rho}_{B}$, 
the following inequality holds
\begin{equation}
    S(\hat{\rho}_{AB}) \leq S(\hat{\rho}_{A}) + S(\hat{\rho}_{B}).
\end{equation}
Equality occurs only when $\hat{\rho}_{AB} = \hat{\rho}_{A} \otimes \hat{\rho}_{B}$. This relation expresses that, when the subsystems are described independently, information about correlations present in the global system is lost \cite{sagawa2022,Bertlmann:Book}.

\end{itemize}

\subsection{Quantum Relative Entropy}

The quantum relative entropy is defined for two quantum states $\hat{\rho}_s$ and $ \hat{\sigma}_s$, and given by
\begin{equation}
    S(\hat{\rho} _s\| \hat{\sigma}_s)=\text{tr}\{\hat{\rho} _s \ln \hat{\rho} _s-\hat{\rho} _s \ln \hat{\sigma}_s\}.
\end{equation}

\noindent This quantity can be interpreted as a measure of distinguishability between two quantum states. However, it is worth noting that $S(\hat{\rho}_s \| \hat{\sigma}_s)$ is not a true distance in the mathematical sense, as it is neither symmetric nor does it satisfy the triangle inequality. Importantly, quantum relative entropy serves as a fundamental tool for characterizing correlations in composite systems and, within the framework of quantum thermodynamics, it plays a key role in quantifying irreversibility in dynamical processes \cite{nakahara2012, sagawa2022}.

\subsection{Properties}
\begin{itemize}[itemsep=-5pt, topsep=0pt]
    \item Quantum relative entropy is non-negative for any pair of quantum states \cite{LandiNotes2018,Bertlmann:Book} \begin{equation}
    S(\hat{\rho} _s \| \hat{\sigma} _s) \geq 0.
\end{equation}
where $S(\hat{\rho}_s \| \hat{\sigma}_s)=0$ only if $\hat{\rho}_s = \hat{\sigma}_s$.

\item Similarly to the von Neumann entropy $S(\hat{\rho}_s)$, quantum relative entropy is invariant under unitary transformations \cite{Bertlmann:Book} \begin{equation}
    S(\hat{\rho} _s \| \hat{\sigma} _s)=S\left(\hat{V}_s \hat{\rho} _s \hat{V}_s^{\dagger} \,\| \, \hat{V}_s \hat{\sigma} _s \hat{V}_s^{\dagger}\right).
\end{equation}

\item Quantum relative entropy can only decrease under the aplication of the same CPTP noisy map to both states \cite{Hayashi:Book}. This fundamental property is known as the monotonicity of $S(\hat{\rho} _s \| \hat{\sigma} _s)$  \begin{equation}
    S(\hat{\rho}_s \| \hat{\sigma}_s) \geq S(\Phi(\hat{\rho}_s) \| \Phi(\hat{\sigma}_s)) .
\end{equation} 

\item For a bipartite quantum state $\rho_{AB}$ with reduced states $\rho_A$ and $\rho_B$, 
quantum relative entropy between $\rho_{AB}$ and the product state $\rho_A \otimes \rho_B$ is given by \cite{Bertlmann:Book}
\begin{equation}
    S(\hat{\rho}_{AB} \,\|\, \hat{\rho}_A \otimes \hat{\rho}_B)
    = S(\hat{\rho}_A) + S(\hat{\rho}_B) - S(\hat{\rho}_{AB}),
\end{equation}
which coincides with the quantum mutual information  $I_{\hat{\rho}_{AB}}(A\,:\,B)$ \cite{sagawa2022}.
\end{itemize}

\section{Maximal Extractable Work - Ergotropy}

The problem of maximal work extraction consists of determining the greatest amount of work that can be obtained from a thermally isolated system via a cyclic external process. In macroscopic systems, this is given by the difference in free energy $F$ between equilibrium states. In finite quantum systems, unitary evolution prevents reaching Gibbs states from an arbitrary initial state, as it preserves the von Neumann entropy. Consequently, the maximum extractable work differs from the classical case \cite{Allahverdyan:04}.
\vspace{0.2cm}\\
\noindent First, let us consider a system $S$ in the macroscopic limit, possessing a large number of degrees of freedom and internal relaxation mechanisms, which can exchange work with macroscopic external sources. Its evolution is governed by a time-dependent Hamiltonian,
\begin{equation}\label{Hfrectrl}
\hat{H}_s(t) = \hat{H}_\text{free} + \hat{H}_\text{ctrl}(t),
\end{equation}
where $\hat{H}_\text{free}$ is the internal Hamiltonian, and $\hat{H}_\text{ctrl}(t)$ is responsible for work exchange. Initially isolated, the system undergoes a cyclic process, being connected to a work source at $t=0$ and disconnected at $t=\tau$. Before and after this interval, we have 
\begin{equation}
\hat{H}_\text{ctrl}(0) = \hat{H}_\text{ctrl}(\tau) = 0.
\end{equation}
Although the system is thermally isolated, its internal parts can exchange energy among themselves. To extract the maximum possible work from $S$, we consider the initial state $\hat{\rho}_{s}(t_0=0)=\hat{\rho}_{s_0}$. According to the closed dynamics of $S$, this evolution is governed by the von Neumann equation. Using the standard formulation of quantum thermodynamics, the infinitesimal work performed on the system is $dW_\text{stand}=\text{tr}\big[\hat{\rho}_{s}(t) \, \frac{d}{dt}\hat{H}_\text{ctrl}(t)\big]$ and the total work done over the time interval $[0,\tau]$ is defined by the change in internal energy \cite{Allahverdyan:04} 
\begin{equation}
    W_\text{stand}=U(\tau)-U(0),
\end{equation}
with $\,U(t)=U\big(\hat{\rho}_{s}(t)\big)= \text{tr}\big[\hat{\rho}_{s}(t) \, \hat{H}_s(t)\big]$. Therefore, we have \begin{equation}
W_\text{stand}= \text{tr}[\hat{\rho}_s(\tau)\, \hat{H}_\text{free}] - U(\hat{\rho}_{s_0}).
\end{equation}
If the system performs work on the surroundings, meaning we extract work from it, we have $U(\tau) < U(0)$ and $W_\text{stand} < 0$. Consequently, the extractable work is given by \begin{equation}
    W_\text{extr}=-W_\text{stand}
\end{equation}
Maximizing the extractable work $W_\text{extr}$ requires finding a final state $\hat{\rho}(\tau)$ that minimizes the final energy. First, the maximal extractable work can be written as \cite{Allahverdyan:04}
\begin{equation}\label{maximoextraible}
W_\text{extr}^\text{max}= \max_{\hat{\rho}_s(\tau)}\left\{U(\hat{\rho}_{s_0})-\text{tr}[\hat{\rho}_s(\tau)\, \hat{H}_\text{free}] \right\}.\\[5pt]
\end{equation}
Equivalently, the maximal extractable work can also be expressed as
\begin{equation}\label{minimoextraible}
    W_\text{extr}^\text{max}=U(\hat{\rho}_{s_0})-\min_{\hat{\rho}_s(\tau)} \left\{\text{tr}[\hat{\rho}_s(\tau)\, \hat{H}_\text{free}]\right\}. \\[5pt]
\end{equation}
According to quantum statistical mechanics, there are two key standard ideas to achieve this maximization. First, in the macroscopic limit, the final state is taken to be a Gibbs state $\,\hat{\rho}_s(\tau)=\hat{\rho}_s^{\, th} = \frac{e^{-\beta \hat{H}}}{Z}$ and second, the von Neumann entropy cannot decrease between the initial and final times
\begin{equation}
    S(\hat{\rho}_{s_0})=S(\hat{\rho}_s^{\,th}).
\end{equation}
Finally, following Ref.~\cite{Allahverdyan:04}, the maximal extractable work for a system in the macroscopic limit satisfying these standard ideas is
\begin{equation}\label{wthmacro}
W_{th} = U(\hat{\rho}_{s_0}) - T S(\hat{\rho}_{s_0}) + T \ln Z,
\end{equation}
This result establishes a fundamental limit since even with the best possible control $\hat{H}_\text{ctrl}(t)$, no more work can be extracted from the system than this value. Using the definition of free energy, $W_{th}$ can be expressed as
\begin{equation}
    W_{th} = F(\hat{\rho}_{s_0}) - F_{th}
\end{equation}
indicating that the maximal extractable work in the macroscopic limit equals the difference between the system’s initial and equilibrium free energies.
\vspace{0.2cm}\\
\noindent In contrast, for finite systems, relaxation mechanisms are not allowed under unitary evolution. According to quantum mechanics and quantum information theory, unitary evolution ensures that the von Neumann entropy $S(\hat{\rho}_s)$ remains constant and that the eigenvalues of the state are preserved, which is characteristic of reversible processes. Therefore, since the evolution of a finite system is unitary, the final state in Eq. \eqref{minimoextraible} is given by 
\begin{equation}
    \hat{\rho}_s(\tau)=\hat{V}_s(\tau)\hat{\rho}_{s_0}\hat{V}_s^\dagger(\tau).
\end{equation}
Consequently, we must minimize $\hat{\rho}_s(\tau)$ over the set of all possible unitary operators ${\hat{V}_s(\tau)}$. This minimization is achieved by imposing a stationarity condition, which, according to Ref.~\cite{Allahverdyan:04}, requires
\begin{equation}
    \big[\hat{\rho}_{s_{\rm min}}(\tau),\hat{H}_\text{free}\big]=0.\\[5pt]
\end{equation}
This commutation relation indicates that $\hat{\rho}_{s_{\rm min}}(\tau)$ must be diagonal in the same basis as $\hat{H}_\text{free}$. Moreover, $\hat{\rho}_{s_{\rm min}}(\tau)$ and $\hat{\rho}_{s_0}$ share the same eigenvalues. To determine the form of $\hat{\rho}_{s_{\rm min}}(\tau)$, we start from the following diagonalized forms
\begin{equation}
    \hat{\rho}_{s_0} = \sum_{j \geq 1} r_j \, \ket{r_j}\bra{r_j}, \quad r_1 \geq r_2 \geq \dots \\[5pt]
\end{equation}
\begin{equation}
    \hat{H}_\text{free} = \sum_{k \geq 1} \varepsilon_k \, \ket{\varepsilon_k}\bra{\varepsilon_k}, \quad \varepsilon_1 \leq \varepsilon_2 \leq \dots \\[5pt]
\end{equation}
Following Refs.~\cite{Pusz78,Lenard78}, the minimization is achieved by assigning the largest $r_j$ to the lowest-energy eigenstates $\ket{\varepsilon_k}$. Therefore, the final state $\hat{\rho}_{s_{\rm min}}(\tau)$ has the following spectral decomposition, which is valid because it is expressed in the eigenbasis of the reference Hamiltonian $\hat{H}_\text{free}$ evaluated at $\hat{H}_s(t=\tau)$
\begin{equation}\label{estadopasivo}
    \hat{\rho}_{s_{\rm min}}(\tau)=\hat{\rho}_{s_0}^\pi= \sum_{j \geq 1} r_j \, \ket{\varepsilon_j}\bra{\varepsilon_j}.\\[5pt]
\end{equation}
The state $\hat{\rho}_{s_0}^\pi$ is called the passive state and satisfies
\begin{equation}\label{pasivosignificado}
    \text{tr}\left[\hat{\rho}_{s_0}^\pi \,\hat{H}_\text{free}\right] \leq \text{tr}\left[ \hat{V}_{s}(t) \hat{\rho}_{s_0}^\pi \hat{V}_{s}^\dagger(t) \,\hat{H}_\text{free} \right].\\[5pt]
\end{equation}


\noindent Physically, Eq.~\eqref{pasivosignificado} indicates that no more work can be extracted from $\hat{\rho}_{s_0}^\pi$ for any unitary $\hat{V}_s(t)$. It is worth noting that for a given state $\hat{\rho}_s(t)$, there is only one unique passive state, denoted $\hat{\rho}_{s}^\pi(t)$. For a geometric visualization of a passive state as part of the ergotropy decomposition, can be seen in Fig.~\ref{fig:ergoTCI}.
\vspace{0.2cm}\\
\noindent With the definition of the passive state in Eq.~\eqref{estadopasivo}, we can substitute it into Eq.~\eqref{minimoextraible}, so that the maximum extractable work for a finite system is given by
\begin{equation}\label{ergotropiatarza}
    \mathcal{E}= W_\text{extr}^\text{max}=\text{tr}\left[\hat{\rho}_{s_0}\, \hat{H}_\text{free})\right] - \text{tr}\left[\hat{\rho}_{s_0}^\pi \hat{H}_\text{free}\right].\\[5pt]
\end{equation}
The $W_\text{extr}^\text{max}$ for a finite system is denoted as $\mathcal{E}(\hat{\rho}_{s_0}, \hat{H}_\text{free})$, commonly called the ergotropy. The minimal energy corresponds to the term $\text{tr}\left[\hat{\rho}_{s_0}^\pi \hat{H}_\text{free}\right]$, and the ergotropy can also be expressed in terms of the internal energy, which is a state function. 
\begin{equation}\label{ergotropiaUinterna}
    \mathcal{E}\big(\hat{\rho}_{s_0}, \hat{H}_\text{free}\big)= U\left(\hat{\rho}_{s_0}, \hat{H}_\text{free}\right)-U\left(\hat{\rho}_{s_0}^\pi, \hat{H}_\text{free}\right).\\[5pt]
\end{equation}
In general, the maximal work extractable satisfies $W_{th} \geq \mathcal{E} \geq 0$, indicating that the work obtainable from a finite system is always non-negative and cannot exceed the macroscopic bound. 
\vspace{0.2cm}\\
\noindent Equation \eqref{ergotropiaUinterna} defines the ergotropy calculated with respect to the state $\hat{\rho}_{s_0}$ at $t=0$, in relation to the energy levels at $t=0$, which, according to Eq.~\eqref{Hfrectrl}, correspond to fixed levels determined by $\hat{H}_\text{free}$. However, if we consider an arbitrary time $t \in [0,\tau]$, the state at that instant, $\hat{\rho}_{s}(t)$, must be evaluated with respect to the instantaneous energy levels defined by $\hat{H}(t)=\hat{H}_\text{free}+\hat{H}_{\rm ctrl}(t)$. In order to capture the ergotropy at any given time, we can introduce the instantaneous ergotropy
\begin{equation}\label{ergotropiainstantanea}
    \mathcal{E}(t)=\text{tr}\left[\hat{\rho}_s(t)\,\hat{H}(t)\right] - \text{tr}\left[\hat{\rho}_s^{\,\pi}(t)\,\hat{H}(t)\right],\\[5pt]
\end{equation}
\begin{equation}\label{ergotropiainstantanea2}
   \mathcal{E}(t)=U\left(\hat{\rho}_s(t)\right)-U\left(\hat{\rho}_s^{\,\pi}(t)\right).
\end{equation}
\subsection{Coherent and Incoherent Contributions}

\noindent In recent years, advances in quantum thermodynamics have highlighted the role of coherence in state functions and in the definitions that arise in thermodynamic processes. In Ref.~\cite{Francica:20}, the incoherent and coherent contributions to the ergotropy of a state $\hat{\rho}_s$ were identified in the following form
\begin{equation}\label{contribucionergotropia}
    \mathcal{E}(\hat{\rho}_s)=\mathcal{E}_I(\hat{\rho}_s)+\mathcal{E}_C(\hat{\rho}_s).\\[5pt]
\end{equation}
To determine the incoherent contribution $\mathcal{E}_I(\hat{\rho}_s)$, we start from the state $\hat{\rho}_s$ expressed in the energy eigenbasis, $\hat{\rho}_s \rightarrow \hat{\rho}_s^{\,\varepsilon}$, and apply the full dephasing map $\Delta(\cdot)$, which removes all coherent elements
\begin{equation}\label{eq:desfasadototal}
     \hat{\rho}_s^{\,\varepsilon-\text{incoh}} = \Delta[\hat{\rho}_s^{\,\varepsilon}] = \sum_k \bra{\varepsilon_k} \hat{\rho}_s^{\,\varepsilon} \ket{\varepsilon_k} \, \ket{\varepsilon_k}\bra{\varepsilon_k}.\\[5pt]
\end{equation}
Here, $\ket{\varepsilon_k}$ denotes the eigenvectors of the Hamiltonian. Note that $\hat{\rho}_s^{\,\varepsilon-\text{incoh}}$ preserves the energy populations $\bra{\varepsilon_k} \hat{\rho}_s^{\,\varepsilon} \ket{\varepsilon_k}$, and therefore its average energy also remains unchanged, $\langle \hat{H} \rangle_{\hat{\rho}_s^{\,\varepsilon-\text{incoh}}} = \langle \hat{H} \rangle_{\hat{\rho}_s^{\,\varepsilon}}$. Once the incoherent state in the energy basis is obtained, the corresponding passive state $\hat{\rho}_s^{\,\varepsilon-\text{incoh}-\pi}$, which is defined by reordering the populations $\bra{\varepsilon_k} \hat{\rho}_s^{\,\varepsilon} \ket{\varepsilon_k}$ in decreasing order with respect to the energy. In other words, the largest population is assigned to the eigenvector $\ket{\varepsilon_k}$ associated with the lowest energy level.
\vspace{0.2cm}\\
\noindent Finally, the incoherent contribution can be computed by evaluating the total ergotropy, as defined in Eq. \eqref{ergotropiatarza}, for the incoherent state $\hat{\rho}_s^{\,\varepsilon-\text{incoh}}$  \begin{equation}\label{ergotropitotalincoherente}
    \mathcal{E}\left(\hat{\rho}_s^{\,\varepsilon-\text{incoh}}\right)=\text{tr} \left[\hat{H}\,\left(\hat{\rho}_s^{\,\varepsilon-\text{incoh}}-\hat{\rho}_s^{\,\varepsilon-\text{incoh}-\pi}\right)\right].\\[5pt]
\end{equation}
Thus, the incoherent contribution to the ergotropy of a state $\hat{\rho}_s$ is defined as the maximum extractable work that depends exclusively on its energy populations. In other words, it corresponds to the total ergotropy of the state expressed in the energy basis, but without coherences \begin{equation}
\mathcal{E}_I(\hat{\rho}_s)=\mathcal{E}(\hat{\rho}_s^{\,\varepsilon-\text{incoh}}).\\[5pt]
\end{equation}
Having established the definition of the incoherent contribution $\mathcal{E}_I$, the coherent part can then be obtained from the following difference
\begin{equation}\label{ergotropicoherente}
    \mathcal{E}_C(\hat{\rho}_s)=\mathcal{E}(\hat{\rho}_s)-\mathcal{E}_I(\hat{\rho}_s).
\end{equation}

\noindent To provide a geometric illustration of ergotropy and of its coherent and incoherent contributions, we consider the simple case of a two level system (qubit) whose evolution is governed by the Hamiltonian in Eq.~\eqref{Hfrectrl}, whose free part satisfies $\hat{H}_{\rm free}\propto -\sigma_z$. Figure \ref{fig:ergoTCI} shows an initial pure state  $\hat{\rho}_{s}$ (black vector) together with its corresponding passive state $\hat{\rho}_{s}^{\, \pi}$ (green vector starting at the center of the sphere), which in this case coincides with the ground state. The total ergotropy $\mathcal{E}$ of the pure state is extracted through an optimal unitary that drives the initial state to the south pole of the Bloch sphere, as indicated by the green arc. 
\vspace{0.2cm}\\
\noindent To visualize the incoherent contribution $\mathcal{E}_I$, we apply the full dephasing map $\Delta$ to the initial pure state, producing the dephased state $ \Delta[\hat{\rho}_s^{\,\varepsilon}]$ (see Eq.\eqref{eq:desfasadototal}), , represented in Fig.~\ref{fig:ergoTCI} by the upward orange vector. Its passive state $ \big(\Delta[\hat{\rho}_s^{\,\varepsilon}]\big)^\pi$, appears as the downward orange vector. The ergotropy extracted from this dephased state corresponds to the incoherent ergotropy $\mathcal{E}_I$ and is obtained through the optimal unitary $\hat{V}_\pi$, illustrated by the orange arc. Since $\mathcal{E}_I \leq \mathcal{E}$, it becomes evident that coherence plays an essential role in the extraction of work from any given state.
\vspace{0.2cm}\\
\noindent By contrast, the coherent contribution $\mathcal{E}_c$ can be visualized by transforming the pure state $\hat{\rho}_{s}$ into a lower-energy state $\hat{\rho}_{s}^{\, \downarrow}$ while preserving its initial coherence. This state is shown as the blue vector in Fig.~\ref{fig:ergoTCI} and is obtained through the same operation $\hat{V}_\pi$. By applying the definition of ergotropy to $\hat{\rho}_{s}^{\, \downarrow}$, we obtain the coherent ergotropy $\mathcal{E}_c$, which in this specific case corresponds to the energy interval between the tip of the downward orange vecto $ \big(\Delta[\hat{\rho}_s^{\,\varepsilon}]\big)^\pi$ and the lower end of the green vector, associated with the ground state $\ket{0}$. The extraction of $\mathcal{E}_c$ follows the blue arc, which overlaps partially with the green arc associated with $\mathcal{E}$ in the lower region of the Bloch sphere.

\begin{figure}[H] 
\centering
\includegraphics[width=0.8\columnwidth]{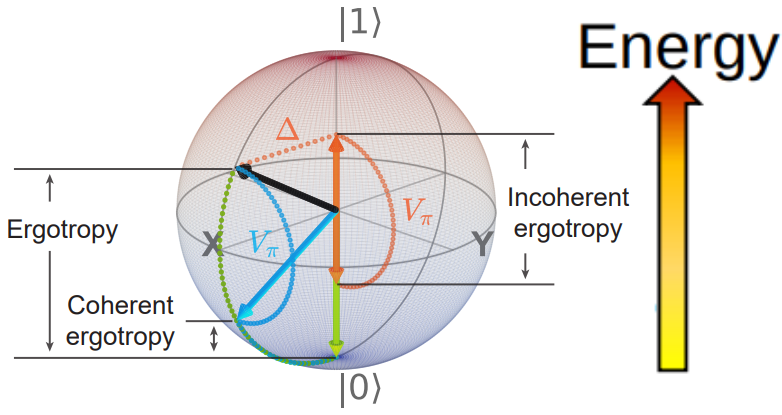}
\caption{(Color online) Geometric representation of the total, incoherent, and coherent ergotropy associated with an initial qubit state on the Bloch sphere, governed by $\hat{H}_{\rm free}=- \hbar\,\omega_0\hat{\sigma}_z/2$.  Adapted from Ref. \cite{xiang2025}  }
\label{fig:ergoTCI}
\end{figure}

\section{Quantum Second Law}

In Ref.~\cite{Landi2021}, G.Landi and M.Paternostro presented a unified formulation of entropy production in open quantum systems. This formulation is obtained by considering the global unitary evolution of the system $S$ and its environment $E$, initially prepared in arbitrary states $\hat{\rho}_s$ and $\hat{\rho}_e$, through a global unitary transformation $\hat{V}_{se}$
\begin{equation}\label{2lawcompuesto}
\hat{\rho}' = \hat{V}_{se} (\hat{\rho}_s \otimes \hat{\rho}_e) \hat{V}_{se}^\dagger.
\end{equation}  

\noindent This description is completely general, as it accommodates weak or strong couplings, time-dependent Hamiltonians, and environments of arbitrary size, including finite ones. All information is encoded in the global unitary evolution. The reduced state of the system is obtained by tracing out the environmental degrees of freedom,
\begin{equation}
\hat{\rho}'_s = \Phi(\hat{\rho}_s)=\text{tr}_e[\hat{\rho}'].
\end{equation}  

\noindent Note that irreversibility conceptually emerges once the information contained in the environment is discarded. This discarded information may either remain locally stored in $E$ or be shared as correlations between $S$ and $E$. The quantity that quantifies this discarded information is referred to as entropy production and is given by,
\begin{equation}\label{produccionentropiainformacional}
\Sigma = I_{\hat{\rho}'}(S:E) + S(\hat{\rho}'_e||\hat{\rho}_e).
\end{equation}  

\noindent Here, one of its contributions corresponds to the mutual information generated between the system and the environment due to their interaction, while the other arises from the displacement of the environment with respect to its initial state, represented by the relative entropy. 
\vspace{0.2cm}\\
\noindent The concept of mutual information comes from information theory and represents an important measure of correlations between two systems (a bipartite system 
$AB$) or variables, expressed as
\begin{equation}\label{informacionmutua}
I_{\hat{\rho}_{AB}}(A:B) = S(\hat{\rho}_A) + S(\hat{\rho}_B) - S(\hat{\rho}_{AB}),
\end{equation}  
while the relative entropy, which quantifies the distinguishability between two states, is defined as
\begin{equation}\label{relativaS}
S(\hat{\rho}||\hat{\sigma}) = \mathrm{tr}[\hat{\rho} \ln \hat{\rho} - \hat{\rho} \ln \hat{\sigma}].
\end{equation}  
Accordingly, using the expressions for mutual information and relative entropy, entropy production $\Sigma$ can be rewritten as
\begin{equation}
\Sigma = S(\hat{\rho}' || \hat{\rho}'_s \otimes \hat{\rho}_e), 
\end{equation}  
which explicitly highlights an asymmetry between the final state of the system and the initial state of the environment $\hat{\rho}'_s \otimes \hat{\rho}_e$. This reflects the loss of information about the environment after the interaction.
\vspace{0.2cm}\\
\noindent Although $\Sigma$ does not initially appear in Clausius form, we can recast it in the standard form of entropy production by proposing
\begin{equation}\label{propoclausius}
\Sigma = \Delta S_s + \textbf{J}, 
\end{equation}
where $\textbf{J}$ denotes the entropy flux from the system to the environment. Since the system and the environment are initially uncorrelated, the mutual information can be expressed as
\begin{equation}
I_{\hat{\rho}'}(S:E) = \Delta S_s + \Delta S_e,
\end{equation}
a consequence of the unitary evolution, which preserves the total entropy of the global state. The entropy flux is therefore defined as
\begin{equation}\label{flujodeentropia}
\textbf{J} = S(\hat{\rho}'_e) - S(\hat{\rho}_e) + S(\hat{\rho}'_e||\hat{\rho}_e) = \mathrm{tr}_e[(\hat{\rho}_e - \hat{\rho}'_e)\ln \hat{\rho}_e].
\end{equation}
We note that \textbf{J} depends solely on the local state of the environment, while $\Delta S_s = S(\hat{\rho}'_s) - S(\hat{\rho}_s)$ depends only on the local state of the system. Thus, entropy production naturally splits into two contributions: $\Delta S_s$, associated with the system, and \textbf{J}, associated with the environment. If we further assume $\hat{\rho}_e=\hat{\rho}_e^{\rm th}$, the entropy flux takes the standard thermodynamic form,\begin{equation}
    \Sigma = \Delta S_s + \beta_e Q_e,
\end{equation}
with $Q_e=\text{tr}\big[\hat{H}_e (\hat{\rho}_e'-\hat{\rho}_e^{\,\rm th})\big]$. It is worth stressing that the assumption of a thermal environment is introduced only to recover the traditional expression of entropy production. The essential point is that we started from a more general, information-theoretic definition of entropy production.
\vspace{0.2cm}\\
\noindent Let us now consider the case without assuming $\hat{\rho}_e$ to be thermal. We focus on scenarios where the map admits a global fixed point $\hat{\rho}^{\rm fixed}_s$ such that
\begin{equation}
\hat{V}_{se}(\hat{\rho}^{\rm fixed}_s \otimes \hat{\rho}_e) \hat{V}_{se}^\dagger = \hat{\rho}^{\rm fixed}_s \otimes \hat{\rho}_e.
\end{equation}  
In this setting, the entropy flux \textbf{J} can be expressed solely in terms of the system’s local states
\begin{equation}
\textbf{J}= \mathrm{tr}_S[(\hat{\rho}'_s - \hat{\rho}_s) \ln \hat{\rho}^{\rm fixed}_s]. 
\end{equation}  
Therefore, entropy production reduces to a fully local expression 
\begin{equation}
\Sigma = S(\hat{\rho}_s || \hat{\rho}^{\rm fixed}_s) - S(\hat{\rho}'_s || \hat{\rho}^{\rm fixed}_s).
\end{equation}  

\noindent The positivity of $\Sigma$ is guaranteed by the contractivity of CPTP maps $\Phi$
\begin{equation}
S(\Phi(\hat{\rho}_s) || \Phi(\hat{\rho}^{\rm fixed}_s)) \le S(\hat{\rho}_s || \hat{\rho}^{\rm fixed}_s), 
\end{equation}  
which ensures that entropy production is non-negative. An important point to mention is that entropy production does not depend on energetic quantities of the system, such as heat or work. Instead, this entropy production relies solely on the amount of information.
\vspace{0.2cm}
\noindent  If we restrict to thermal maps, where the global fixed point corresponds to the thermal state $\hat{\rho}_s^{\rm th}$, the entropy production reduces to
\begin{equation}
\Sigma = S(\hat{\rho}_s || \hat{\rho}^{\rm th}_s) - S(\hat{\rho}'_s || \hat{\rho}^{\rm th}_s), 
\end{equation}  
a fully local expression for the system. Writing the system’s thermal state in terms of the Helmholtz free energy,
\begin{equation}
\hat{\rho}_s^{\mathrm{th}} = e^{-\beta_s \left( \hat{H}_s - F \right)} ,
\end{equation}
we can write the entropy production can be expressed in terms of the internal energy,
\begin{equation}\label{produccionenergiainterna}
    \Sigma = \Delta S_{s} - \beta_s \Delta U \geq 0,
\end{equation}
Alternatively, using the first law of thermodynamics in the weak-coupling regime, we can write
\begin{equation}
\Sigma = \Delta S_{s} - \beta_s \big( Q+ W\big) \geq 0.
\end{equation}
where $Q$ y $W$ represent heat and work, respectively. In this way, the interpretation that follows from equation \eqref{produccionenergiainterna} indicates that entropy production, besides being a purely informational quantity (see Eq. \eqref{produccionentropiainformacional}), can also be expressed in energetic terms, such as the change in internal energy. In the context of thermal maps, the distinction between heat and work becomes less relevant. In other words, entropy production is independent of the specific formulation of quantum thermodynamics, although each definition of heat and work provides its own perspective on the process.

\end{chapter}
\begin{chapter}{NON-MARKOVIANITY THROUGH ENTROPY-BASED QUANTUM THERMODYNAMICS}
\label{cap4}

This chapter explores the connections between quantum thermodynamics, formulated in terms of the von Neumann entropy, and the non-Markovian dynamics of open quantum systems. In particular, we propose an original approach to characterize non-Markovianity through thermodynamic variables such as heat and work, which exhibit monotonic behavior under Markovian evolution.
\vspace{0.2cm}\\
\noindent The main motivation of this chapter is to employ the entropy-based formulation of the first law of thermodynamics to define a measure of non-Markovianity based on the heat flow during the evolution of a single-qubit system. This approach is particularly suitable for unital dynamical maps that preserve the sign of the internal energy, allowing for a natural connection with quantum coherence in incoherent maps. In addition, we present applications to both dissipative and non-dissipative processes, highlighting the role of thermodynamic variables as effective indicators of non-Markovianity.
\vspace{0.2cm}\\
\noindent This chapter is based on the first work developed during the PhD~\cite{john}, published in Physical Review A. Its inclusion in this thesis reflects our original contribution at the interface between quantum thermodynamics and quantum information theory.

\section{Quantum thermodynamics}

According to the standard framework for quantum thermodynamics~\cite{alicki}, the internal energy $U$ of a system described by a density operator $\hat{\rho}_s$ is provided  
by the expected value of its Hamiltonian $\hat{H}_s$, i.e., $U=\mathrm{tr}\left[\hat{\rho}_s \hat{H}_s\right]$. In this formalism, the first law of thermodynamics emerges from an infinitesimal change 
in the internal energy 
\begin{equation}
    dU=\delta Q_\text{stand}+\delta W_\text{stand},
\end{equation}
with 
\begin{equation}
 \delta Q_\text{stand}=\mathrm{tr}\left[d\hat{\rho}_s \hat{H}_s\right]\quad \text{and}\quad
 \delta W_\text{stand}=\mathrm{tr}\left[\hat{\rho}_s d\hat{H}_s\right]\\[2.5pt]   
\end{equation}
defining the heat absorbed by system and 
the work performed on system, respectively. Considering the density operator as expressed in its spectral decomposition, we have 
\begin{equation}
    \hat{\rho}_s=\sum_k r_k \left|r_k\right\rangle\left\langle r_k\right|,
\end{equation} 
where $\left|r_k\right\rangle$ denotes an eigenvector of $\hat{\rho}_s$ and $r_k$ the corresponding eigenvalue. Then, 
the thermodynamic quantities $U$, $\delta Q_\text{stand}$, and $\delta W_\text{stand}$ can be rewritten as
\begin{equation}\label{Eq:energy1}
U=\sum_k r_k \left\langle r_k\right|\hat{H}_s\left| r_k\right\rangle,\\[5pt]
\end{equation}
\begin{equation}\label{Eq:heat}
\delta Q_\text{stand}=\sum_k dr_k \left\langle r_k\right|\hat{H}_s\left| r_k\right\rangle+\sum_k r_k \left(\left\langle r_k\right|\hat{H}_s\,\,\,(d\left|r_k\right\rangle)+(d\left\langle r_k\right|)\,\,\hat{H}_s\,\left|r_k\right\rangle\right),\\[5pt]
\end{equation}
\begin{equation}\label{Eq:work}
\delta W_\text{stand}=\sum_k r_k \left\langle r_k\right|d\hat{H}_s\left|r_k\right\rangle.\\[5pt]
\end{equation}
Note that the first term on the right-hand-side of Eq.~(\ref{Eq:heat}) is the responsible for changes in the von Neumann entropy,  $S= -k_{B}\mathrm{tr}\left[\hat{\rho}_s\mathrm{ln}\hat{\rho}_s\right]$, since 
\begin{equation}
 dS= -k_{B}\sum_k  dr_k \mathrm{ln}r_k.\\[5pt]
\end{equation} 
Therefore, in order to connect the heat flow with the entropy change as in classical thermodynamics, an entropy-based formulation of quantum thermodynamics has recently been introduced in Refs.~\cite{alipour1,alipour2,Ahmadi2020}. 
In this framework, heat and work are redefined through 
\begin{equation}\label{Eq:heat2}
\delta\mathbb{Q}_\text{entro}=\delta Q_\text{stand} -\delta \mathbb{W}^*
\end{equation}	
and
\begin{equation}\label{Eq:work2}
\delta \mathbb{W}_\text{entro}=\delta W_\text{stand} +\delta \mathbb{W}^*,\\[2.5pt]
\end{equation}
where $\delta \mathbb{W}^*$ is an additional work contribution given by
\begin{equation}\label{Eq:workd*}
\delta \mathbb{W}^*=\sum_k r_k \left(\,\, \left\langle r_k\right|H\,\,(d\left|r_k\right\rangle)+(d\left\langle r_k\right|)\,\,\,H\,d\left|r_k\right\rangle\,\right).\\[2.5pt]
\end{equation}
The work $\delta \mathbb{W}^*$ is related to the variation $d\left|r_k\right\rangle$ of the density operator eigenvectors. Notice that the entropy-based formalism satisfies the first law of thermodynamics, 
i.e., 
\begin{equation}
    dU=\delta \mathbb{Q}_\text{entro}+\delta \mathbb{W}_\text{entro},\\[2.5pt]
\end{equation} being then equivalent to the standard framework for $\delta \mathbb{W}^{*}=0$. Remarkably, it can be shown that the existence of quantum coherence in $\hat{\rho}_s$ 
in the energy eigenbasis $\{\left|h_k\right\rangle\}$ is a necessary ingredient for a non-zero work $\delta\mathbb{W}^*$ if the energy eigenvectors are fixed (i.e., for  
$d\left|h_k\right\rangle= 0$, $\forall k$).
\vspace{0.2cm}\\
\noindent Here, we will define coherence through the $l_1$-norm \cite{cramer} of $\hat{\rho}_s$ in the energy eigenbasis, reading
\begin{equation}\label{Eq:coherence1}
C(\hat{\rho}_s)=\sum_{k\neq l}\left|\left\langle h_k\right|\hat{\rho}_s\left|h_l\right\rangle\right|.\\[5pt]
\end{equation}
Indeed, observe that, if $\hat{H}_s$ is constant and $\hat{\rho}_s$ and $\hat{H}_s$ have a common basis of eigenvectors, then $d\left|h_k\right\rangle = d\left|r_k\right\rangle = 0$. This leads to $ \delta \mathbb{W}^*=0$. 
Moreover, since a common basis for $\hat{H}_s$ and $\hat{\rho}_s$ implies that $\hat{\rho}_s$ is diagonal in the energy eigenbasis $\{\left|h_k\right\rangle\}$, we will have $C(\hat{\rho}_s)=0$. 
\vspace{0.2cm}\\
\noindent For an arbitrary  single qubit system, the density operator can be written in the Pauli basis $\{\hat{\mathbb{I}},\vec{\hat{\sigma}}\}$ as
\begin{equation}
\hat{\rho}_s=\frac{1}{2} \left(\hat{\mathbb{I}}+\vec{r}\cdot\vec{\hat{\sigma}}\right),\\[2.5pt]
\end{equation}
where  $\vec{r}=(x,y,z)$ is the Bloch vector.
For the Hamiltonian, we have $\hat{H}_s=-\vec{h}\cdot\vec{\hat{\sigma}}$.
We observe that $\vec{r}=(x,y,z)$ can be interpreted as a classical magnetic dipole moment immersed in an external magnetic field $\vec{h}$. 
Indeed, the quantities in Eqs.~(\ref{Eq:energy1}) - (\ref{Eq:coherence1}) reduce to \cite{vallejo}:
\begin{equation}\label{Eq:energyq}
U=-\vec{h}\cdot\vec{r},\\[5pt]
\end{equation}
\begin{equation}\label{Eq:heatq}
\delta Q_\text{stand}=-\vec{h}\cdot d\vec{r},\\[5pt]
\end{equation}
\begin{equation}\label{Eq:workq}
\delta W_\text{stand}=-\vec{r}\cdot d\vec{h},\\[5pt]
\end{equation}
\begin{equation}\label{Eq:heat2q} 
\delta\mathbb{Q}_\text{entro}=\frac{U}{r}dr,\\[5pt]
\end{equation}
\begin{equation}\label{Eq:work2q} 
\delta \mathbb{W}_\text{entro}=rd\left(\frac{U}{r}\right),\\[5pt]
\end{equation}
\begin{equation}\label{Eq:work*q}
\delta \mathbb{W}^{*}=-rh\,\hat{h}\cdot d\hat{r},\\[5pt]
\end{equation}
and
\begin{equation}\label{coherenceq}
C=r\sqrt{1-\frac{(U_{r})^2}{h^2}},\\[5pt]
\end{equation}
where $r\equiv |\vec{r}|$ is the purity, $U_r\equiv U/r$ is the internal energy per unit of purity, $h\equiv |\vec{h}|$ is the positive energy eigenvalue, $\hat{r}\equiv\vec{r}/r$ is 
the dipole direction, and $\hat{h}\equiv\vec{h}/h$ is the external field direction. Note that heat and work in the entropy-based qubit formalism, namely, $\delta\mathbb{Q}_\text{entro}$ and $\delta\mathbb{W}_\text{entro}$, 
are necessarily associated with changes in $r$ (and consequently in $S$) and $U_r$, respectively. On the other hand, the standard contributions for these thermodynamic quantities, 
$\delta Q_\text{stand}$ and $\delta W_\text{stand}$, are required to be associated with variations in $\vec{r}$ and $\vec{h}$, respectively. \\
\vspace{0.2cm}\\
\noindent An interpretation for $\delta \mathbb{W}^*$ in terms of the behavior of $\vec{r}$ can be obtained through 
the relation between $\delta \mathbb{W}^*$ and $C(\hat{\rho}_s)$. First, a change in the eigenvectors of $\hat{\rho}_s$ is required for non-zero $\delta \mathbb{W}^*$. By imposing $d\hat{h}=0$ (i.e., a fixed energy eigenbasis), 
we can express $\delta \mathbb{W}^*$ in Eq.~(\ref{Eq:work*q}) as
\begin{equation}
    \delta \mathbb{W}^*=h\,C\,d\theta,
\end{equation}
where $h\,C=|\vec{r}\times\vec{h}|$ denotes the absolute value of the torque 
on $\vec{r}$ induced by $\vec{h}$ and $\theta=\text{arcos}(\hat{h}\cdot\hat {r})$ is the angle between $\vec{r}$ and $\vec{h}$~\cite{vallejo}. Therefore, $\delta \mathbb{W}^{*}$ is 
equivalent to the energy cost required to rotate a magnetic dipole moment immersed in an external magnetic field, being proportional to coherence. 
More generally, $\delta \mathbb{W}^{*}$ represents the departure from the quasistatic dynamics~\cite{alipour1,alipour2}.

\section{Characterizing non-Markovianity}
\label{sec:III}

Let us suppose an open-system dynamical evolution governed by a time-local master equation 
\begin{equation} \label{Eq:mastereq}
\frac{d}{dt}\hat{\rho}_s(t) = {\cal L}_t \hat{\rho}_s(t) = -i[\hat{H}_s(t),\hat{\rho}_s(t)] + \sum_i \gamma_i(t) \Big( \hat{L}_i(t)\hat{\rho}_s(t) \hat{L}^\dagger_i(t) - \tfrac{1}{2} \{ \hat{L}^\dagger_i(t)\hat{L}_i(t), \hat{\rho}_s(t)\} \Big).
\end{equation}
where ${\cal L}_t$ is the time-dependent generator, $\hat{H}_s(t)$ is the effective Hamiltonian of the system, $\hat{L}_i(t)$ are the Lindblad operators. 
By taking $\gamma_i(t) \ge 0$, ${\cal L}_t$ assumes the Lindblad form at each instant of time \cite{Gorini1976, Lindblad1976}. Consequently, the master equation solution $\hat{\rho}_s(t)=\Phi_{t,\,\tau}\hat{\rho}_s(\tau)$ is obtained through a completely positive trace-preserving (CPTP) map $\Phi_{t,\,\tau} = {\hat{\mathcal {{T}}}} \exp \left(\int_{\tau}^{t} dt^\prime{\cal L}_{t^\prime} \right)$, with ${\hat{\mathcal T}}$ representing the chronological time-ordering operator. In this case, the dynamical map $\Phi_{t,\,\tau}$ satisfies the divisibility condition, i.e.,  $\Phi_{t,\,\tau}=\Phi_{t,\,r}\Phi_{r,\,\tau}$ $(t \ge r \ge \tau \ge 0$), which characterizes the Markovianity of the 
dynamical evolution. On the other hand, for  $\gamma_i(t) < 0$, the corresponding dynamical map $\Phi_{t,\,\tau}$ may not be CPTP for
intermediate time intervals and the divisibility property of the overall CPTP dynamics is violated, which characterizes a non-Markovian behavior~\cite{Rivas2010, Breuer2009, Chruscinski2014}. 
\vspace{0.2cm}\\
\noindent Now, assume $F_{\alpha}(t) = F_{\alpha}(\hat\rho(t))$ represents an arbitrary monotonic function of $t$ under divisible dynamical maps, where $\alpha=+1$ and $\alpha=-1$ indicate increasing and decreasing behaviors, 
respectively. Then, a sign change in $\frac{d}{dt}{F}_{\alpha}(t)$ works as a witness of non-Markovianity. From this breakdown of monotonicity, we propose a measure of non-Markovianity as 
\vspace{0.5cm}
\begin{equation}\label{Eq:NF}
N_{F_{\alpha}}[\Phi]=\max_{\hat{\rho}_{s}(t_0)}\int_{\text{sgn}\frac{d}{dt}{F}_{\alpha}=-\alpha}{\left|\frac{d}{dt}{F}_{\alpha}(t)\right|dt}.\\[5pt]
\end{equation}
The maximization in Eq.~(\ref{Eq:NF}) is performed over all sets of possible initial states $\hat{\rho}_s(t_0)$, and the integration  extends over all time intervals for which the sign of $\,\frac{d}{dt}{F}_{\alpha}(t)\,$ is $\text{sgn}\frac{d}{dt}{F}_{\alpha}=-\alpha$.  If $\{(t_i^k, t_f^k)\}$ represents the set of all time intervals for which $\text{sgn}\frac{d}{dt}{F_{\alpha}}=-\alpha$, then we can write
\begin{equation}\label{Eq:NF2}
N_{F_{\alpha}}[\Phi]=\max_{\hat{\rho}_s(t_0)}\sum_{k\, :\, \text{sgn}\frac{d}{dt}{F}_{\alpha}=-\alpha}\left|F_{\alpha}(t_f^k)-F_{\alpha}(t_i^k)\right|.\\[5pt]
\end{equation}
We can employ different functions $F_{\alpha}$ depending on the dynamical map. First, let us consider the case of operations over incoherent states. By choosing a fixed basis $\{|i\rangle\}$ in a $d$-dimensional 
Hilbert space, incoherent states are defined by density operators $\hat{\rho}_s^{\,\text{incoh}}$ that are diagonal when expressed in the basis $\{|i\rangle\}$, namely, 
\begin{equation}
 \hat{\rho}_s^{\,\text{incoh}}=\sum_i c_i \, |i\rangle\langle i|   ,\\[2.5pt]
\end{equation}
with $c_i$ denoting an  
arbitrary complex amplitude. Then, it follows the notion of an  incoherent map, which is a dynamical map 
leading any incoherent state to another incoherent state. For the case of incoherent quantum maps, it can be shown that quantum coherence $C(\hat\rho(t))$ can witness non-Markovianity through  $F_{-1}(t)=C(\hat\rho(t))$~\cite{titas}. 
We can also consider the case of a unital map, which maps the identity operator to itself, $\Phi(\hat{\mathbb{I}})=\hat{\mathbb{I}}$. In this case, it can be shown 
that the von Neumann entropy can witness non-Markovianity through $F_{+1}(t)=S(\hat\rho(t))$~\cite{haseli}. 
\vspace{0.2cm}\\
\noindent An operational approach to determine whether or not a dynamical map is unital or incoherent can be established through the sufficient conditions~\cite{titas} obtained from Eq.~(\ref{Eq:mastereq}) 
\begin{itemize}
    \item[\hypertarget{caso 4i}{\textcolor{blue}{\textbf{(i)}}}] if $[\hat{L}_i,L_i^{\dagger}]=0$ then $\Phi$ is a {\it{unital map}}.
    \item[\hypertarget{caso 4ii}{\textcolor{blue}{\textbf{(ii)}}}] If $\left\langle h_n\right|\hat{L}_i\left|h_k\right\rangle\left\langle h_k\right|\hat{L}^{\dagger}_i\left|h_m\right\rangle=0$ for all $k$ and $n\neq m$ then $\Phi$ is an {\it{incoherent map}} in the energy eigenbasis $\{\left|h_k\right\rangle\}$.
\end{itemize} 
There are several well known quantum maps that are both incoherent and unital, such as phase flip, bit flip, bit-phase flip, among others~\cite{Nielsen-Book}. For further applications of Eq.~(\ref{Eq:NF2}) as a non-Markovianity measure and its relationship with correlation measures, see also Ref.~\cite{Paula2016}. Notice that, in this generalized approach, $\alpha$ may depend on the initial state, enabling the use of thermodynamic quantities such as internal energy, heat, and work to characterize non-Markovianity.
\vspace{0.2cm}\\
\noindent The use of the von Neumann entropy to witness non-Markovianity for unital maps suggests that heat flow, as defined by the entropy-based formulation of quantum thermodynamics, is also able to detect non-Markovian processes for unital maps. 
Indeed, the expression $\frac{d}{dt}{\mathbb{Q}_\text{entro}}=U_r\,\frac{d}{dt}{r}$ obtained from Eq.~(\ref{Eq:heat2q}) reveals that heat is {\textit{monotonically}} related to the purity (consequently to the entropy) for single qubit energy sign-preserving dynamics i.e., quantum 
evolutions such that the internal energy $U$ is either a non-negative or a non-positive function of the time $t$. Notice that the purity does not increase under unital Markovian quantum processes~\cite{Streltsov2018}.~Then, we can establish the following measure of 
non-Markovianity:
\begin{itemize}
\item [(\textbf{a})] \textit{$N_{\mathbb{Q}_\text{entro}}[\Phi]$ is a measure of non-Markovianity if $\Phi$ is a single-qubit energy sign-preserving unital map.}
\end{itemize}
In this case, heat can witness non-Markovianity via $F_{\alpha}(t)=\mathbb{Q}_\text{entro}(t)$ with $\alpha=\text{sgn}\,U\neq 0$.  As an illustration, the sign of $U$ does not change under \textit{isochoric} processes, characterized by $\frac{d}{dt}\left(U_r\right) = 0$, with heat a monotonic function of the purity
\begin{equation}\label{Eq:isochoric}
\mathbb{Q}_\text{entro}=\Delta U=\frac{U}{r}\left(r-r_0 \right),
\end{equation}
where $r(t_0)=r_0$ represents the initial purity. Then, we can employ either $\mathbb{Q}_\text{entro}$ or $U$ to quantify the degree of non-Markovianity for \textit{isochoric unital maps}. 
\vspace{0.2cm}\\
\noindent Other examples include \textit{non-dissipative}  processes, characterized by $\frac{d}{dt}{U}=0$,  where heat and work are monotonically related to the quantum coherence,
\begin{equation}\label{Eq:nondissipative}
\mathbb{Q}_\text{entro}=-\mathbb{W}_\text{entro}=U\,\mathrm{ln}\,\,\sqrt{\frac{C^2+U^2/h^2}{C_{0}^2+U^2/h^2}},
\end{equation}
being $C_0$ the initial coherence. Consequently, $\mathbb{Q}_\text{entro}$ or $\mathbb{W}_\text{entro}$ can characterize the non-Markovianity of \textit{non-dissipative unital maps} as well as \textit{non-dissipative incoherent maps}. 
\vspace{0.2cm}\\
\noindent Regardless of $\mathbb{Q}_\text{entro}$, we can use $U$ and $\mathbb{W}_\text{entro}$ for characterization of non-Markovianity if $\vec{h}(t)\cdot \vec{r}(t)$ and $U_r(t)$ are monotonic functions of $t$ for $\gamma_i (t)\geq 0$ ($\forall \, t\geq 0$), respectively. 

\noindent As a special case, let us provide sufficient conditions for the thermodynamical variables to witness non-Marovianity for a time-independent Hamiltonian. In this scenario, we take $\hat{H}_s=\omega_0\hat{\sigma}_z$ and 
denote $\vec{r}(t)=[x(t),y(t),z(t)]$. Moreover, let's define $z_r(t)\equiv \frac{z(t)}{r(t)} = \text{cos}\,\theta(t)$. Then, it follows that
\begin{equation}\label{Eq:energy-z}
U(t)=\omega_0z(t),\\[5pt]
\end{equation}
\begin{equation}\label{Eq:heatflow-z}
\frac{d}{dt}{\mathbb{Q}_\text{entro}}(t)=\omega_0 \,{z}_{r}(t)\frac{d}{dt}{r}(t),\\[5pt]
\end{equation}
\begin{equation}\label{Eq:workflow-z}
\frac{d}{dt}{\mathbb{W}_\text{entro}}(t)=\frac{d}{dt}{\mathbb{W}}^*(t)=\omega_0\,r(t)\frac{d}{dt}\Big({z}_{r}(t)\Big).\\[5pt]
\end{equation}
Hence, from Eq.~(\ref{Eq:heatflow-z}), we can establish that: 
\begin{itemize}
\item [(\textbf{b})] \textit{$N_{\mathbb{Q}_\text{entro}}[\Phi]$ is a measure of non-Markovianity for a time-independent Hamiltonian $\hat{H}_s=-\omega_0\hat{\sigma}_z$ if $\Phi$ is a single-qubit unital map that does not invert the sign of $z(t)$.}
\end{itemize}
Furthermore, it follows from Eq.~(\ref{Eq:energy-z}) and Eq.~(\ref{Eq:workflow-z}) that $N_U$ and $N_{\mathbb{W}_\text{entro}}$ are measures of non-Markovianity if $z(t)$ and $z_r(t)$  are monotonic functions of time for $\gamma_i (t)\geq 0$ ($\forall\, t \geq 0$), respectively.


\section{Applications}

\subsection{A dissipative quantum evolution}

Let us first consider  a dissipative single-qubit dynamics described by a time-local master equation given by Eq.~\eqref{Eq:mastereq}, with
\begin{equation}
    \hat{H}_s=\omega_0\hat{\sigma}_z, \quad\hat{L}_i =\delta_{1,i}\, \hat{\sigma}_x,\,\,\, \text{and} \,\,\,\gamma_i=\delta_{1,i}\,\gamma,
\end{equation} such that $\omega_0>0$ and $\gamma>0$. Then
\begin{equation} \label{Eq:mastereq-markov1}
    \frac{d\hat{\rho}_s(t)}{dt}=-i\omega_0\left[\hat{\sigma}_z, \hat{\rho}_s(t)\right]+\gamma
    \left[\hat{\sigma}_{x}\hat{\rho}_s(t)\hat{\sigma}_{x}-\hat{\rho}_s(t)\right].\\[1pt]
\end{equation}

\noindent This master equation generates a Markovian quantum process $\hat{\rho}_s(t)=\Phi_{M}\,\hat{\rho}_s(t_0)=(\hat{\mathbb{I}}+\vec{r}(t)\cdot\vec{\hat{\sigma}})/2$, where $\vec{r}(t)=[x(t),y(t),z(t)]$, 
whose solution for the Bloch vector is given by
\begin{equation} 
x(t) =\frac{e^{-\gamma t}}{2 \omega}\left[\alpha_x e^{\omega t}+\beta_x e^{-\omega t} \right], \label{xx1}\\[5pt]
\end{equation}
\begin{equation} 
y(t) =\frac{e^{-\gamma t}}{2 \omega}\left[\alpha_y e^{\omega t}+\beta_y e^{-\omega t} \right], \label{yy1}\\[5pt]
\end{equation}
\begin{equation} 
z(t)= z_0 e^{-2\gamma t},  \label{Eq:BlochVectorExample1}\\[5pt]
\end{equation} 
where
\begin{equation} 
\alpha_x= \omega x_0 + \gamma x_0 - 2 \omega_0 y_0\quad \quad,\quad \quad \beta_x= \omega x_0 - \gamma x_0 + 2 \omega_0 y_0, \\[2.5pt]
\end{equation}
\begin{equation} 
\alpha_y= \omega y_0 - \gamma y_0 + 2 \omega_0 x_0\quad \quad,\quad \quad \beta_y= \omega y_0 + \gamma y_0 - 2 \omega_0 x_0, \\[3.5pt]
\end{equation}
\noindent with 
\begin{equation}
\,\omega=\sqrt{\gamma^2 - 4 \omega_0^2}\,\end{equation}
and $\,\vec{r}(t_0)=[x_0,y_0,z_0]\,$ denoting the initial state. Figure~ \ref{fig:paper1app1} illustrates the evolution of the Bloch vector for an specific initial state.

\begin{figure}[H] 
\centering
\includegraphics[width=0.42\columnwidth]{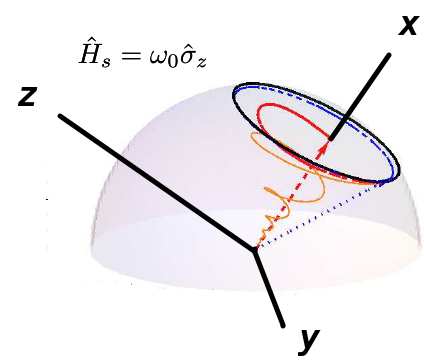}
\caption{(Color online) Evolution of the Bloch vector under the master equation \eqref{Eq:mastereq-markov1}. The Hamiltonian $\omega_0 \hat{\sigma}_z$ causes rotation around $z$, while dissipation via $\hat{\sigma}_x$ reduces the vector’s amplitude and drives the state toward the $x$-axis, eventually reaching the maximally mixed state $\hat{\rho}_s(t \to \infty) = \mathbb{I}/2$. Adapted from Ref.\cite{Lombardo2014}.
}
\label{fig:paper1app1}
\end{figure}

\noindent The Markovian map $\Phi_M$ is both unital and incoherent, since the Lindblad operators $\hat{A}_i =\delta_{1,i}\, \hat{\sigma}_x$ satisfy the conditions \hyperlink{caso 4i}{\textbf{(i)}} and \hyperlink{caso 4ii}{\textbf{(ii)}} described in Sec.~\ref{sec:III}. Moreover, $\Phi_M$ is also a map that preserves the sign of $z(t)$ (note that $\text{sgn}\,z(t)=\text{sgn}\,z_0$ $\forall\, t\geq 0$). Consequently, we can use the monotonicity of $\mathbb{Q}_\text{entro}(t)$ or $C(t)$ as functions of $t$ to observe the Markovianity of Eq.~(\ref{Eq:mastereq-markov1}). Then
\begin{equation}\label{Eq:NF1}
N_{F_{\alpha}}[\Phi_M]=0,
\end{equation}
where $F_{\alpha}(t)= \mathbb{Q}_\text{entro}(t)$, with $\alpha=\text{sgn}\,z_0\neq 0$ or $F_{\alpha}(t)=C(t)$ with $\alpha=-1$. Indeed, Fig. \ref{fig:1} illustrates the sign preservation of $\frac{dF_{\alpha}(t)}{dt}$ for the initial state $\vec{r}(t_0)=[1/2,0,1/2]$, 
where the heat and coherence flows reduce to
\vspace{0.5cm}
\begin{equation}
\frac{d}{dt}\mathbb{Q}_\text{entro}(t)=\frac{\omega_0 \gamma \left[2\omega_0^2(1-\cosh{(2\omega t)}) -\omega^2 e^{-2\gamma t}\right]e^{-2 \gamma t}}{\omega^2 e^{-2 \gamma t}+\gamma^2 \cosh{(2 \omega t)}+\omega \gamma \sinh{(2\omega t)}-4 \omega_0^2},
\end{equation}
\vspace{0.1cm}
\begin{equation}
\frac{d}{dt}{C}(t)=\frac{2\omega_0^2 \gamma e^{-\gamma t}\left(1-\cosh{2 \omega t}\right)}{\omega \sqrt{\gamma^2 \cosh{(2 \omega t)}+\omega \gamma \sinh{(2\omega t)}-4\omega_0^2}}.
\end{equation}
\begin{figure}[ht!]
\centering
\includegraphics[scale=0.47]{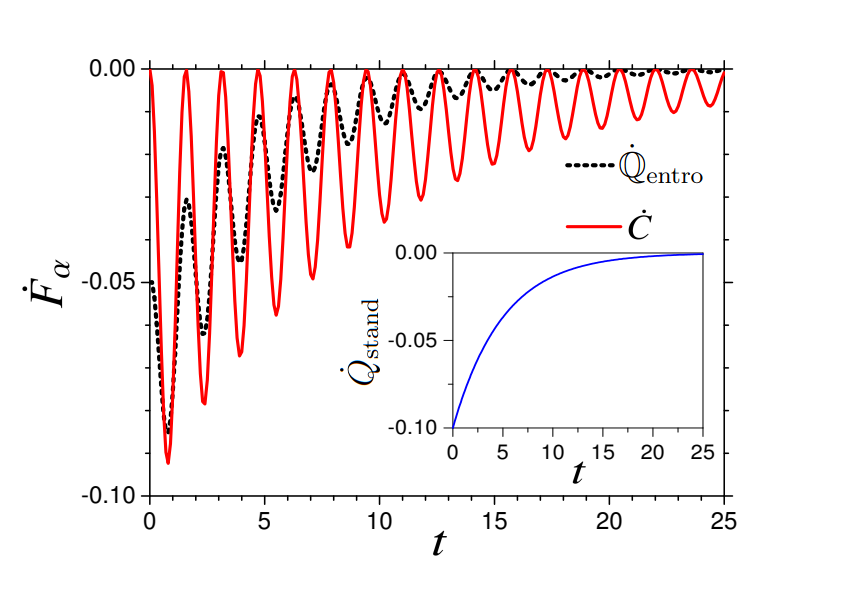}
\caption{(Color online) Heat (dotted black line) and quantum coherence (solid red line) flows as a function of time for $\vec{r}_0=[1/2,0,1/2]$, in units such that $\gamma=0.1$ and $\omega_0=1$. Inset: $\dot{Q}_\text{stand}$ as a function of $t$. Dots denote time derivatives: $\dot{X} \equiv dX/dt$.
}
\label{fig:1}
\end{figure}
\vspace{0.05cm}\\
\noindent The Markovianity of Eq.~(\ref{Eq:mastereq-markov1}) can also be witnessed from the behavior of the internal energy and work  (see plots for $U(t)$ and $\mathbb{W}_\text{entro}(t)$ for $\vec{r}(t_0)=[1/2,0,1/2]$ in Ref.~\cite{alipour1}). 
Concerning the standard thermodynamic quantities, we have that the conventional heat flow $\frac{d}{dt}Q_\text{stand}(t)$ is equal to the internal energy flow $\frac{d}{dt}U(t)$. This holds for an arbitrary time-independent Hamiltonian, since  the conventional work flow $\frac{d}{dt}W_\text{stand}(t)=0$ in this case. From Eqs.~(\ref{Eq:energy-z}) and~(\ref{Eq:BlochVectorExample1}), we then obtain 
\begin{equation}
    \frac{d}{dt}{Q}_\text{stand}(t)=\frac{d}{dt}{U}(t)=-2\gamma \, z_0 \, \omega_0 \,e^{-2\gamma t}.\\[2.5pt]
\end{equation} 
Since  the map $\Phi_M$ preserves the sign of $\frac{d}{dt}{Q}_\text{stand}(t)\,$ (see the inset plot in Fig. \ref{fig:1}), we can also conclude that $N_{F_{\alpha}}[\Phi_M]=0$ for $F_{\alpha}(t)= Q_\text{stand}(t)$ with $\alpha=\text{sgn}\,z_0\neq 0$. 
From the inset in Fig. \ref{fig:1}, observe that the monotonicity between the heat flow and the coherence flow is lost in the conventional framework, but the identification of the Markovian behavior still works.
\vspace{0.2cm}\\
\noindent In order to investigate a non-Markovian scenario, we may take temporary negative constant rates $\gamma$. In this case, Eqs.~(\ref{xx1}),~(\ref{yy1}), and (\ref{Eq:BlochVectorExample1}) 
are  kept for either positive or negative piecewise constant $\gamma$. 
This fact implies that the signs of internal energy, heat, and coherence flows will not be preserved throughout the dynamics, then identifying the non-Markovian behavior.  
Notice that the equivalence observed in this example between the 
conventional and entropy-based formalisms to characterize non-Markovianity is far from general. Indeed, we will consider next a non-dissipative quantum evolution for a time-indepedent Hamiltonian. 
As we will show for that case, both internal energy and conventional heat flow fail as non-Markovianity witnesses, since that $\frac{d}{dt}{Q}_\text{stand}(t)=\frac{d}{dt}{U}(t)=0.$

\subsection{A non-dissipative quantum evolution}

Let us consider now the dynamics of a single qubit under \textit{dephasing}, whose master equation is given by 
\begin{equation}\label{Eq:mastereq3} 
    \frac{d}{dt}{\hat{\rho}_s}(t)=-i\omega_0 \left[\hat{\sigma}_z, \hat{\rho}_s(t)\right]+\gamma
    \left[\hat{\sigma}_{z}\hat{\rho}_s(t)\hat{\sigma}_{z}-\hat{\rho}_s(t)\right],\\[2.5pt]
\end{equation}
This master equation can be derived from Eq.~\eqref{Eq:mastereq} by taking 
\begin{equation}
 \hat{H}_s=\omega_0\hat{\sigma}_z,\quad \hat{L}_i =\delta_{1,i}\, \hat{\sigma}_z,\,\, \text{and}\,\,\, \gamma_i=\delta_{1,i}\,\gamma.
\end{equation}
The solution is given by a map $\Phi_{D}$ such that $\hat{\rho}_s(t)=\Phi_{D}\hat{\rho}_s(t_0)=(\hat{\mathbb{I}}+\vec{r}(t)\cdot\vec{\hat{\sigma}})/2$ with \cite{titas}
\begin{equation} \label{rtnondissipative}
\vec{r}(t)=[x_0\Gamma (t), y_0\Gamma (t),z_0],\\[2.5pt]
\end{equation}
where $\,\Gamma(t)=\exp[-\int_{t_0=0}^t{\gamma(t) dt}]
\,$. Figure~\ref{fig:paper2app21} shows the Bloch vector evolution for an arbitrary initial state.

\begin{figure}[H] 
\centering
\includegraphics[width=0.40\columnwidth]{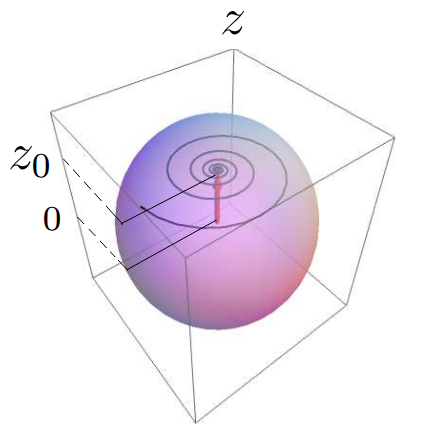}
\caption{(Color online) Evolution of the Bloch vector under the master equation \eqref{Eq:mastereq3}. The dynamics occurs in planes $z(t) = z_0$, while the $x(t)$ and $y(t)$ components decay exponentially according to \eqref{rtnondissipative}. In the long-time limit $t \to \infty$, the initial state evolves toward a fixed point of the dephasing map $\vec{r}(\infty)=(0,0,z_0)$. Adapted from Ref.\cite{koch450}}
\label{fig:paper2app21}
\end{figure}

\noindent In this dynamics, the internal energy and the quantum coherence are given by 
\begin{equation}\label{Eq:U-deph}
U(t)=U(t_0=0)=U_0, 
\end{equation}
with $\,U_0=\omega_0z_0\,$ and 
\begin{equation}\label{Eq:C-deph}
C(t)=C_0\Gamma (t) 
\end{equation}
with $\, C(t_0=0)=C_0=\sqrt{r_{0}^2-z_{0}^2},\,$ respectively. The  map $\Phi_D$ is both \textit{non-dissipative unital} and \textit{non-dissipative incoherent}. Consequently,  we can use $\mathbb{Q}_\text{entro}(t)$, $\mathbb{W}_\text{entro}(t)$, or $C(t)$ to 
characterize the non-Markovianity of Eq.(\ref{Eq:mastereq3}). Inserting Eqs.~(\ref{Eq:U-deph}) and (\ref{Eq:C-deph}) in Eq.~(\ref{Eq:nondissipative}), we can express heat in the form
\begin{equation}\label{Eq:Q-deph}
\mathbb{Q}_\text{entro}(t)=\omega_0\,\frac{z_0}{r_0}\,r_0\ln \sqrt{\Gamma(t)^2+\left(1-\Gamma(t)^2\right)\left(\frac{z_0}{r_0}\right)^2}.\\[5pt]
\end{equation}
\noindent The non-Markovianity measure in Eq.~(\ref{Eq:NF2}) then reads 
\begin{eqnarray} \label{Eq:Nmax}
&&\hspace{-0.5cm}N_{\mathbb{Q}_\text{entro}}[\Phi_D]=\max_{\hat{\rho}_{s}(t_0)}\,\left\{\sum_{k:\text{sgn}\,\frac{d}{dt}{\mathbb{Q}_{entro}}=\text{sgn}\,U_0}\left|\mathbb{Q}_\text{entro}(t_f^k)-\mathbb{Q}_\text{entro}(t_i^k)\right|\right\} \nonumber \\[5mm]
&&\hspace{-0.5cm}=\omega_0\,\max_{|\frac{z_0}{r_0}|,r_0}\,\,\left\{r_0\sum_{k:\gamma < 0}\left|\frac{z_0}{r_0}\right|\ln \sqrt{\frac{\Gamma(t_f^k)^{2}+(1-\Gamma(t_f^k)^{2})\left|\frac{z_0}{r_0}\right|^2}{\Gamma(t_i^k)^{2}+(1-\Gamma(t_i^k)^{2})\left|\frac{z_0}{r_0}\right|^2}}\right\} \nonumber \\[5mm]
&&\hspace{-0.5cm}=\omega_0\,\max_{|z_{0}|}\,\,\left\{\sum_{k:\gamma < 0}|z_{0}|\ln \sqrt{\frac{\Gamma(t_f^k)^{2}+(1-\Gamma(t_f^k)^{2})|z_{0}|^2}{\Gamma(t_i^k)^{2}+(1-\Gamma(t_i^k)^{2})|z_{0}|^2}}\right\}.\\
&&\quad \nonumber
\end{eqnarray}
Note that a pure initial state (i.e., $r_0=1$) results from the maximization procedure in Eq.~(\ref{Eq:Nmax}). In order to enable a numerical comparison with the coherence-based
measure of non-Markovianity, let us consider a typical example of a zero-temperature bosonic reservoir with an Ohmic-like spectral density, where the time-dependent dephasing 
rate presents the specific form \cite{titas, Haikka2013, Shekhar2015}
\begin{equation}\label{Eq:gamma}
\gamma(t,s)=[1+(\omega_c t)^2]^{-s/2}\Gamma_{eu} [s]\sin[s\arctan(\omega_c t)],
\end{equation}
being $s\geq 0$ the ohmicity parameter, $\Gamma_{eu}[s]$ the Euler Gamma function, and $\omega_c$ the cutoff spectral frequency. In this case, $\gamma(t,s)<0$ occurs for 
$t_i^k=t_{2k-1}$ and $t_f^k=t_{2k}$, where \begin{equation}
    t_{k}=\frac{\tan[k\pi/2s]}{\omega_c},
\end{equation}
being $k$ an integer such that $0\leq k \leq \lfloor s\rfloor$ ($\lfloor s\rfloor$  is the floor function of $s$). 
Thus, we can conclude that $0\leq s \leq 2$ and $s> 2$ correspond to the Markovian and non-Markovian regimes of this model, respectively (see Fig.~\ref{fig:paper1app2backflow} for a schematic representation). 
\begin{figure}[H]
\centering
\includegraphics[width=0.77\columnwidth]{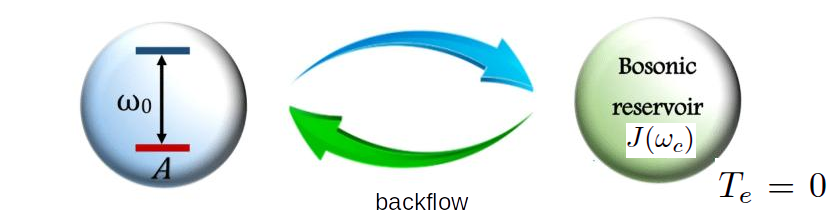}
\caption{(Color online) Qubit of frequency $\omega_0$ interacting with a bosonic reservoir with spectral density $J(\omega_c)$ and $T_e=0$. Arrows show information flow. For $s < 2$ only the blue arrow appears; for $s > 2$, both arrows are present. Adapted from Ref.\cite{hadadi2022}.
}
\label{fig:paper1app2backflow}
\end{figure}

\noindent In terms of the ohmicity parameter, $N_{\mathbb{Q}_\text{entro}}[\Phi_D]$ takes the form 
\begin{equation}\label{Eq:NQ-s} 
N_{\mathbb{Q}_\text{entro}}(s)=\omega_0\ln\left[\prod_{k=1}^{\lfloor s/2 \rfloor}\frac{\Gamma(t_{2k})^{2}+(1-\Gamma(t_{2k})^{2})z_{max}^2}{\Gamma(t_{2k-1})^{2}+(1-\Gamma(t_{2k-1})^{2})z_{max}^2}\right]^{\frac{z_{max}}{2}},\\[5pt]
\end{equation}
where $z_{max}$ represents the value of $|z_0|$ that maximizes the last expression in Eq.~(\ref{Eq:Nmax}). 

\vspace{0.15cm}

\noindent The coherence-based measure $N_{C}[\Phi]$ has been employed 
to quantify the degree of non-Markovianity of the incoherent map $\Phi_D$ for the dephasing rate described in Eq.~(\ref{Eq:gamma}) \cite{titas}. As function of the $s$-parameter, $N_C[\Phi_D]$ reduce to

\begin{equation*}
      N_{C}(s)=\max_{\hat{\rho}_{s}(t_0)}\,\left\{\sum_{k:\text{sgn}\,\frac{d}{dt}{C}(t)=+1}\left|C(t_f^k)-C(t_i^k)\right|\right\}
\end{equation*}
\vspace{0.2cm}
\begin{equation}\label{Eq:NC-s}
  N_{C}(s)=\sum_{k=1}^{\lfloor s/2 \rfloor}\left|\Gamma(t_{2k})-\Gamma(t_{2k-1})\right|,\\[5pt]
\end{equation}

\noindent where a maximally coherent state (i.e., $C_0=1$) emerges from the maximization process. In order to witness non-Markovianity, Fig.~\ref{fig:2} shows the time evolution of the heat $\mathbb{Q}_\text{entro}(t)$ for an initial pure state under the single-qubit dephasing channel in both Markovian and non-Markovian regimes. Note that $\mathbb{Q}_\text{entro}(t)$ is a monotonically decreasing function of time for $s = 1.5$. On the other hand, a non-monotonic behavior arises due to the backflow of heat from the environment to the system for $s = 3.5$. 
\begin{figure}[H]
\centering
\includegraphics[scale=0.63]{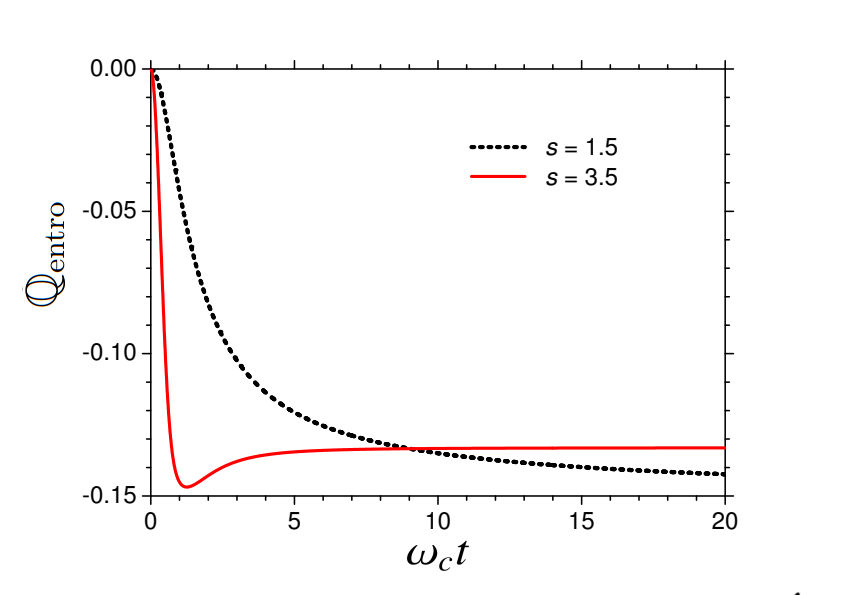}
\caption{(Color online)  $ \mathbb{Q}_\text{entro}$ as a function of $\omega_c t$ for $s = 1.5$ (black dashed line) and $s = 3.5$ (red solid line), with $\omega_0=1$, $r_0=1$, and $z_0=0.05$.}
\label{fig:2} 
\end{figure}
\noindent The behaviors of the non-Markovianity measures $N_{\mathbb{Q}_\text{entro}}$ and $N_C$  as functions of $s$ are illustrated in Fig.~\ref{fig:3}, with the inset showing $z_{max}$ versus $s$. These measures are monotonically related, i.e.,  
\begin{equation}
    N_C(s)\approx 2.0 \, N_{\mathbb{Q}_\text{entro}}(s),
\end{equation}
with both assuming non-zero values only for $s > 2$, a maximum value at $s=3.2$, and negligible values for $s > 5$. From the behavior of $z_{max}(s)$, 
notice that a maximally coherent state (where $z_0=0$) does not optimize the expression in Eq.~(\ref{Eq:NQ-s}) for all $s$.
\begin{figure}[H]
\centering
\includegraphics[scale=0.63]{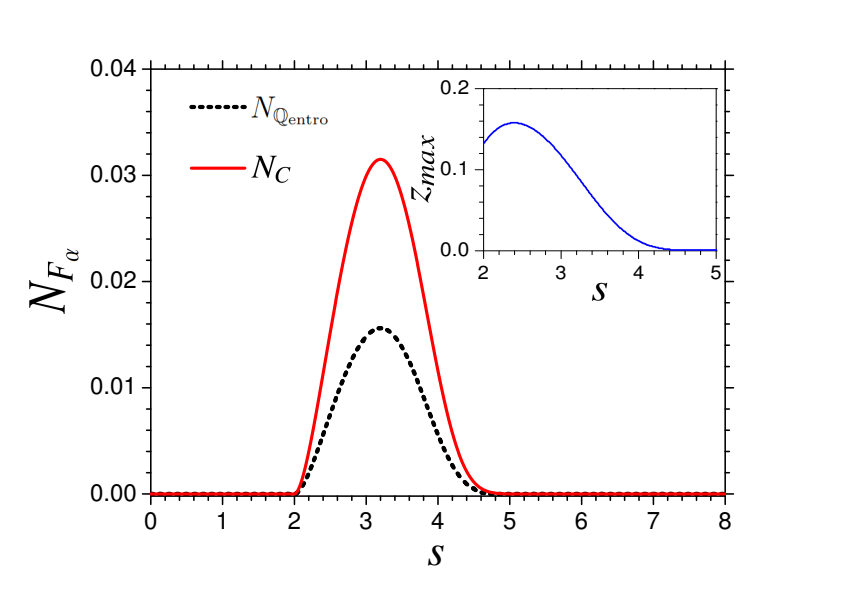}
\caption{(Color online) $N_{\mathbb{Q}_\text{entro}}$ (green solid line) and $N_C$ (blue dashed line) as functions of $s$ for $\omega_0=1$ and $\omega_c =1$. Inset: $z_{max}$ as a function of $s$.}
\label{fig:3}
\end{figure}

\end{chapter}

\begin{chapter}{QUBIT DYNAMICS OF ERGOTROPY AND ENVIRONMENT-INDUCED WORK}
\label{cap5}
\vspace{-1cm}
This chapter is focused to the study of ergotropy as an energy resource in open quantum systems, analyzing its behavior under both Markovian and non-Markovian dynamics. In particular, we establish an analytical relationship between ergotropy and environment-induced work, using the framework of entropy-based quantum thermodynamics.
\vspace{0.2cm}\\
\noindent The main motivation of this work is to explore how the interaction with the environment can change the ergotropy of the system, even when the Hamiltonian stays constant. To do this, we formulate the ergotropy of arbitrary single-qubit states in terms of energy and coherence. This allow us to identify specific conditions for the occurrence of ergotropy freezing and sudden death phenomena, in a way similar to what happens with quantum correlations. Based on this formulation, we show that environment-induced work can be interpreted as a variation in the system’s ergotropy, up to a limit determined by the energy cost of transition between the initial and final passive states.
\vspace{0.2cm}\\
\noindent This chapter is based on the second research work developed during the PhD \cite{Choquehuanca2024}, published in Physical Review A.  Its inclusion in this thesis reflects our original contribution to the understanding of how ergotropy evolves under open system dynamics, and clarifies the role that ergotropy plays in the entropy-based formulation of quantum thermodynamics.

\section{Ergotropy and its quantum dynamics}
The ergotropy is defined as the maximum amount of energy that can be extracted from a quantum system via cyclic unitary operation \cite{Allahverdyan:04}, i.e., 
\begin{equation}
 \mathcal{E}(\hat{\rho}_s)=\max_{\hat{V}\in \mathcal{U}}\,\{U(\hat{\rho}_s)-U(\hat{V}\hat{\rho}_s \hat{V}^{\dagger})\}   
\end{equation}
where $U(\hat{\rho}_s)=\text{tr}\left[\hat{\rho}_s \hat{H}_s\right]$ represents the internal energy, with $\hat{H}_s$ and $\hat{\rho}_s$ denoting the Hamiltonian and the density operator, respectively, and $\mathcal{U}$ the set of all unitary transformations. These transformations are required to be cyclic with respect to $\hat{H}_s$. Assuming the spectral decomposition $\hat{\rho}_s=\sum_nr_n\ket{r_n}\bra{r_n}$ and $\hat{H}_s=\sum_n\varepsilon_n\ket{\varepsilon_n}\bra{\varepsilon_n}$, with the eigenstates reordered so that $r_0\ge r_1\ge ...$ and $\varepsilon_0\le \varepsilon_1\le...$, a close expression for the ergotropy can be obtained in terms of the passive state, $\rho_s^{\,\pi}=\sum_nr_n\ket{\varepsilon_n}\bra{\varepsilon_n}$:
\begin{equation}
\mathcal{E}(\hat{\rho}_s)=U(\hat{\rho}_s)-U(\hat{\rho}_{s}^{\,\pi}).
\label{W}
\end{equation}
In order to explore the role played by quantum coherence, let us consider the $l_1$-norm of coherence, $C(\hat{\rho}_s)=\min_{\hat{\delta}_s \in \mathcal{I}}\lVert \hat{\rho}_s - \hat{\delta}_s \rVert_{l_1}$, with $\mathcal{I}$ representing the set of all incoherent states (i.e., diagonal states) in the basis $\{\ket{\varepsilon_n}\}$.  The minimization leads to $C(\hat{\rho}_s)=\lVert \hat{\rho}_s - \Delta \hat{\rho}_{s} \rVert_{l_1}$, where $\Delta \hat{\rho}_{s} = \sum_n\bra{\varepsilon_n}\hat{\rho}_s\ket{\varepsilon_n}\ket{\varepsilon_n}\bra{\varepsilon_n}$ denotes the dephased state~\cite{cramer}. In terms of $\Delta \hat{\rho}_{s}$, we can define the incoherent ergotropy, i.e., the maximum amount of energy that can be extracted from $\rho$ without altering its quantum coherence,
\begin{equation}
\mathcal{E}_I(\hat{\rho}_s)=\mathcal{E}(\Delta \hat{\rho}_{s}),
\label{WI}
\end{equation}
as well as the coherent ergotropy,
\begin{equation}
\mathcal{E}_C(\hat{\rho}_s)=\mathcal{E}(\hat{\rho}_s)-\mathcal{E}(\Delta \hat{\rho}_{s}).
\label{WC}
\end{equation}
Note that $\mathcal{E}(\hat{\rho}_s)=\mathcal{E}_I(\hat{\rho}_s)+\mathcal{E}_C(\hat{\rho}_s)$~\cite{Francica:20}. Now, let us consider a two-level quantum system 
governed by dimensionless Hamiltonian $\hat{H}_s=-\hat{\sigma}_z$ represented through its spectral decomposition as $\hat{H}_s=\ket{\varepsilon_1}\bra{\varepsilon_1}-\ket{\varepsilon_0}\bra{\varepsilon_0}$, with associated eigenvalues $\varepsilon_0=-1$ and $\varepsilon_1=1$. 
In this case, quantum coherence and energy are, respectively, given by
\begin{equation}\label{Eq:coherence}
C(\hat{\rho}_s)=2\lvert \bra{\varepsilon_0}\hat{\rho}_s\ket{\varepsilon_1}\rvert,
\end{equation}
\begin{equation}\label{Eq:energy}
U(\hat{\rho}_s)=1-2\bra{\varepsilon_0}\hat{\rho}_s\ket{\varepsilon_0},\\[5pt]
\end{equation}
where $-1\leq U \leq 1$ and $0\leq C \leq 1$ such that $U^2+C^2 \leq 1$. From Eqs.~\eqref{W}, \eqref{WI}, and~\eqref{WC}, we can express the ergotropy for an arbitrary two-level system as a function of $C$ and $U$, yielding
\begin{equation}
\mathcal{E}(C,U)=\sqrt{C^2+U^2}+U.
\label{W2}
\end{equation}
For the incoherent and coherent contributions for ergotropy, we obtain
\begin{equation*}
    \mathcal{E}_I(U)=2\max\{0,U\}
\end{equation*}
\begin{equation}\label{Eq:WI}
\mathcal{E}_I(U)=\mathcal{E}(0,U)
\end{equation}
and
\begin{equation*}
    \mathcal{E}_C(C,U)=\sqrt{C^2+U^2}-|U|
\end{equation*}
\begin{equation}\label{Eq:WC}
\mathcal{E}_C(C,U)=\mathcal{E}(C,-|U|),\\[5pt]
\end{equation}
respectively. Note that $\mathcal{E}(C,0)=\mathcal{E}_C(C,0)=C$. 
According to Eq.~\eqref{W2}, the dynamics of ergotropy will depend on the dynamics of both coherence $C$ and internal energy $U$. 
We can then determine explicit conditions for $U$ and $C$ that ensure peculiar behaviors for the ergotropy as a function of time. 
\vspace{0.2cm}\\
\noindent In particular, it is evident that,
\vspace{0.2cm}\\
\hypertarget{caso a}{\textcolor{blue}{\textbf{(a)}}} $\mathcal{E}(t)=\mathcal{E}_0\,\,$ iff $\,\,C(t)=C_0\,\,$ and $\,\,U(t)=U_0\,\,$\,\,$(\forall\,t)$. 
\vspace{0.5cm}\\
\hypertarget{caso b}{\textcolor{blue}{\textbf{(b)}}} $\mathcal{E}(t)=0$ iff  $C(t)=0$ and $U(t)\leq 0 $ $(\forall\,t)$, where $\mathcal{E}_0=\mathcal{E}(0)$, $C_0=C(0)$, and $U_0=U(0)$.\\[7pt]

\begin{figure}[H] 
\centering
\includegraphics[width=0.49\columnwidth]{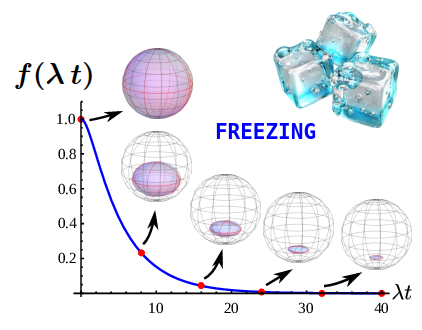}
\caption{(Color online) Visual representation of the “freezing” behavior, where the function $f(\lambda t)$ monotonically decays until it reaches a stationary constant value. In this example, it saturates close to zero around $\lambda t \approx40$. The sequence of Bloch spheres illustrates the progressive loss of coherence and populations, as the Bloch vector shrinks toward the center and eventually remains frozen at a minimal asymmetric radius for later times. Adapted from Ref.\cite{Lorenzo2013}.}
\label{fig:paper2freezing}
\end{figure}

\begin{figure}[H] 
\centering
\includegraphics[width=0.67\columnwidth]{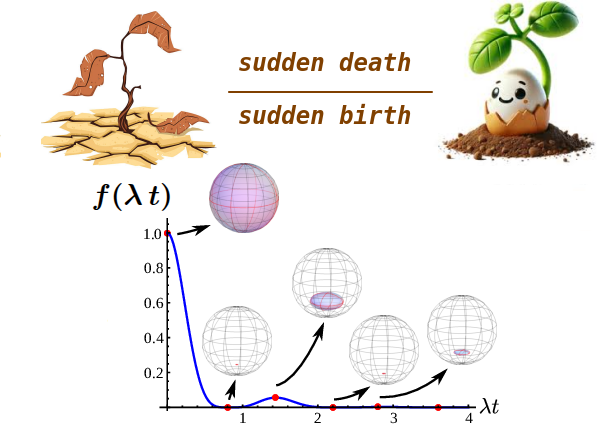}
\caption{(Color online) Illustration of the “sudden death / sudden birth” behavior, where the function $f(\lambda t)$ rapidly vanishes at specific instants but reappears at later times, repeating cycles of disappearance and resurgence before finally decaying permanently. In this figure, the Bloch spheres show the collapse of coherence and population (sudden death) when the Bloch vector momentarily reaches the center, followed by the re-expansion of the Bloch vector (sudden birth), evidencing temporary revivals of coherence and population. Adapted from Ref.\cite{Lorenzo2013}.}
\label{fig:paper2sudden}
\end{figure}

\noindent The properties \hyperlink{caso a}{\textbf{(a)}} and \hyperlink{caso b}{\textbf{(b)}} reveal that the dynamics of ergotropy can exhibit freezing and sudden death effects, respectively, as illustrated in Figs. \ref{fig:paper2freezing} and \ref{fig:paper2sudden}. To explore these two phenomena, we start by defining the dynamics of a quantum system interacting with an environment in terms of Kraus representation as $\hat{\rho}_s(t) = \sum_{i} \hat{K}_i(t)\hat{\rho}_s(t_0=0) \hat{K}_i(t)^\dagger$, where the Kraus operators $\{\hat{K}_i(t)\}$ satisfy the completeness relation $\sum_{i} \hat{K}_{i}(t)^{\dagger} \hat{K}_{i}(t) = \hat{\mathbb{I}}$ and allows the environment characterization in both Markovian and non-Markovian regimes.

{\subsection{Freezing} \label{II(A)}}
\noindent According to condition \hyperlink{caso a}{{\bf (a)}}, the ergotropy freezing can be observed in non-dissipative quantum processes, such as the phase damping (PD) map, when the coherence remains unchanged. The Kraus operators for a non-Markovian PD map are defined as \cite{Utagi:20,TYu:10}
\begin{equation}\label{Eq:PD}
    \hat{K}_0(t) = \sqrt{\frac{1+e^{-q(t)}}{2}}\, \hat{\mathbb{I}}\quad ,\quad  \hat{K}_1(t) = \sqrt{\frac{1-e^{-q(t)}}{2}}\, \hat{\sigma}_z\,,
\end{equation} 
 where
 \begin{equation}
     q(t)=\frac{\gamma}{2}\{t + \frac{1}{\Gamma}(\exp(-\Gamma t )-1) \}.
 \end{equation}
 Here $\gamma^{-1}$ and $\Gamma^{-1}$  define the qubit relaxation time and the reservoir correlation time, respectively, with the Markovian regime in the limit $\Gamma\rightarrow\infty$. For this non-Markovian map, the time evolution of the coherence and energy are, respectively, given by 
\begin{equation}
    C(t)=e^{-q(t)}C_0\,\quad\text{and}\quad\,U(t)=U_0. 
\end{equation}
Thus, from Eq.~(\ref{W2}), a freezing ergotropy at a nonvanishing value $2U_0$ is accomplished for zero initial coherence and positive initial energy, i.e.,
\begin{equation}\label{Eq: Ergotropy}
\mathcal{E}(t)=2U_0\,\,\,\forall\, t \,\,\, \text{iff} \,\,\,C_0=0\,\,\,\text{and}\,\,\, U_0>0.
\end{equation}
According to Eqs.~(\ref{Eq:WI}) and~(\ref{Eq: Ergotropy}), $\mathcal{E}_{I}(t)=2U_0$  $\forall \, t$ iff $U_0>0$ (for all $C_0$). To illustrate this phenomenon, we show $\mathcal{E}$, $\mathcal{E}_I$, and $\mathcal{E}_C$ in Fig.~\ref{fig:PD} as functions of $\gamma t$ for $C_0=0.5$ and $U_0=0.5$, where the initial state is taken as $\rho_{00}=0.25$, $\rho_{11}=1-\rho_{00}$, and $|\rho_{01}|=0.25$.
\vspace{0.2cm}\\
\noindent In both non-Markovian and Markovian regimes, notice that the incoherent component freezes for all $\gamma t$, while the coherent part exhibits a monotonic decay. In terms of the total ergotropy, the contribution of its coherent component allows the ergotropy to decay until it achieves a steady state. Additionally, it is clear to see in Fig.~\ref{fig:PD} a delayed ergotropy decay in the non-Markovian regime in comparison with the Markovian regime. 

\begin{figure}[H]
\centering
\includegraphics[scale=0.25]{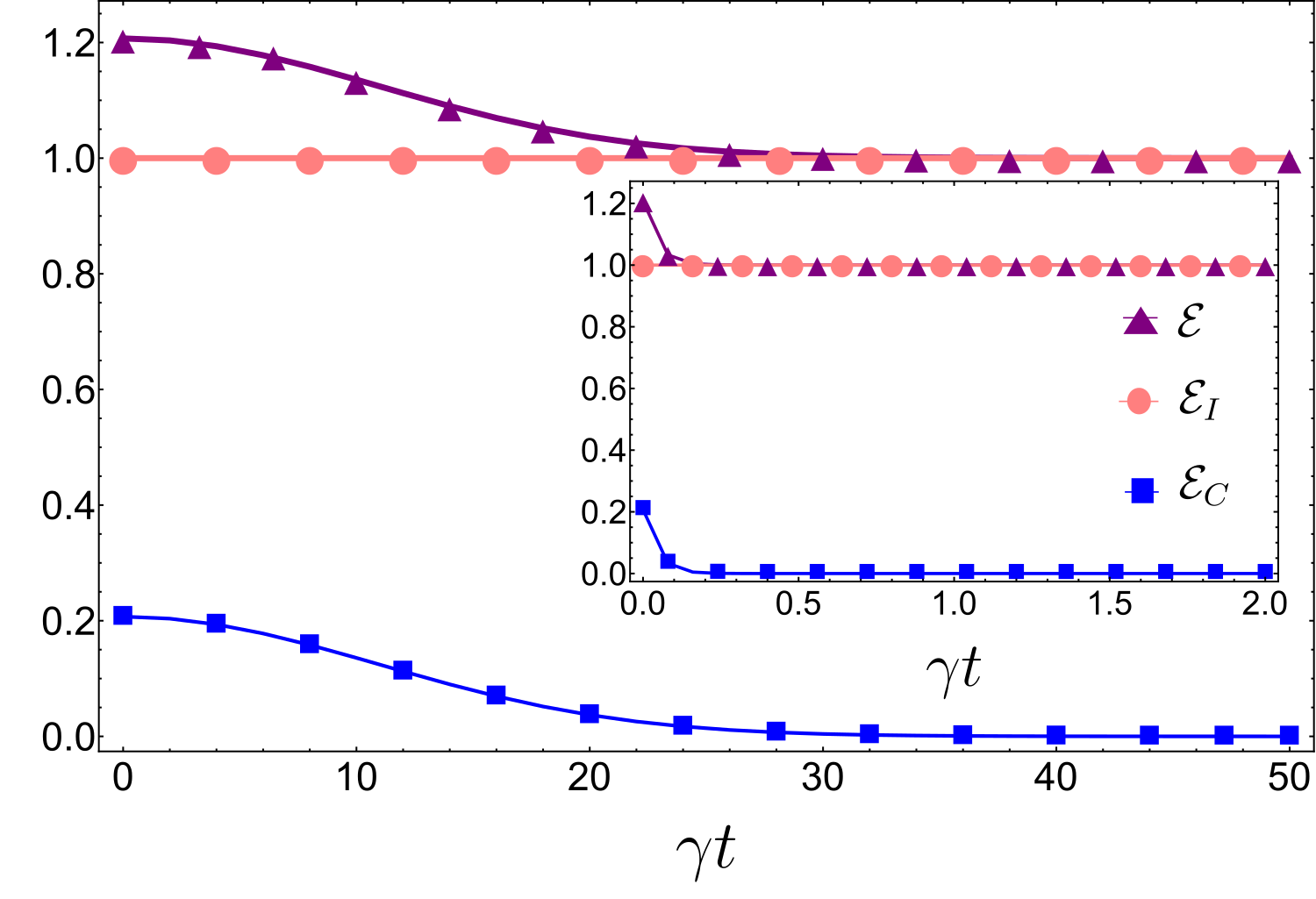}
\caption{(Color online) Dynamics of the ergotropy $\mathcal{E}$, incoherent ergotropy $\mathcal{E}_I$, and coherent ergotropy $\mathcal{E}_C$ as functions of $\gamma t$ under a non-Markovian PD map with $\Gamma=0.01\gamma$.  Inset: Same functions under a Markovian PD map ($\Gamma\rightarrow\infty$).  The initial conditions in both cases are $C_0=0.5$ and $U_0=0.5$.}
\label{fig:PD}
\end{figure}
{\subsection{Sudden death}\label{II(B)}}
\noindent Condition \hyperlink{caso b}{\textbf{(b)}} ensures that the sudden death phenomenon can be achieved in dissipative quantum processes, such as the amplitude damping (AD) map, when coherence is absent. The Kraus operators for a non-Markovian AD map are given by \cite{Bellomo:07} \\[0.1pt]\begin{equation}
    \hat{K}_0(t)=\left[
\begin{array}{cc}
1 & 0 \\
0 & \sqrt{q(t)} 
\end{array} \right]\quad,\
 \, \, \hat{K}_1(t)=\left[
\begin{array}{cc}
0 & \sqrt{1-q(t)} \\
0 & 0 
\end{array} \right],\
\label{ADsd}\\[5pt]
\end{equation}
where 
\begin{equation}
    q(t)=e^{-\Gamma t}\left[\cos\left(\frac{dt}{2}\right)+\frac{\Gamma}{d}\sin\left(\frac{dt}{2}\right)\right]^2,\\[5pt]
\end{equation}
with $d=\sqrt{2\gamma\Gamma -\Gamma^2}$. The spectral width $\Gamma$ and the coupling strength $\gamma$ are related to the reservoir correlation time ($\Gamma^{-1}$) and the qubit relaxation time ($\gamma^{-1}$), respectively.  The dynamics of coherence and energy for this non-Markovian map are, respectively, determined by 
\begin{equation}
    C(t)=\sqrt{q(t)}C_0\,\quad\text{and} \quad\,U(t)=(1+U_0)q(t)-1. \\[5pt]
\end{equation}
As $U(t)<U_0$ for all $t>0$, ergotropy collapses and revivals can be observed if the initial coherence is zero and the initial energy is positive, in agreement with the property \hyperlink{caso b}{\textbf{(b)}}. These sudden changes occur when the energy changes its sign during the quantum process. Consequently, the sudden change times $\{t_n\}$ satisfy the condition
\begin{equation}
    q(t_n)=\frac{1}{1+U_0}\,\,\,\quad\text{with}\quad\,\,\,t_1<t_2<...<t_{sd},
\end{equation}
where odd and even values of $n$ indicate sudden deaths and births, respectively, with $t_{sd}$ characterizing an eternal death. Thus, we conclude that
\begin{equation}
\mathcal{E}(t)=0\quad\forall\, t\geq t_{sd} \quad \text{iff} \quad C_0=0\quad \text{and}\quad U_0>0.
\end{equation}
Consequently, $\mathcal{E}_I(t)=0$ $\forall\, t\geq t_{sd}$ iff $U_0>0$ (for all $C_0$). In the Markovian regime, $\Gamma\rightarrow\infty$, the emergence of eternal death is determined by 
\begin{equation}
t_{sd}=t_1=\frac{\ln(1+U_0)}{\gamma}.
\end{equation}
In this limit, there are no temporary collapses and revivals.
\vspace{0.2cm}\\ 
\noindent The Fig.~\ref{fig:AD} shows the ergotropy, as well as its incoherent and coherent parts, as functions of $\gamma t$ for the initial conditions $C_0 = 0.5$, $U_0 = 0.5$, and initial state given by $\rho_{00}=0.25$, 
$\rho_{11}=1-\rho_{00}$, and $|\rho_{01}|=0.25$. Remarkably, the incoherent component exhibits a non-monotonic (monotonic) decay until the eternal death time $t_{sd} \approx 297/\gamma$ ($t_{sd} \approx 0.405/\gamma$) for the non-Markovian (Markovian) regime, with $\Gamma=0.001\gamma$ ($\Gamma \rightarrow \infty$). Besides, notice that the coherent part contributes to the inhibition of this phenomenon. 

\begin{figure}[H]
\centering
\includegraphics[scale=0.25]{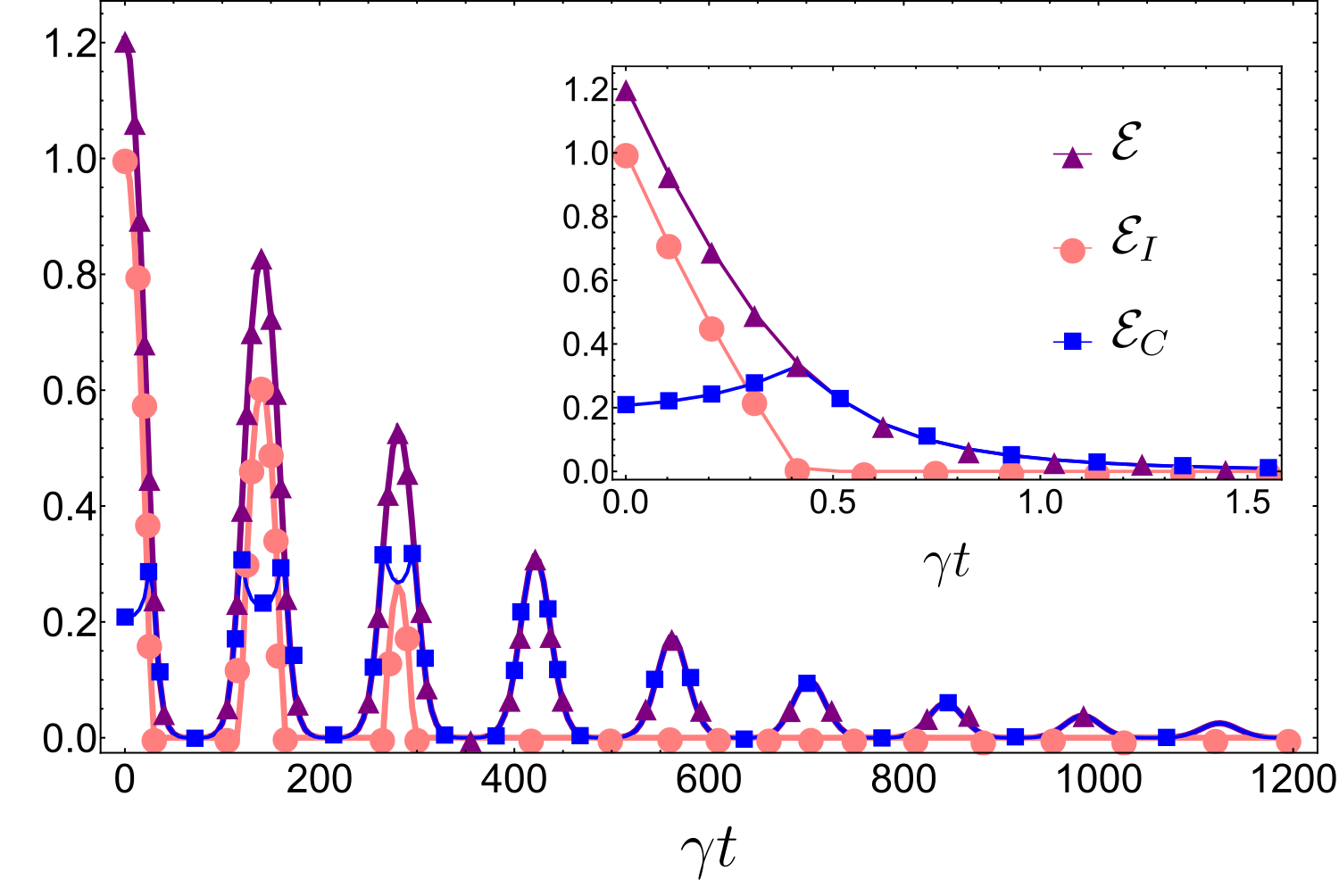}
\caption{(Color online) 
 Dynamics of the ergotropy $\mathcal{E}$, incoherent ergotropy $\mathcal{E}_I$, and coherent ergotropy $\mathcal{E}_C$ as functions of $\gamma t$ under a non-Markovian AD map with $\Gamma=0.001\gamma$.  Inset: Same functions under a Markovian AD map ($\Gamma \rightarrow \infty$).  The initial conditions in both cases are $C_0=0.5$ and $U_0=0.5$.}
\label{fig:AD}
\end{figure}

{\section{Environment-induced work}}
\noindent In order to use ergotropy as a resource, we will investigate how to explicitly extract energy in the form of work via the system-environment interaction. 
The energy balance in a thermodynamic process is ruled by the first law of thermodynamics. In its quantum version, a standard formulation can be written 
as~\cite{alicki}
\begin{equation}
dU = \delta {Q}_\text{stand}  + \delta {W}_\text{stand},
\end{equation}
where $dU$ is the infinitesimal internal energy change, 
$\delta {Q}_\text{stand}$ is the infinitesimal heat exchanged in the process, 
and $\delta {W}_\text{stand}$ is the infinitesimal work performed by (or on) the system, 
with $U = \text{tr} (\hat{H}_s \hat{\rho}_s)$, $\delta {Q}_\text{stand} = \text{tr} (\hat{H}_s \, d\hat{\rho}_s)$, and $\delta {W}_\text{stand} = \text{tr} (d\hat{H}_s \, \hat{\rho}_s)$.
\vspace{0.2cm}\\
\noindent As an alternative formulation, we can modify the definition of $\delta {Q}_\text{stand}$ so that heat is directly linked with the entropy variation. 
In this entropy-based framework for quantum thermodynamics~\cite{alipour1}, heat and work are defined through 
$\delta \mathbb{Q}_\text{entro}=\delta {Q}_\text{stand}- \delta \mathbb{W}^*$ and $\delta \mathbb{W}_\text{entro}=\delta {W}_\text{stand}+\delta \mathbb{W}^*$, so that 
the first law of thermodynamics is
\begin{equation} \label{Eq:1-law}
    dU = \delta \mathbb{Q}_\text{entro} +\delta \mathbb{W}_\text{entro},  
\end{equation}
with $\delta \mathbb{W}^*$ introduced as an environment-induced work~\cite{alipour1}
\begin{equation}\label{Eq:work*}
\delta \mathbb{W}^*= \sum_n r_n \left(\left\langle r_n\right|\hat{H}_s\,\,\,(d\left|r_n\right\rangle)+ (d\left\langle r_n\right|)\,\,\,\hat{H}_s\,\left|r_n\right\rangle\right),\\[1pt]
\end{equation}
\noindent where $\ket{r_n}$ represents an eigenvector of the density operator $\rho$ and $r_n$ the corresponding eigenvalue.
Notice that the first law of thermodynamics is preserved, with the internal energy infinitesimal $dU$ kept unchanged 
due to the new definitions of heat and work. From this point on, unless stated otherwise, 
heat and work will refer to the entropy-based formulation of quantum thermodynamics. A utility for the definition of heat $\mathbb{Q}_\text{entro}$ as a witness of non-Markovianity for unital quantum maps has been recently provided~\cite{john}.
\vspace{0.2cm}\\
\noindent Now, we will provide an operational meaning for $\mathbb{W}^*$ in terms of ergotropy variation. We consider a quantum system described by an initial density operator $\hat{\rho}_s(t_0=0) \to \hat{\rho}_{s_0} $ and governed by a constant Hamiltonian $\hat{H}_s$. In this scenario, the conventional work ${W}_\text{stand}$ is null. The system interacts with an external environment and is taken to a final density operator $\hat{\rho}_s(t_c) \to \hat{\rho}_{s_c}$ at a specific characteristic time $t_c$ such that 
the total heat $\mathbb{Q}_\text{entro}$ exchanged with the environment is vanishing. For this effective adiabatic process, the environment-induced work is the only contribution to the energy balance 
in the first law of thermodynamics (see Fig.~\ref{fig:paper2adiabatic} for a visual illustration of this scenario), i.e.,
\vspace{-0.1cm}
\begin{equation}
\mathbb{Q}_\text{entro}(\hat{\rho}_{s_c})=0 
\end{equation}
\vspace{-0.2cm}
and, consequently,
\begin{equation}
\mathbb{W}^{*}(\hat{\rho}_{s_c})=\Delta U(\hat{\rho}_{s_c})=U(\hat{\rho}_{s_c})-U(\hat{\rho}_{s_0}).
\label{Energ_var}
\end{equation}
By examining the ergotropy variation,
\begin{equation}
\Delta \mathcal{E}(\hat{\rho}_{s_c})= \mathcal{E}(\hat{\rho}_{s_c})- \mathcal{E}(\hat{\rho}_{s_0}),\\[0.1pt]
\label{Erg_var}
\end{equation} by using Eq.~(\ref{W}), we can write \begin{equation}
\Delta \mathcal{E}(\hat{\rho}_{s_c})= [U(\hat{\rho}_{s_c})-U(\hat{\rho}_{s_{c}}^\pi)] - [U(\hat{\rho}_{s_0})-U(\hat{\rho}_{s_{0}}^\pi)],
\label{Erg_var2}
\end{equation} where $\hat{\rho}_{s_{c}}^\pi$ and $\hat{\rho}_{s_{0}}^\pi$ are the passive states associated with $\hat{\rho}_{s_{c}}$ and $\hat{\rho}_{s_{0}}$, respectively.

\begin{figure}[H] 
\centering
\includegraphics[width=0.7\columnwidth]{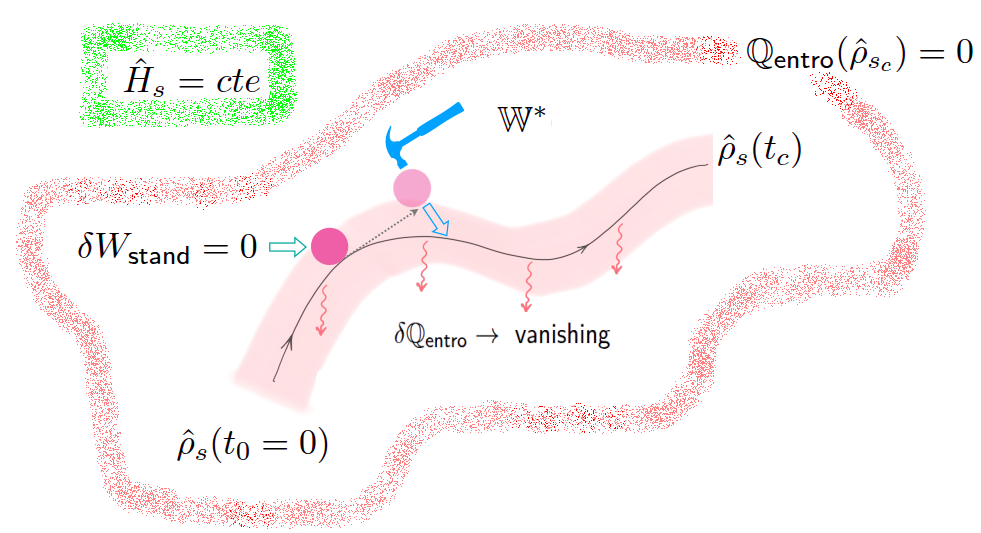}
\caption{(Color online) Schematic representation of an effective adiabatic evolution from $\hat{\rho}_s(t_0)$ to $\hat{\rho}_s(t_c)$ under a constant Hamiltonian. Along this dynamical trajectory, the environment–induced work $\mathbb{W}^{*}$ becomes the only relevant energetic contribution, directly captured through the ergotropy variation. Adapted from Ref.\cite{Alipour2019arxiv}.}
\label{fig:paper2adiabatic}
\end{figure}

\noindent Finally, by using Eq.~(\ref{Energ_var}), and defining the passive energy variation,
\begin{equation}
\Delta U_\pi(\hat{\rho}_{s_c}) = U(\hat{\rho}_{s_{c}}^\pi)- U(\hat{\rho}_{s_{0}}^\pi),
\end{equation}
we obtain
\begin{equation}
\Delta \mathcal{E}(\hat{\rho}_{s_c})= \mathbb{W}^*(\hat{\rho}_{s_c}) - \Delta U_\pi(\hat{\rho}_{s_c}).\\[5pt]
\label{Erg_var3}
\end{equation}
The contribution $\Delta U_\pi(\hat{\rho}_{s_c})$ for the ergotropy can be interpreted as the energy cost to transit between the initial and final passive states $\hat{\rho}_{s_{0}}^\pi$ and $\hat{\rho}_{s_{c}}^\pi$, respectively. We observe that Eq.~(\ref{Erg_var3}) agrees with the discussion about the energetics of the ergotropy 
in Ref.~\cite{binder}, with $\Delta U_{\pi}$ defined there as an operational heat. Here, we can then directly connect the 
environment-induced work $\mathbb{W}^*$ with the variation of ergotropy for constant Hamiltonians, reinforcing the interpretation of $\mathbb{W}^*$ as an effective work extracted due to the system-environment interaction.
\vspace{0.2cm}\\
\noindent For two-level systems, the Hamiltonian is given by $\hat{H}_s=-\hat{\sigma}_z$ 
and the density matrix, in terms of the Bloch sphere, can be written as $\hat{\rho}_s(t)=\left(\hat{\mathbb{I}}+\vec{r}(t)\cdot\vec{\hat{\sigma}}\right)/2$, where $\vec{r}(t)=[x(t),y(t),z(t)]$ is the Bloch vector and $\hat{\mathbb{I}},\vec{\hat{\sigma}}$ are Pauli operators. In this case, coherence and internal energy are given by $C(t)=\sqrt{x(t)^2+y(t)^2}$ and $U(t)=- z(t)$, respectively. Since $\delta \mathbb{Q}_\text{entro} =(U/r)dr$ ~\cite{john}, the characteristic 
adiabatic time $t_c$ can be obtained through the condition
\begin{equation}
\mathbb{Q}_\text{entro}(t_c)=-\int_{t_0=0}^{t_c}\frac{z(t)}{r(t)}\frac{d{r}(t)}{dt} \,dt=0.\\[5pt]
\label{Eq:Qzero}
\end{equation}
For the components of $\Delta\mathcal{E}(t_c)$ in Eq.~(\ref{Erg_var3}), 
we have
\begin{equation}
 \mathbb{W}^*(t_c)= -\Delta z(t_c),
\end{equation}
\begin{equation}
 \Delta U_{\pi}(t_c)= -\Delta r(t_c).\\[5pt]
\end{equation}
Thus, we conclude that the work $\mathbb{W}^*(t_c)$ and the passive energy variation $-\Delta U_{\pi}(t_c)$ are associated with the ergotropic cost of rotation and scale transformation 
(dilation or contraction) of the Bloch vector $\vec{r}$, respectively (see Fig.~\ref{fig:paper2rotarcontraer} for visualization).

\begin{figure}[H] 
\centering
\includegraphics[width=0.49\columnwidth]{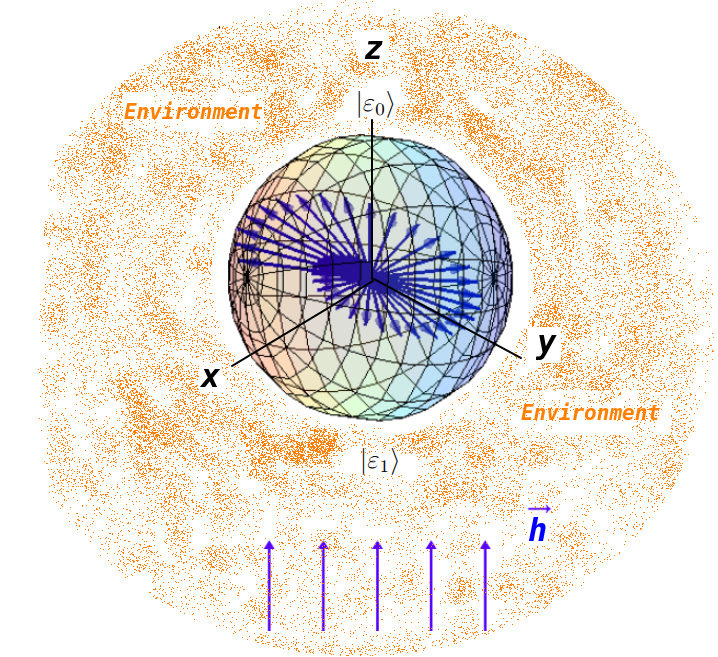}
\caption{(Color online) Visual representation of a qubit interacting with an environment under a magnetic field $\vec{h}$, which acts as the ``internal engine'' defining its Hamiltonian. The Bloch vector $\vec{r}(t)$, shown in blue inside the sphere, contracts due to the energetic cost $\Delta U_\pi$. The rotation around the axis defined by $\vec{h}$ arises from the work induced by the environment through an ergotropic rotational cost $\mathbb{W}^*$. Adapted from Ref.\cite{lang2010}.}
\label{fig:paper2rotarcontraer}
\end{figure}
\noindent According to Eq.~(\ref{Eq:WI}), the incoherent ergotropic variation, $\Delta\mathcal{E}_I$, vanishes for quantum processes with constant or non-positive energy. In these cases, the ergotropy variation is purely coherent, i.e.,
\begin{equation}
 \Delta \mathcal{E}(t)=\Delta \mathcal{E}_C(t) \quad \text{if} \quad    \frac{d{U}(t)}{dt}=0 \,\,\,\,\,\text{or} \,\,\,\,U(t)\leq 0,\,\,\, \forall t
\end{equation}
For non-dissipative quantum processes,
\begin{equation}
\Delta \mathcal{E}_C(t_c)= \mathbb{W}^*(t_c)=\Delta U_\pi(t_c)=0\,\quad\,\left(dU/dt=0\right).
\end{equation}
Thus, the PD process discussed in Sec. \ref{II(A)} is unable to extract the available quantum resource $\mathcal{E}_C$ through environment-induced work. In other words, there is no effective adiabatic process with $\mathbb{W}^*(t_c)\neq0$ for PD maps.
\vspace{0.2cm}\\
\noindent On the other hand, the extraction is possible for dissipative quantum processes such as the AD map described in Sec. \ref{II(B)}. Here, we will illustrate the connection between $\mathbb{W}^*$ and $\Delta \mathcal{E}$ by considering the paradigmatic model of the decay of an excited state of a two-level atom interacting with an environment by spontaneous emission~\cite{Nielsen-Book, Breuer2009} (a Markovian AD process).  The spontaneous emission process is governed by the Markovian master equation
\begin{equation} \label{Eq:mastereq-markov}
        \frac{d}{dt}\hat{\rho}_s(t)=i\left[\hat{\sigma}_z, \hat{\rho}_s(t)\right]+\gamma
    \left[\hat{\sigma}^{-}\hat{\rho}_s(t)\hat{\sigma}^{+}-\frac{1}{2}\left\{\hat{\sigma}^{+}\hat{\sigma}^{-},\hat{\rho}_s(t)\right\}\right],
\end{equation}

\noindent where $\gamma$ is the dissipation rate of spontaneous emission and 
\begin{equation}
\hat{\sigma}^{+}=(\hat{\sigma}_x -i \hat{\sigma}_y)/2,
\end{equation}
\begin{equation}
\hat{\sigma}^{-}=(\hat{\sigma}_x +i \hat{\sigma}_y)/2
\end{equation}
are the raising and lowering operators for a two-level atom. 
Notice that the ground state of $\hat{H}_s$ is the computational state $|0\rangle$ in the north pole of the Bloch sphere, which is the expected long time limit after energy loss in the spontaneous 
emission dynamics (see Fig.~\ref{fig:paper2ADadiabatico} for an intuitive illustration). The solution of Eq. (\ref{Eq:mastereq-markov}) with an arbitrary initial state $\Vec{r}_0=[x_0,y_0,z_0]$ is given by $\Vec{r}(t)=[x(t),y(t),z(t)]$ with
\begin{equation}\label{Eq:rx}
x(t) =e^{-\gamma t/2}\left[x_0\cos{2 t} + y_0\sin{2 t} \right], 
\end{equation}
\vspace{-0.9cm}
\begin{equation} \label{Eq:ry}
y(t) =e^{-\gamma t/2}\left[y_0\cos{2 t} -x_0\sin{2 t} \right],
\end{equation}
\vspace{-0.9cm}
\begin{equation} \label{Eq:rz}
z(t) =e^{-\gamma t}\left[-1+z_0+ e^{\gamma t} \right].
\end{equation}
\begin{figure}[H] 
\centering
\includegraphics[width=0.42\columnwidth]{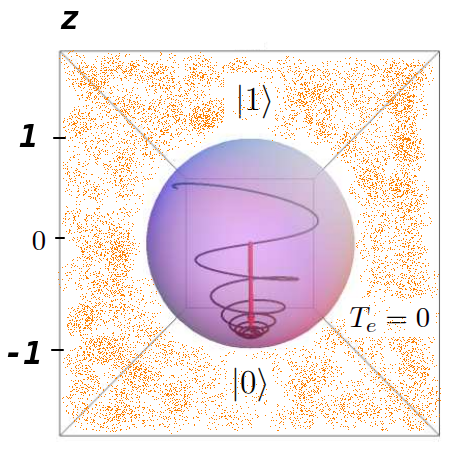}
\caption{(Color online) Visualization of spontaneous emission for a qubit with $\hat{H}_s = -\sigma_z$, interacting with a zero-temperature environment via an amplitude damping map. The blue trajectory inside the sphere represents the evolution of the qubit’s state as given by the solution of the Markovian master equation Eq. \eqref{Eq:mastereq-markov}, illustrating relaxation toward the ground state $\ket{0}$ due to dissipative dynamics. Adapted from Ref.\cite{koch450}.}
\label{fig:paper2ADadiabatico}
\end{figure}
\noindent According to this solution and Eq.~(\ref{Eq:Qzero}), the dimensionless characteristic adiabatic time $\tau_c=\gamma t_c$ is a function only of the initial parameters $r_0=(x_0^2+y_0^2+z_0^2)^{1/2}$ and $\theta_0=\text{arccos}[z_0/r_0]$, where $0\leq r_0\leq 1$ and $0\leq \theta_0 \leq \pi$. We numerically investigate the characteristic adiabatic time for arbitrary initial states $\vec{r}_0=(r_0,\theta_0)$ in Fig. \ref{fig:tc}. Notice that $\tau_c$ is not negligible for initial states close to $\theta_0=\pi/2$  or $r_0=1$, mainly for $\theta_0\geq \pi/2$ (north hemisphere), where $\Delta \mathcal{E}=\Delta \mathcal{E}_C$. 
\begin{figure}[H]
		\centering		\includegraphics[scale=0.44]{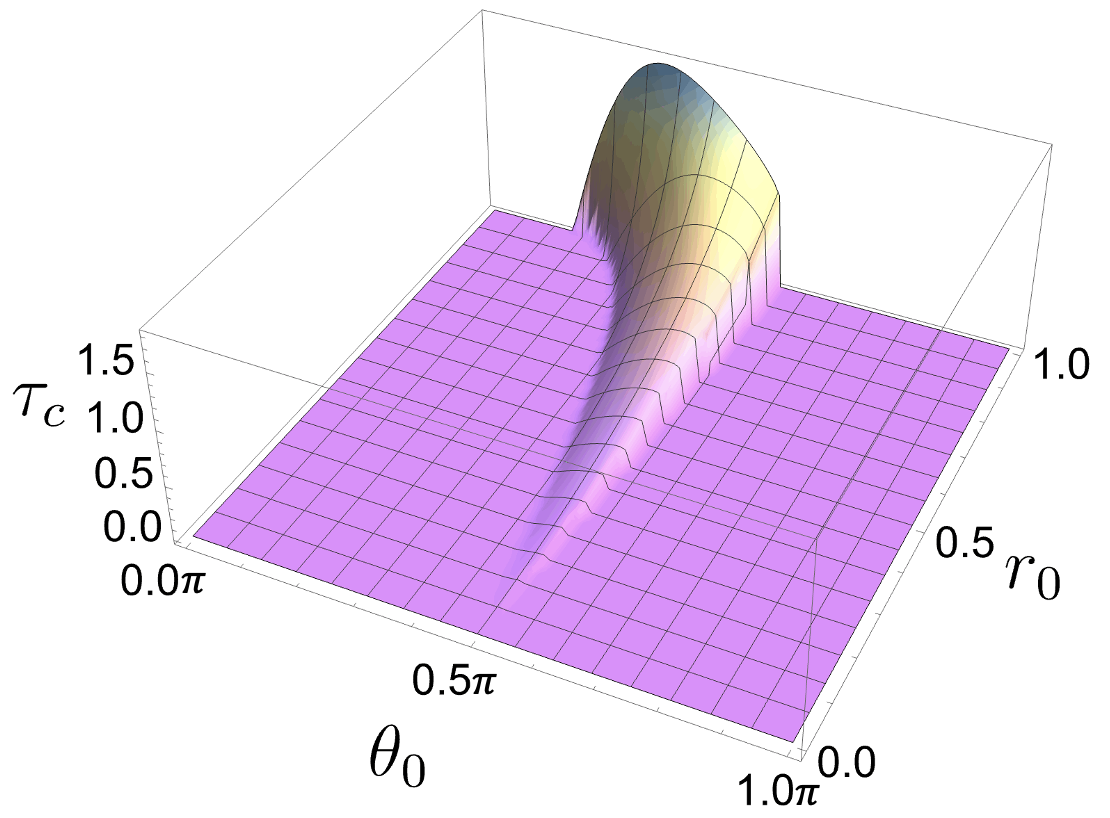}
  \caption{(Color online) Dimensionless characteristic time $\tau_c$ as a function of the initial state $(r_0,\,\theta_0)$ for $0\leq r_0\leq 1$ and $0\leq \theta_0 \leq \pi$.}
  \label{fig:tc}
\end{figure}
\noindent We also study the energy cost $\Delta U_\pi$ as a function of $r_0$ and $\theta_0$. This is exhibited in Fig.~\ref{fig:DeltaUpi}. Notice that $\Delta U_\pi\geq 0$ for all initial states. Therefore, from Eq.~(\ref{Erg_var3}), the work $\mathbb{W}^*(t_c)$ performed on the system by the environment is not greater (in absolute value) than the ergotropy variation $\Delta \mathcal{E}(t_c)$. This result implies that the environment cannot provide more energy to the system than it can be extracted via the definition of extractable work through a variation of ergotropy. \\

\begin{figure}[H]
		\centering
		\includegraphics[scale=0.56]{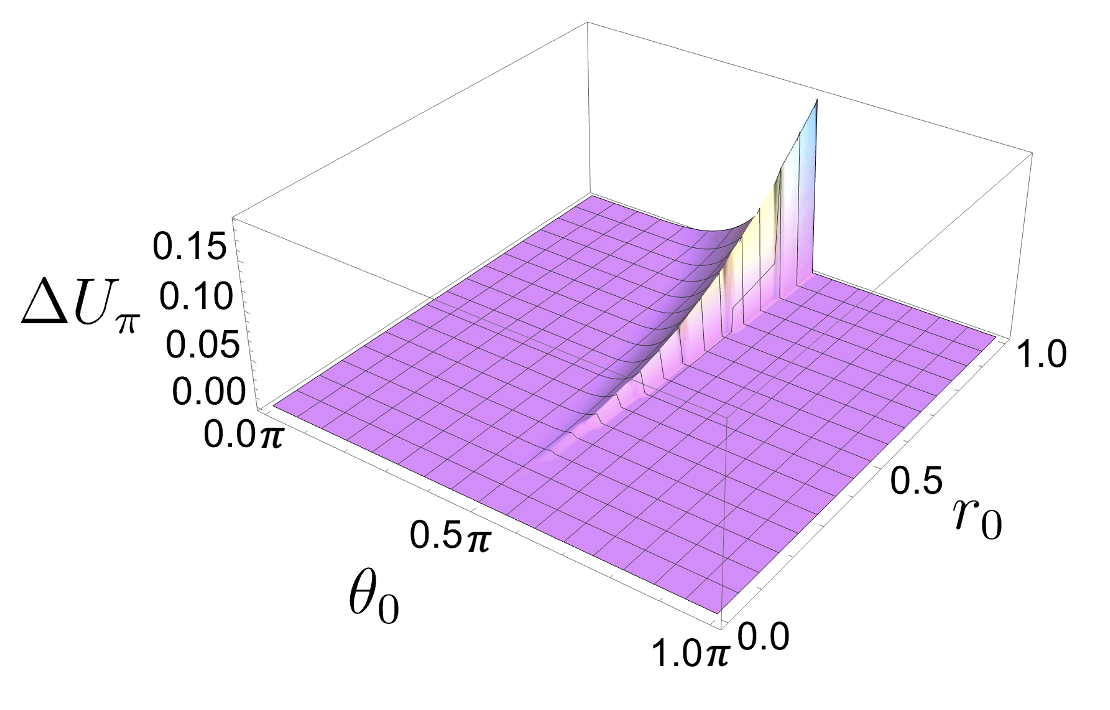}
  \caption{(Color online) Passive energy cost $\Delta U_\pi$ as a function of the initial state $(r_0,\,\theta_0)$ for $0\leq r_0\leq 1$ and $0\leq \theta_0 \leq \pi$.}
  \label{fig:DeltaUpi}
	\end{figure} 
\noindent Finally, we plot $\Delta \mathcal{E}(t_c)$ and $\mathbb{W}^*(t_c)$ for two particular types of families of initial states: a family of mixed states located on the equatorial plane of the Bloch sphere (see Fig.~\ref{fig:fammix}) and a family of pure states located on the upper surface of the Bloch sphere (see Fig.~\ref{fig:pure}), where $t_c\neq 0$. As a by-product, we can also determine the energy cost associated with the remaining contribution $\Delta U_\pi(t_c)$ for the ergotropy variation, as shown in the inset of Fig.~\ref{fig:fammix} and Fig.~\ref{fig:pure}. In theses cases, it is evident that 
\vspace{0.5cm}
\begin{equation}
    \frac{\mathbb{W}^*(t_c)}{\Delta \mathcal{E}(t_c)}\leq 1,
\end{equation}
\vspace{0.5cm}\\
where $\Delta \mathcal{E}(t_c)=\Delta \mathcal{E}_C(t_c)$. Notice also that the singular behavior of the dimensionless characteristic time shown in Fig. \ref{fig:tc} is not manifested in Figs.~\ref{fig:fammix} and~\ref{fig:pure}. This occurs because we have kept the dynamics in the north hemisphere of the Bloch sphere, with $\theta_0 \le \pi/2$.
\vspace{0.2cm}
\noindent We can also consider a more general dynamics, such as the non-Markovian case. This can be analyzed using the physical process as in Eq.~(\ref{ADsd}), from which Eq.~(\ref{Eq:mastereq-markov}) follows as a Markovian limit. For each initial state, instead of only a single $t_{c}$, a set of characteristic times emerges, which is denoted by $\{t_{nc}\}$. By looking at these characteristic times,
we show that, by taking the largest $t_{nc}$ for each state, we obtain results very close to the Markovian case, as shown in Fig.~\ref{fig:fammix} and 
Fig.~\ref{fig:pure}. This behavior holds independently of the ratio $\Gamma/\gamma$. 
\begin{figure}[H]
\vspace{-5cm}
\centering
\includegraphics[scale=0.49]{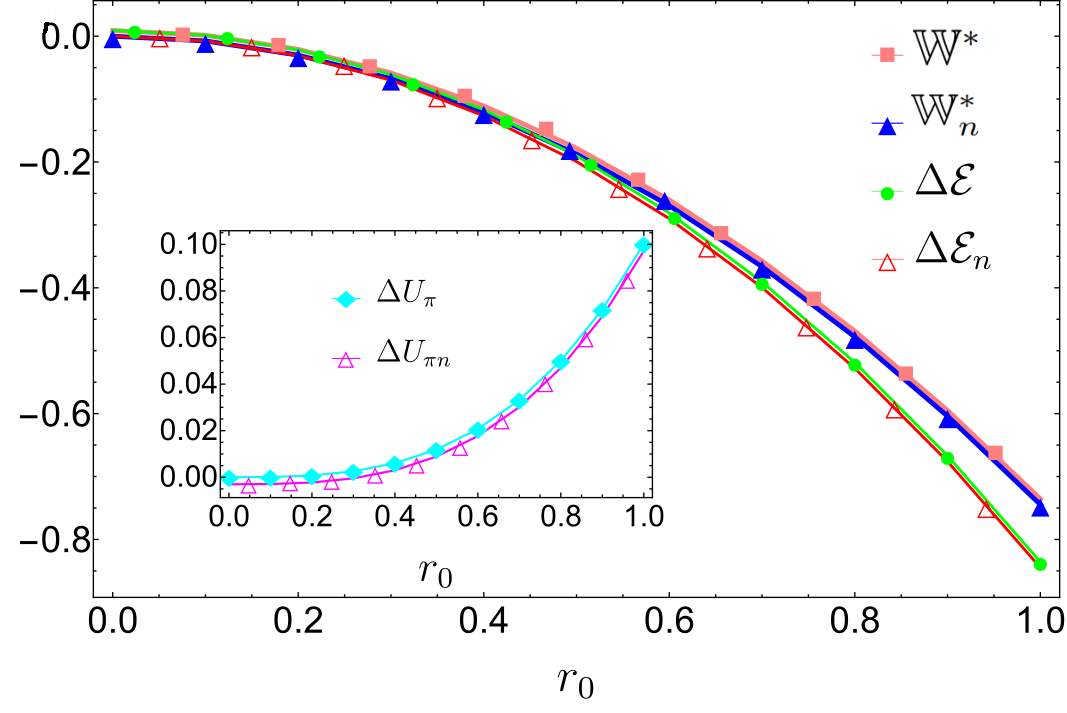}
\caption{(Color online) Environment-induced
work in the Markovian $\mathbb{W}^*$ and non-Markovian $\mathbb{W}^{*}_{n}$ regimes, as well as ergotropy variation in the Markovian $\Delta \mathcal{E}$ and non-Markovian $\Delta \mathcal{E}_{n}$ regimes, as functions of $r_0$ ($0\leq r_0 \leq 1$) for $\theta_0 = \pi/2$. For the non-Markovian dynamics, we adopted $\Gamma=0.01 \gamma$. Inset: Passive energy cost for both Markovian $\Delta U_{\pi}$ and non-Markovian $\Delta U_{\pi n}$ regimes.}
\label{fig:fammix}
\end{figure}

\begin{figure}[H]
\centering
\includegraphics[scale=0.31]{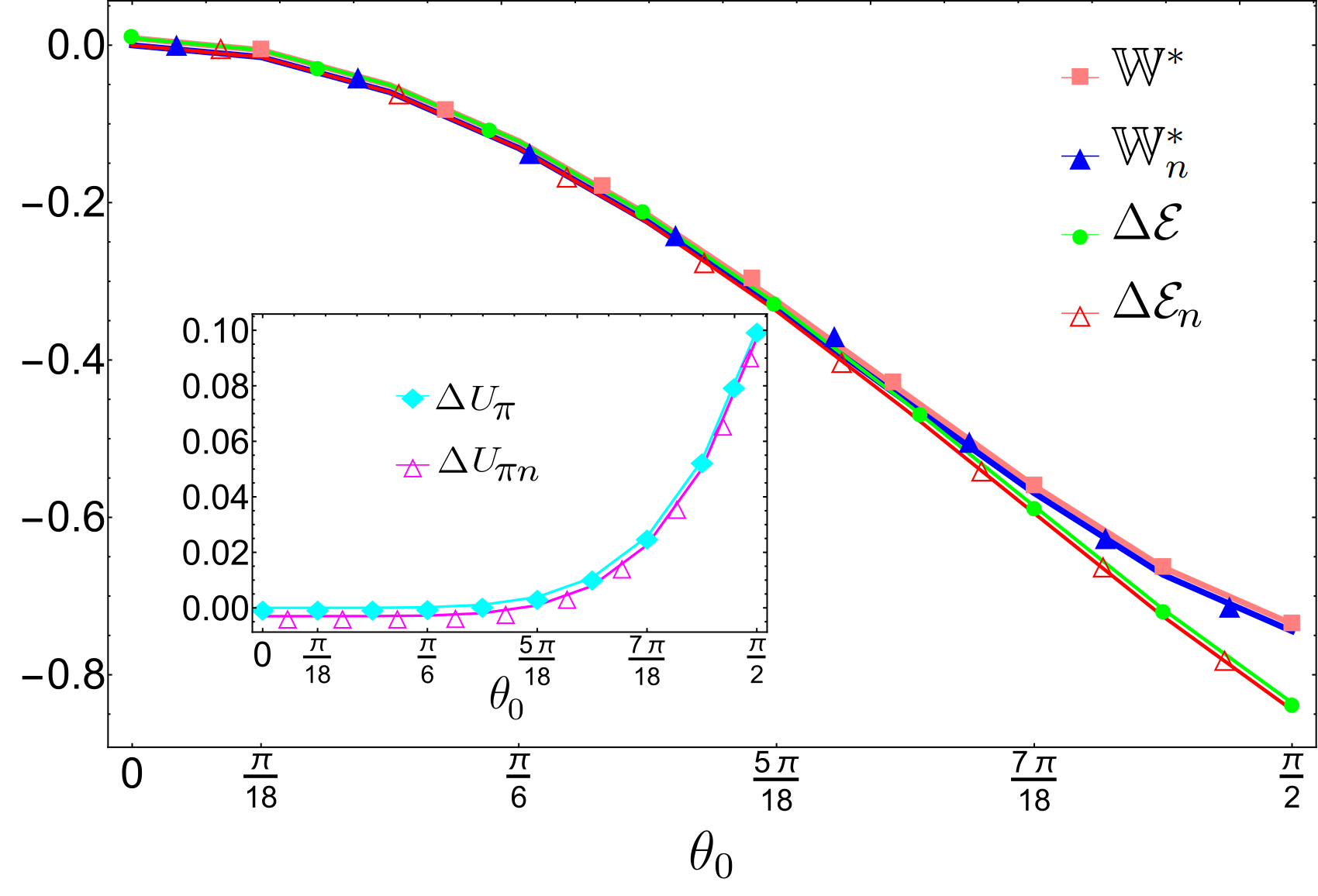}
\caption{(Color online)  Environment-induced
work in the Markovian $\mathbb{W}^*$ and non-Markovian $\mathbb{W}^{*}_{n}$ regimes, as well as ergotropy variation in the Markovian $\Delta \mathcal{E}$ and non-Markovian $\Delta \mathcal{E}_{n}$ regimes, as functions of $\theta_0$ ($0\leq \theta_0 \leq \pi/2$) for $r_0 = 1$. For the non-Markovian dynamics, we adopted $\Gamma=0.01 \gamma$. Inset: Passive energy cost for both Markovian $\Delta U_{\pi}$ and non-Markovian $\Delta U_{\pi n}$ regimes.}
\label{fig:pure}
\end{figure}

\end{chapter}

\begin{chapter}{ERGOTROPY-BASED QUANTUM THERMODYNAMICS}
\label{cap6}
This chapter focuses on the development of a new formulation of quantum thermodynamics based on ergotropy. Unlike other approaches, such as the standard or entropy-based formulations, this proposal allows us to reinterpret the infinitesimal heat flow as the infinitesimal change of the passive state associated with the system’s state. In this way, the concept of heat acquires a structural invariance under passive transformations.
\vspace{0.2cm}\\
\noindent The main goal of this work is to establish a stronger connection between average heat and von Neumann entropy. One of its contributions is the application to the study of non-Markovianity in unital maps. In this context, it is shown that the ergotropy-based average heat can be used as a better measure of non-Markovian. Furthermore, this formulation introduces a semi-definite positive temperature even in out-of-equilibrium regimes. In addition, we presents applications to generalized amplitude damping and dephasing maps, both in dissipative and non-dissipative regimes, under Markovian and non-Markovian evolutions. The results highlight the advantages of the ergotropy-based approach.
\vspace{0.2cm}\\
\noindent This chapter is based on the third work developed during the PhD \cite{john2025}, published in Physical Review A, and presents the earlier version of the work prior to its publication. Its inclusion in this thesis reflects our original contribution to the understanding of irreversible thermodynamic processes from a perspective centered on ergotropy.

\section{First law of quantum thermodynamics}

 First, we consider an arbitrary quantum system described by a density operator whose spectral decomposition is  \begin{equation}
     \hat{\rho}_s(t)=\sum_n r_n(t)\ket{r_n(t)}\bra{r_n(t)}
 \end{equation} and a time-dependent Hamiltonian whose spectral decomposition is 
 \begin{equation}
     \hat{H}_s(t)=\sum_{n} \varepsilon_n(t)\ket{\varepsilon_n(t)}\bra{\varepsilon_n(t)}.
 \end{equation} In both decompositions it is satisfied that $r_n(t)\geq r_{n+1}(t)$ and $\varepsilon_n(t)\leq \varepsilon_{n+1}(t)$. On the other hand, the ergotropy of the state $\hat{\rho}_s(t)$ is defined as the maximum amount of energy that can be extracted from our quantum system via unitary cyclic operation \cite{Allahverdyan:04}, i.e. $\mathcal{E}\big(\hat{\rho}_s(t)\big)=\max_{V(t) \in \mathcal{U}}\big\{\,U(\,\hat{\rho}_s(t)\,)-U\big(\,V(t)\,\hat{\rho}_s(t)\, V^{\dagger}(t)\,\big)\,\big\}$  being $U\big(\hat{\rho}_s(t)\big)=\text{tr}[\hat{\rho}_s(t)\, \hat{H}_s(t)]$ the internal energy of the system, $\mathcal{U}$ the set of all unitary transformations. In this way, the ergotropy can be written as follows 
 \begin{equation}\label{eq6:ergotropy}
\mathcal{E}\big(\,\hat{\rho}_s(t)\,\big)=U\big(\,\hat{\rho}_s(t)\,\big)-U\big(\,\hat{\rho}_{s}^{\,\pi}(t)\,\big),
\end{equation}
where $\hat{\rho}_{s}^{\,\pi} (t)$ is known as passive state or optimized state and it is defined as $\hat{\rho}_{s}^{\,\pi}(t)=\sum_n r_n(t)\ket{\varepsilon_n(t)}\bra{\varepsilon_n(t)}$. From Eq. \eqref{eq6:ergotropy}, we can write
\begin{equation*}
    dU\big(\,\hat{\rho}_s(t)\,\big)=dU\big(\,\hat{\rho}_{s}^{\,\pi}(t)\,\big)+d\mathcal{E}\big(\,\hat{\rho}_s(t)\,\big),
\end{equation*}
where
\begin{equation}
    dU\big(\,\hat{\rho}_{s}^{\,\pi}(t)\,\big)=\text{tr}[d \hat{\rho}_{s}^{\,\pi}(t)\, \hat{H}_s(t)]+\text{tr}[\hat{\rho}_{s}^{\,\pi}(t) \, d\hat{H}_s(t)]
\end{equation}

\noindent Thus, we can establish a first law of quantum thermodynamics based on concept of ergotropy as 
\begin{equation}\label{eq6:1ley}
    dU(t)=\delta \mathbbm{Q}_\text{ergo}(t) + \delta \mathbbm{W}_\text{ergo}(t)
\end{equation}
with
\begin{equation}
    \delta \mathbbm{Q}_\text{ergo}(t)\equiv \text{tr}\left[\delta \hat{\rho}_{s_\pi} (t)\, \hat{H}_s(t)\right]\quad , \quad \delta \mathbbm{W}_\text{ergo}(t) \equiv \text{tr}\left[\hat{\rho}_{s_\pi} (t)\, \delta \hat{H}_s(t)\right]+d\mathcal{E}(t).\label{heatwork}
\end{equation}

\noindent Here, $\delta$ denotes an inexact differential. Note that the work $\delta \mathbbm{W}_\text{ergo}(t)$demands a time-dependent Hamiltonian or ergotropy variation. In particular, ergotropy can be decomposed into incoherent and coherent parts~\cite{Francica:20, Choquehuanca2024,Shi2022}, with the coherent contribution playing a potentially significant role in the performance of quantum thermal devices. 
\vspace{0.2cm}\\
\noindent On the other hand, we can observe that our heat $\delta \mathbbm{Q}_\text{ergo}(t)$ is invariant under unitary transformations and requires a change of the passive state. Consequently, $\delta \mathbbm{Q}_\text{ergo}(t)$ is directly related to the variation  $dS(t)$  of the von Neumann entropy, defined as
\begin{equation}
 S(t)=\text{tr}[\hat{\rho}_s(t)\, \hat{\mathbb{S}}(t)]   ,
\end{equation}
where $\hat{\mathbb{S}}(t)$ is the so-called entropy operator. This is given by
\begin{equation}
    \hat{\mathbb{S}}(t)=-k_B\,\text{ln}\,\hat{\rho}_s(t).
\end{equation}
In particular, the passive contribution of any functional $f\big(\,\hat{\rho}_s(t)\,\big)$ can be defined as 
\begin{equation}
 f_{\pi}\big(\,\hat{\rho}_s(t)\,\big) \equiv f\big(\,\hat{\rho}_{s}^{\,\pi}(t)\,\big)   
\end{equation}
In this context, both heat and entropy variations satisfy
\begin{align}
 \delta \mathbbm{Q}_\text{ergo}(t)&=\delta \mathbbm{Q}_{\text{ergo}_\pi}(t)\\[3.5pt]   
 dS(t)=dS_\pi(t)&=\text{tr}[\delta \hat{\rho}_{s}^{\,\pi}(t)\,\hat{\mathbb{S}}^{\,\pi}(t)]
\end{align}
Note that both $\delta \mathbbm{Q}_\text{ergo}(t)$ and $dS(t)$ necessarily depend on $d \hat{\rho}_{s}^{\,\pi}(t)$, being quantities invariant under the passive transformation $\hat{\rho}_s(t) \to \hat{\rho}_{s}^{\,\pi}(t)$. Indeed, the quantity $\text{tr}[d\hat{\rho}_{s}^{\,\pi}(t) \hat{H}_s(t)]$ has been identified in Ref.~\cite{Niedenzu2018} as the fraction of the exchanged energy between a quantum system and a bath that necessarily causes an entropy change. In the ergotropy-based formulation, this passive energy is interpreted as total heat, in analogy with equilibrium classical thermodynamics.

\section{Temperature}

A closed expression for temperature, denoted here as ${T}_\text{stand}$, in a general nonequilibrium quantum system has been obtained by taking the partial derivative of the von Neumann entropy with respect to the internal energy~\cite{alipour1, Alipour2021}:
\begin{equation}\label{temperature}
\frac{1}{{T}_\text{stand}}\equiv \left(\frac{\partial S}{\partial U}\right)_{\{x_i\}_{i=2}^{d^2-1}}=\frac{\text{Cov}(\hat{H}_s,\hat{\mathbb{S}})}{\text{Cov}(\hat{H}_s,\hat{H}_s)},\\[5pt]
\end{equation}
where $\{x_i\}$ is a set of independent parameters kept constant in the partial derivative and 
\begin{equation}
\text{Cov}(\hat{X},\hat{Y})=\frac{\text{tr}[\hat{X}\hat{Y}]}{d}-\frac{\text{tr}[\hat{X}]\text{tr}[\hat{Y}]}{d^2}\\[5pt]
\end{equation}is the covariance between the operators $\hat{X}$ and $\hat{Y}$ evaluated with respect to the maximally mixed state ${\hat{\mathbb{I}}/d}$, with $d$ denoting the dimension of the associated Hilbert space. We take $x_i = \text{tr}[\hat{\rho}_s\, \hat{O}_i]$, which represent the mean values of traceless
orthonormal observables $\{\hat{O}_i\}_{i=2}^{d^2-1}$. This set can be made complete by adding $\hat{O}_0=\hat{\mathbb{I}}/\sqrt{d}$, which denotes the
normalized identity operator, and $\hat{O}_1=(\hat{H}_s-\text{tr}[\hat{H}_s]\hat{\mathbb{I}}/d)/\sqrt{\text{Cov}(\hat{H}_s,\hat{H}_s)d}$, which is the operator associated with the Hamiltonian.  
\vspace{0.2cm}\\
\noindent As $dU_{\pi}(t) =\delta \mathbbm{Q}_{ergo}(t)$ for a zero-work process and $dS_{\pi}(t)=dS(t)$, a definition of temperature, denoted by $\mathbbm{T}_\text{ergo}$, compatible with the relationship between heat and entropy in the ergotropy-based formulation  is given by the passive part of ${T}_\text{stand}$,
\begin{equation}\label{ptemperature}
\mathbbm{T}_\text{ergo}\Big(\hat{\rho}_s(t)\Big)\equiv\frac{\text{Cov}(\hat{H}_s,\hat{H}_s)}{\text{Cov}(\hat{H}_s,\hat{\mathbb{S}}_{\pi})}.\\[5pt]
\end{equation}
Note that Eq.~(\ref{ptemperature}) is obtained from Eq.~(\ref{temperature}) by taking the passive transformation $\hat{\mathbb{S}} \to \hat{\mathbb{S}}_{\pi}$ over the entropy operator. Since $\hat{\mathbb{S}}_{\pi}$ is a passive state, the  functional $\mathbbm{T}_\text{ergo}\Big(\hat{\rho}_s(t)\Big)$ satisfies the properties (see Ref.~\cite{Alipour2021}):
\begin{itemize}
    \item[(a)] Positivity: $\mathbbm{T}_\text{ergo}\Big(\hat{\rho}_s(t)\Big)\geq 0$, $\forall\,\hat{\rho}_s(t)$.
    \item[(b)] Nullity for pure states: $\mathbbm{T}_\text{ergo}(\ket{\psi(t)}\bra{\psi(t)})=0$.
    \item[(c)] Divergence for maximally mixed states: $\mathbbm{T}_\text{ergo}(\hat{\mathbb{I}}/d)\rightarrow\infty$.
    \item[(d)] Invariance under unitary operations: $\mathbbm{T}_\text{ergo}(\hat{V}(t)\hat{\rho}_s(t_0) \hat{V}^{\dagger}(t))=\mathbbm{T}_\text{ergo}(\hat{\rho}_s(t_0))$.
    \item[(e)] Consistency with the Gibbsian state: $\mathbbm{T}_\text{ergo}(\hat{\rho}_s^{th})=1/k_B\beta$ where $\hat{\rho}_s^{th}=(1/Z)e^{-\beta \hat{H}_s}$ with $Z=\text{tr}[e^{-\beta \hat{H}_s}]$.
\end{itemize}

\section{Second law of quantum thermodynamics}
Let us consider a CPTP evolution dictated by a dynamical map with a fixed point $\hat{\rho}_s^*$, i.e.,
\begin{equation}
\hat{\rho}_s(t)=\Phi_t\big(\hat{\rho}_s(t_0=0)\big) \,\,\,\,\,\text{such that }\,\,\,\,\,\Phi_t(\rho_s^*)=\rho_s^*.
\end{equation}
In this scenario, for an infinitesimal time interval $dt$, the entropy production is given by \cite{Landi2021}
\begin{equation}
 \delta\Sigma(t)= S(\hat{\rho}_s(t)||\hat{\rho}_s^*)-  S(\hat{\rho}_s(t+dt)||\hat{\rho}_s^*) \equiv -\delta S(\hat{\rho}_s(t)||\hat{\rho}_s^*)\geq 0, 
\end{equation}
where  $S(\hat{\rho}_s(t)||\hat{\rho}_s^*)=k_B\text{tr}[\hat{\rho}_s(t)(\ln{\hat{\rho}_s(t)}-\ln{\hat{\rho}}_s^*)]$ provides the relative entropy between $\hat{\rho}_s(t)$ and $\hat{\rho}_s^*$. This inequality is a direct consequence of the contractivity of the relative entropy under CPTP maps, i.e.,
\begin{equation*}
 S\big(\Phi_t(\hat{\rho}_s(0))\,||\,\Phi_t(\hat{\rho}_s^*)\big)\leq S\big(\hat{\rho}_s(0)||\hat{\rho}_s^*\big)   
\end{equation*}
Similarly to the variation of internal energy, entropy production does not depend on the specific definitions of heat and work, exhibiting passive $\delta\Sigma_{\pi}(t)$ and non-passive $\delta\Sigma_{n\pi}(t)$ components:
\begin{equation}
\delta\Sigma(t)=\delta\Sigma_{\pi}(t)+\delta\Sigma_{n\pi}(t)
\end{equation}
where
\begin{equation}
\delta\Sigma_{\pi}(t)=-\delta S(\hat{\rho}_{s}^{\,\pi}(t)||\hat{\rho}_s^*)
\end{equation}
and
\begin{equation}
\delta\Sigma_{n\pi}(t)=\delta [S(\hat{\rho}_{s}^{\,\pi}(t)\,||\,\hat{\rho}_s^*)-S(\hat{\rho}_s(t)||\hat{\rho}_s^*)].
\end{equation}
This decomposition reveals a refined structure underlying the second law of quantum thermodynamics. In the equilibrium regime (i.e., when $\hat{\rho}_s(t)=\hat{\rho}_{s}^{\,\pi}(t)$), note that the passive component completely determines entropy production, while the non-passive term contributes exclusively in nonequilibrium conditions.
\vspace{0.2cm}\\
\noindent We can rewrite the second law in a generalized Clausius form by considering thermal maps, i.e., quantum evolutions for which the fixed point is a Gibbs state at some equilibrium temperature $T_e$:
\begin{equation}
\hat{\rho}_s^*=\hat{\rho}_e=\frac{e^{-\hat{H}_s/k_BT_e}}{\text{tr}[e^{-\hat{H}_s/k_BT_e}]} .  
\end{equation}
In this case, we obtain
\begin{equation}
\delta\Sigma(t)=dS(t)+\frac{\delta \mathbbm{Q}_{e}(t)}{T_e}\geq0,
\end{equation}
where 
\begin{equation}
 \delta \mathbbm{Q}_e(t)\equiv -dU(t)+\text{tr}[\hat{\rho}_e d \hat{H}_s(t)]   
\end{equation}
defines an effective heat associated with the environment. The passive and non-passive parts of the entropy production are given by 
\begin{equation}
 \delta\Sigma_{\pi}(t)=-\delta S(\hat{\rho}_s^{\,\pi}(t)||\rho_e)=dS(t)+\frac{\delta \mathbbm{Q}_{e \pi }(t)}{T_e}  
\end{equation}
and
\begin{equation}
 \delta\Sigma_{n\pi}(t)=-\frac{d\mathcal{E}(t)}{T_e}, 
\end{equation}
where we identify $S(\hat{\rho}_s^{\,\pi}(t)||\hat{\rho}_e)$ as the classical relative entropy~\cite{Sone2021}, with $\delta \mathbbm{Q}_{e \pi }(t)$ representing the passive part of the effective heat. This reduces to $\delta \mathbbm{Q}_{e \pi}(t)=- \delta \mathbbm{Q}_\text{ergo}(t)$ for time-independent Hamiltonians, with $\delta\Sigma_{\pi}(t)$ then leading to the classical Clausius inequality ($\delta\Sigma_{n\pi}(t)=0$ in the equilibrium regime).

\section{Comparison with previous formulations}
In the standard  quantum formulation \cite{alicki}, heat and work are defined by changes in $\hat{\rho}_s(t)$ and $\hat{H}_s(t)$, respectively:
\begin{equation}
 \delta {Q}_\text{stand}(t)\equiv \text{tr}[d\hat{\rho}_s(t)\, \hat{H}_s(t)] \\[5pt]  
\end{equation}
and
\begin{equation}
\delta {W}_\text{stand}(t) \equiv \text{tr}[\hat{\rho}(t)_s \, d\hat{H}_s(t)],\\[5pt]
\end{equation}
such that $dU(t)=\delta {Q}_\text{stand}(t)+\delta {W}_\text{stand}(t)$. Since $\delta {Q}_\text{stand}(t)$ is not invariant under passive transformation, i.e., $\delta {Q}_\text{stand}(t)\neq \delta {Q}_{\text{stand}_\pi}(t)$, the conventional heat is not necessarily connected with $dS$. In the entropy-based framework \cite{alipour1}, an additional work $\mathbb{W}^{*}(t)$ narrows the connection between heat and entropy variation: 
\begin{equation}
 \delta\mathbb{Q}_\text{entro}(t) \equiv\delta {Q}_\text{stand}(t)-\delta\mathbb{W}^{*}(t)   
\end{equation}
and 
\begin{equation}
\delta\mathbb{W}_\text{entro}(t)\equiv \delta{W}_\text{stand}(t)+\delta\mathbb{W}^{*}(t),    
\end{equation}
where 
\begin{equation}
\delta \mathbb{W}^*(t)\equiv \text{tr}[\delta\hat{\rho}_s^{\,ep}(t) \hat{H}_s(t)] 
\end{equation}
with 
\begin{equation}
\delta \hat{\rho}_s^{ep}(t)=\sum_n r_n(t) \,d\,(\ket{r_n(t)}\bra{r_n(t)})   \\[5pt] 
\end{equation}
representing the change in $\hat{\rho}_s(t)$
due to eigenprojector variations. Indeed, we can write $\delta\mathbb{Q}_\text{entro}(t)=\text{tr}[\delta \hat{\rho}_s^{ev}(t)\,\hat{H}_s(t)]$ and $dS(t)=\text{tr}[\delta \hat{\rho}_s^{ev}(t) \,\hat{\mathbb{S}}(t)]$, where the equation $\delta \hat{\rho}_s^{ev}(t)=\sum_n dr_n (t)\ket{r_n(t)}\bra{r_n(t)}$ denotes the change in $\hat{\rho}_s(t)$ due to eigenvalue variations (note that $d\hat{\rho}_s(t)=\delta\hat{\rho}_s^{ev}(t)+\delta\hat{\rho}_s^{ep}(t)$). However, although both $\delta\mathbb{Q}_\text{entro}(t)$ and $dS(t)$ depend on $\delta\hat{\rho}_s^{ev}(t)$, we have $\delta\mathbb{Q}_\text{entro}(t)\neq \delta\mathbb{Q}_{\text{entro}_{\pi}}(t)$ and, consequently, the entropy-based heat is also not completely linked to entropy variation. Since $\mathbb{W}^{*}(t)$ and $\mathcal{E}(t)$ are purely non-passive quantities, we have the following connections among the three formulations:
\begin{equation}
\delta \mathbbm{Q}_\text{ergo}(t)=\delta {Q}_{\text{stand}_\pi}(t)=\delta \mathbb{Q}_{\text{entro}_{\pi}}(t)   
\end{equation}
and 
\begin{equation}
\delta \mathbbm{W}_{\text{ergo}_\pi}(t)=\delta {W}_{\text{stand}_\pi}(t)=\delta \mathbbm{W}_{\text{entro}_\pi}(t). 
\end{equation}
 There is also an operational formulation involving ergotropy, where energy variation is divided into three parts~\cite{Binder2015}: 
 \begin{equation}
 \Delta U(t)=Q_{op}(t)+W_{ad}(t)+\Delta\mathcal{E}(t)   
 \end{equation}
 for a general and finite quantum process $\Big(\hat{\rho}_s(t_0),\hat{H}_s(t_0)\Big)\rightarrow \Big(\hat{\rho}_s(t),\hat{H}_s(t)\Big)$, where 
 \begin{equation}
 Q_{op}(t)\equiv \text{tr}\left[\hat{\rho}_s^{m\pi}(t)\, \hat{H}_s(t_0)\right]-\text{tr}\left[\rho_{s}^{\pi}(t_0)\, \hat{H}_s(t_0)\right]    
 \end{equation}
 and 
 \begin{equation}
 W_{ad}(t)\equiv\text{tr}\left[\rho_{s}^{\pi}(t)\, \hat{H}_s(t)\right]-\text{tr}\left[\hat{\rho}_s^{m\pi}(t) \hat{H}_s(t_0)\right]    
 \end{equation}
 define the operational heat and the adiabatic work, respectively, with
 \begin{equation}
 \hat{\rho}_s^{m\pi}(t)\equiv \sum_{n} r_{n}(t)\ket{\varepsilon_{n}(t_0)}\bra{\varepsilon_{n}(t_0)}    
 \end{equation}
 corresponding to an auxiliary state. Note that 
 \begin{equation}
\Delta U_{\pi}(t)=\mathbbm{Q}_\text{ergo}(t)+\mathbbm{W}_{\text{ergo}_\pi}(t)=Q_{op}(t)+W_{ad}(t).
 \end{equation}
 In particular, we have $\mathbbm{Q}_\text{ergo}(t)=Q_{op}(t)$ and $\mathbbm{W}_{\text{ergo}_\pi}(t)=W_{ad}(t)=0$ when $d \hat{H}_s(t)=0$. Thus, assuming $W_{ad}(t)+\Delta\mathcal{E}(t)$ as the total work, the operational and the ergotropy-based formulations are equivalent for time-independent Hamiltonians. However, the equivalence between the two formulations fails for time-dependent Hamiltonians. For example, we have $Q_{op}(t)=0$,  $\forall \,\hat{H}_s(t)$, such that the initial Hamiltonian $\hat{H}_s(0)=0$.

\section{Qubit thermodynamics}
As already mentioned, the qubit is a fundamental resource in quantum technologies, so we study ergotropy-based quantum thermodynamics for the case of a qubit. First, let us consider an arbitrary qubit system, where $\hat{\rho}_s(t)=(\hat{\mathbb{I}}+\vec{r}(t)\cdot \vec{\hat{\sigma}})/2$ and a general Hamiltonian $\hat{H}_s(t)=-\vec{h}(t)\cdot \vec{\hat{\sigma}}$, with $\vec{r(t)}=(\,x(t),y(t),z(t)\,)$ representing the Bloch vector, $\vec{\hat{\sigma}}=(\sigma_x,\sigma_y,\sigma_z)$ the Pauli operators, and $\vec{h}(t)=\big(\,h_x(t),h_y(t),h_z(t)\,\big)$ the local field.  In this scenario~\cite{vallejo,john,Vallejo2020},  we have for the conventional formulation
\begin{align}
 \delta Q_\text{stand}(t)&=-\vec{h}(t)\cdot d\vec{r}(t) \\[2.5pt]
 \delta W_\text{stand}(t)&=-\vec{r}(t) \cdot d\vec{h}(t)
\end{align}
For the entropy-based formulation 
\begin{align}
\delta \mathbb{Q}_\text{entro}(t)&=\frac{U(t)}{r(t)}\,dr(t)\\[2.5pt]
\delta \mathbb{W}_\text{entro}(t)&=r\, d\left(\frac{U(t)}{r(t)}\right)
\end{align}
In both formulations with
\begin{equation}
 U(t)=-\vec{h}(t)\cdot \vec{r}(t)   
\end{equation}
For the ergotropy-based framework, from Eq.~(\ref{heatwork}), we then obtain
\begin{equation}\label{eq:Qergotropy}
\delta \mathbbm{Q}_\text{ergo}(t)=-h(t) \,dr(t)
\end{equation}
and
\begin{equation}
\delta \mathbbm{W}_\text{ergo}(t)=-r(t)\,dh(t)+d\,\mathcal{E}(t),   
\end{equation}
where 
\begin{equation}
 \mathcal{E}(t)=U(t)+h(t)\,r(t).   
\end{equation}
The expressions for the temperature of a qubit have been obtained for the conventional and entropy-based formulations through the derivative of the von Neumann entropy with respect to energy in a zero work
process \cite{vallejo,Vallejo2020}: 
\begin{equation}
T_\text{stand}(t)=\frac{h^2(t)\, r(t)}{k_B (\vec{h}(t)\cdot\vec{r}(t))\,\text{tanh}^{-1}r(t)}   
\end{equation}
and
\begin{equation}
 \mathbb{T}_\text{entro}(t)=\frac{\vec{h}(t)\cdot\vec{r}(t)}{k_B r(t)\,\text{tanh}^{-1}r(t)},   
\end{equation}
with the conventional temperature ${T}_\text{stand}(t)$ compatible with the temperature defined in Eq.~(\ref{temperature}). From Eq.~(\ref{ptemperature}), we obtain the ergotropy-based temperature
\begin{equation}\label{eq:ergoT}
\mathbbm{T}_\text{ergo}(t)=\frac{h(t)}{k_B \text{tanh}^{-1}r(t)}.
\end{equation}
Note that 
\begin{equation}
\mathbbm{T}_\text{ergo}(t)={T}_{\text{stand}}^{\,\pi}(t)=\mathbb{T}_{\text{entro}}^{\,\pi}(t).    
\end{equation}
Furthermore, since 
\begin{equation}
dS(t)=-k_B\text{tanh}^{-1}r(t)\,dr(t),
\end{equation}
we can write
\begin{equation}
 dS(t)=\frac{\delta \mathbbm{Q}_\text{ergo}(t)}{\mathbbm{T}_\text{ergo}(t)}.   
\end{equation}
Remarkably, the ergotropy-based formulation applied to a qubit system leads to an expression  that resembles the well-known classical relation between entropy and heat.

\section{Applications}
\subsection{Qubit under generalized amplitude damping}

Consider the Markovian quantum master equation for a generalized amplitude damping (GAD) process (we adopt $\hbar=1$), 
\begin{equation}\label{eqpaper3app1}
\frac{d\hat{\rho}_s(t)}{dt}=-i[\hat{H}_s(t),\hat{\rho}_s(t)]\,+\mathcal{D}^-[\hat{\rho}_s(t)]\,+\mathcal{D}^+[\hat{\rho}_s(t)],\\[5pt]   
\end{equation}
which describes a qubit interacting with a bosonic thermal reservoir at finite temperature $T_e$ (see Fig. \ref{fig:paper3app1a} for the corresponding schematic representation)~~\cite{Breuer:Book,Alipour2016,Yang2024,camati2019,cherian2019,zeng2024},
where 
\begin{align}
\mathcal{D}^{-}[\hat{\rho}_s(t)] &= \gamma^{-} \Big( \hat{\sigma}^{-} \hat{\rho}_s(t) \hat{\sigma}^{+} 
- \frac{1}{2} \{\hat{\sigma}^{+}\hat{\sigma}^{-}, \hat{\rho}_s(t)\} \Big), \\[7pt]
\mathcal{D}^{+}[\hat{\rho}_s(t)] &= \gamma^{+} \Big( \hat{\sigma}^{+} \hat{\rho}_s(t) \hat{\sigma}^{-} 
- \frac{1}{2} \{\hat{\sigma}^{-}\hat{\sigma}^{+}, \hat{\rho}_s(t)\} \Big),
\end{align}
describes the emission and absorption process, with $
    \gamma^- =\gamma_0(N+1) \,\,  \text{and} \,\, \gamma^+=\gamma_0 N.$
The ladder operators are
$\hat{\sigma}^+ = \hat{\sigma}_x + i \hat{\sigma}_y\,,\, \hat{\sigma}^- = \hat{\sigma}_x - i \hat{\sigma}_y,$
and $N=1/(e^{\beta_e  \omega_0}-1)$ is the Planck distribution at frequency $\omega_0$, and $\beta_e=1/(k_B T_e)$ is the inverse temperature of the bath.

\begin{figure}[H]
\centering
\includegraphics[width=0.42\columnwidth]{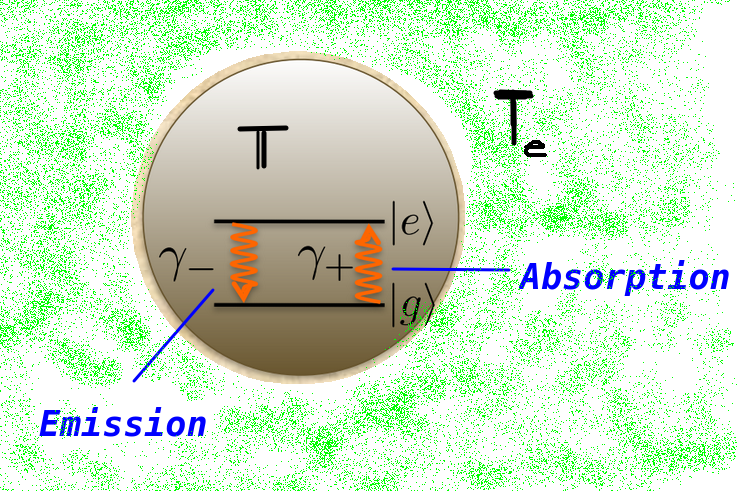}
\caption{(Color online) Qubit with energy levels $|g\rangle$ (ground) and $|e\rangle$ (excited) interacting with a bosonic bath at temperature $T_e$. The decay rates $\gamma^-$ and $\gamma^+$ correspond to emission and absorption processes, respectively. The symbol $\mathbbm{T}$ inside the qubit represents its temperature, and converges to $T_e$ over time. Adapted from Ref.\cite{SantosCBPF2024}.}
\label{fig:paper3app1a}
\end{figure}

\noindent Assuming $\vec{h}=(0,0,-\omega_0/2)$ and $k_{B}T_e=10\omega_0$, we numerically solve the master equation for the qubit initially prepared in the mixed state $\vec{r}_{\pm}(0)=(0.45,\,0.00,\,\pm0.80)$ (upper and lower hemispheres of the Bloch sphere, see Fig. \ref{fig:paper3app1b} for visualization). 

\begin{figure}[H] 
\centering
\includegraphics[width=0.37\columnwidth]{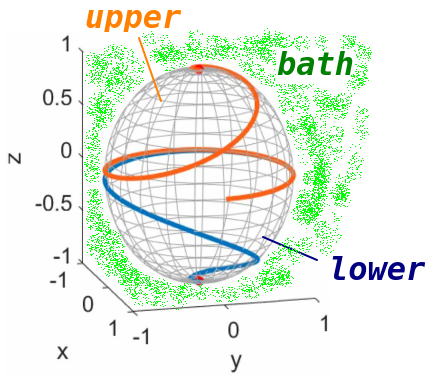}
\caption{(Color online) Qubit interacting with a thermal bath. The orange curve represents the state dynamics within the upper hemisphere of the Bloch sphere, while the blue curve represents the evolution in the lower hemisphere. Adapted from Ref.\cite{Singh2024}. }
\label{fig:paper3app1b}
\end{figure}
\noindent Figure \ref{fig:GAD} illustrates the dynamical behaviors of the three different temperature definitions ($\,{T}_\text{stand}(t)$, $\mathbb{T}_\text{entro}(t)$, and $\mathbbm{T}_\text{ergo}(t)\,$). All temperature quantifiers converge to the environment temperature at long times. Observe that the conventional temperature exhibits nonanalytical behavior at $\omega_0 t \approx0.2195$ for the initial state $\vec{r}_+(0)$. At this time, the state exhibits the Bloch vector component $z=0$. Note also that, as shown before, the ergotropy-based temperature is always positive.
\begin{figure}[H]
\centering
\includegraphics[width=0.56\columnwidth]{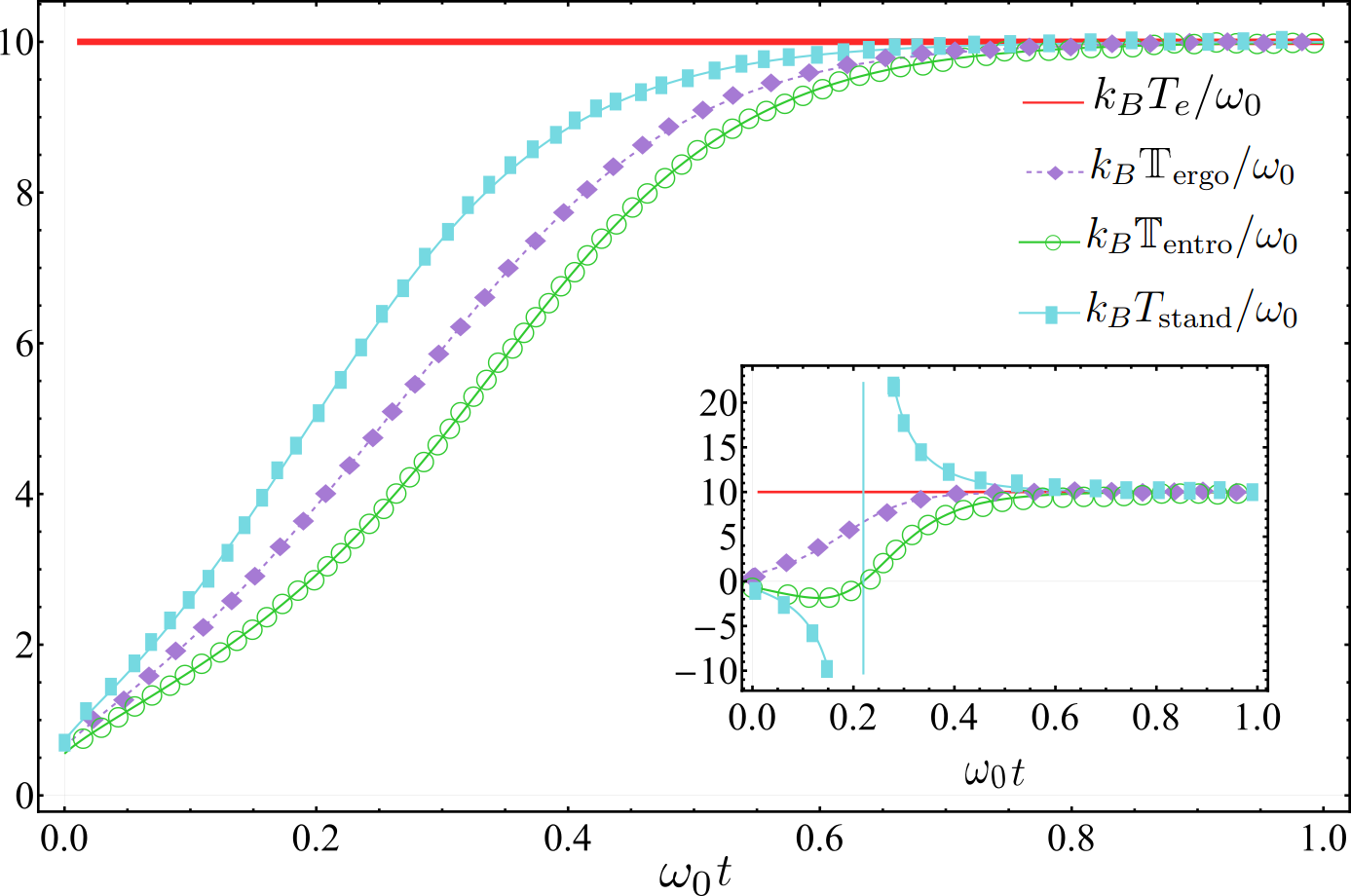}
\caption{(Color online) Dimensionless temperatures $k_BT_e/\omega_0$,  $k_B \mathbbm{T}_\text{ergo}/\omega_0$, $k_B\mathbb{T}_\text{entro}/\omega_0$, and $k_B{T}_\text{stand}/\omega_0$ as functions of the dimensionless time $\omega_0t$  for a qubit under a Markovian GAD process with $\vec{r}_{-}(0)=(0.45,\,0.00,\,-0.80)$. Inset: Same functions for the initial state $\vec{r}_{+}(0)=(0.45,\,0.00,\,0.80)$. We have used $\gamma_0=1$.}
\label{fig:GAD}
\end{figure}

\subsection{Qubit under phase damping}

Let us consider a qubit under a phase damping (PD) dynamics~\cite{john,Yang2024,Marcantoni2017,Garcia2022},
\begin{equation}\label{PD}
\frac{d\hat{\rho}_s(t)}{dt}=-i[\hat{H}_s(t),\hat{\rho}_s(t)]\,+ \gamma(t)(\hat{\sigma}_z \hat{\rho}_s(t)\hat{\sigma}_z - \hat{\rho}_s(t)) ,   
\end{equation}
assuming a time-dependent Hamiltonian~\cite{scopa2019}, with
\begin{equation}
    \vec{h}(t)=\left[0,0,-\frac{\omega_0}{2}(1-\cos{\omega t})\right],
\end{equation}
See Fig. \ref{fig:paper3app2a} for a schematic visualization of this Hamiltonian behavior. Here, we assume a time-independent decoherence
rate $\gamma(t)=\gamma$ (i.e., Markovian regime), and an arbitrary initial state $\vec{r}(0)=(x_0,y_0,z_0)$. In this case, the solution is given by $\vec{r}(t)=[x(t),y(t),z(t)]$\\

\begin{figure}[H] 
\centering
\includegraphics[width=0.6\columnwidth]{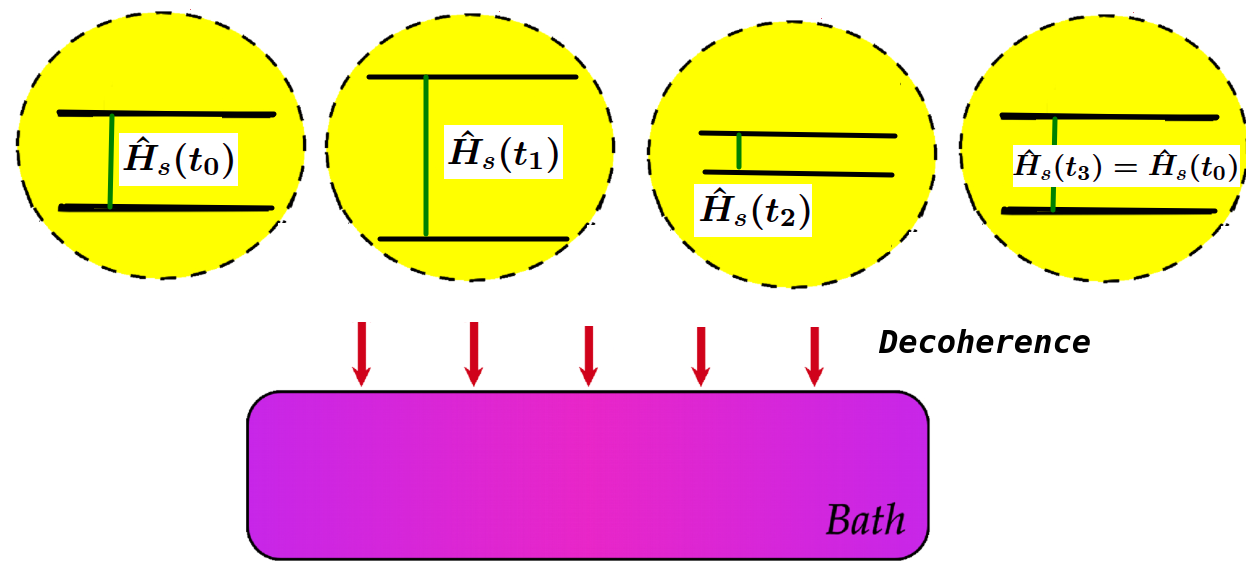}
\caption{(Color online) Qubit in contact with a bath under a Markovian master equation \eqref{PD}. The Hamiltonian $\hat{H}_s=\frac{\omega_0}{2}(1-\cos{\omega t})\,\hat{\sigma}_z$ acts as an external control, defining the qubit energy levels and inducing oscillatory behavior. Adapted from Ref.\cite{QubitRabi2023}. }
\label{fig:paper3app2a}
\end{figure}

\noindent with
\begin{equation}
    x(t)=e^{-2t\,\gamma} \left( x_0 \cos{\alpha} - y_0 \sin{\alpha} \right),
\end{equation}
\begin{equation}
 y(t)=e^{-2t\, \gamma} \left( y_0 \cos{\alpha} + x_0 \sin{\alpha} \right),   
\end{equation}
\begin{equation}
    z(t)=z_0,
\end{equation}
where 
\begin{equation}
\alpha=\frac{\omega_0}{\omega}( \omega t -\sin \omega t).    
\end{equation}

\noindent To demonstrate the stronger connection of the ergotropy-based heat with entropy over other heat formulations, we consider the evolution in the $xy$-plane of the Bloch sphere ($z_0=0$; see Fig.\ref{fig:paper3app2b} for visualization), where only the ergotropy-based heat is non-vanishing and monotonically related to von Neumann entropy, as shown in Fig.~\ref{fig:PDM}.

\begin{figure}[H] 
\centering
\includegraphics[width=0.4\columnwidth]{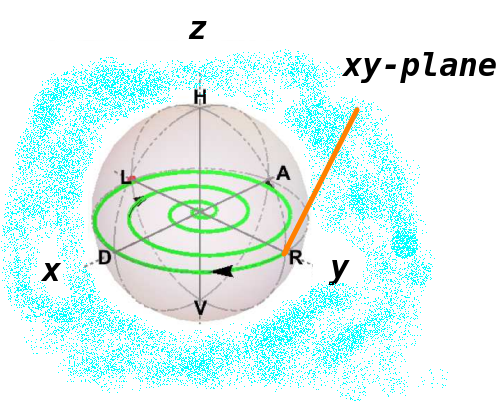}
\caption{(Color online) Qubit under Markovian dephasing, where the state dynamics is represented by the green trajectory confined to the $xy$-plane of the Bloch sphere. The loss of coherence can be interpreted as informational heat. Adapted from Ref.\cite{Urrego2018}}
\label{fig:paper3app2b}
\end{figure}

\noindent In particular, $\mathbbm{Q}_\text{ergo}(t)$ resembles the behavior of  classical heat in a reversible process~\cite{prasad2016,sears1975,Tahir2020,cheng2006}. It can be viewed as an informational heat, quantifying decoherence through 
\begin{equation}
\Delta S(t)=\int \frac{\delta \mathbbm{Q}_\text{ergo}(t)}{\mathbbm{T}_\text{ergo}(t)},   
\end{equation}
which increases as coherence is lost. Meanwhile, $\mathbbm{T}_\text{ergo}(t)$ acts as an internal parameter controlling the degradation of quantum information as a function of time.

\begin{figure}[h!]
\centering
\includegraphics[width=0.56\columnwidth]{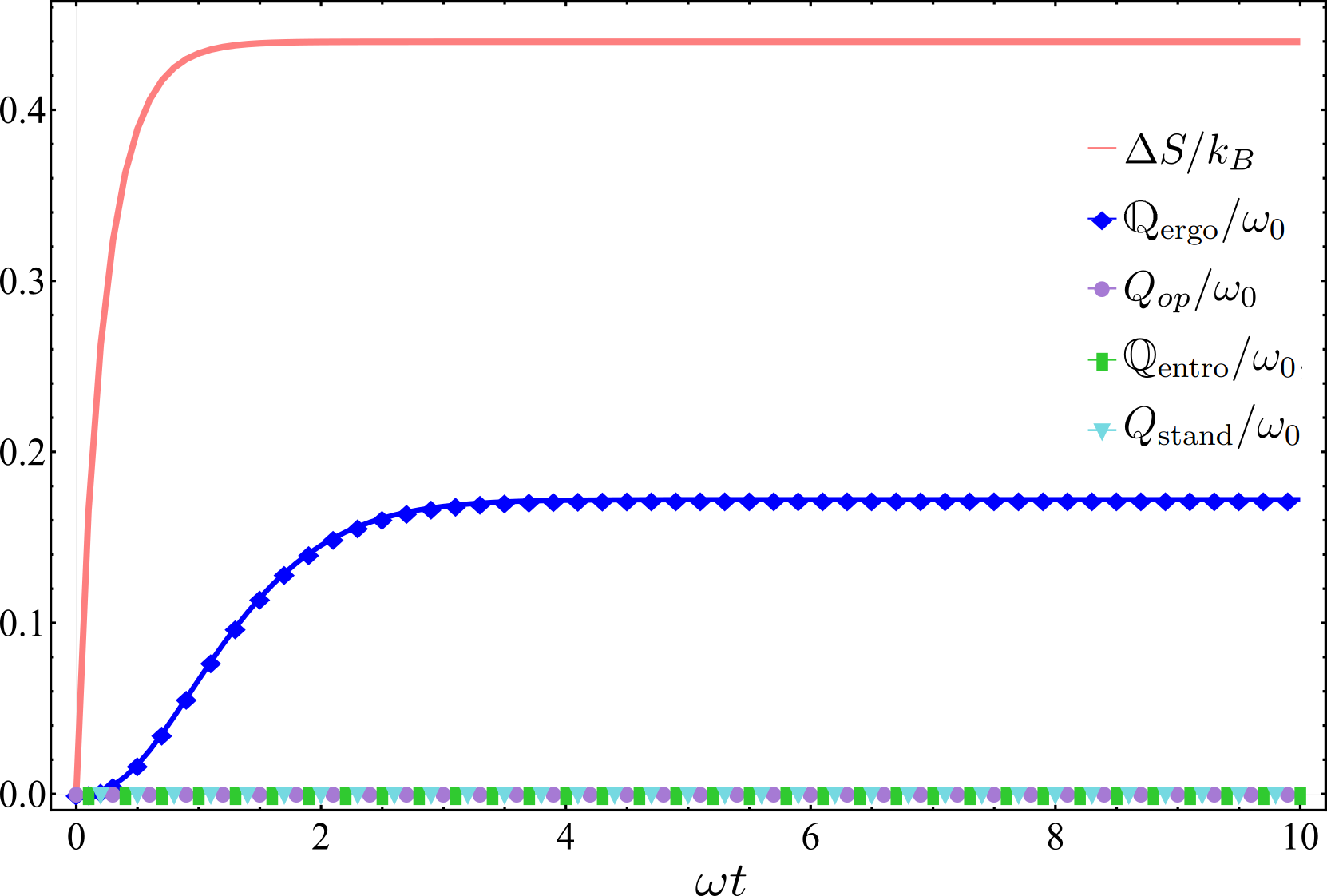}
\caption{(Color online) Dimensionless heats $\mathbbm{Q}_\text{ergo}/\omega_0$, $Q_{op}/\omega_0$, $\mathbb{Q}_\text{entro}/\omega_0$, $Q_\text{stand}/\omega_0$, and entropy variation $\Delta S/k_B$  as functions of the dimensionless time $\omega t$ for a qubit under a Markovian PD process with $\vec{r}(0)=(0.5,\,0.7,\, 0.0)$ and $\gamma=\omega$.}
\label{fig:PDM}
\end{figure}


\subsection{Quantifying non-Markovianity via heat}

We now explore a scenario in which a qubit is coupled to a non-Markovian PD noise governed by Eq.~(\ref{PD})~\cite{Breuer:Book, Haikka2011}. We assume a zero-temperature bosonic reservoir with an Ohmic-like spectral density (see Fig. \ref{fig:paper3app3a} for the corresponding schematic representation), where the time-dependent decoherence rate is 
\begin{equation}
\gamma(t,s)=\left[1+(\omega_c t)^2\right]^{-s/2}\, \Gamma_e[s]\, \sin{[s\,\arctan(\omega_c t)]}    
\end{equation}
with $\Gamma_e[x]$ denoting the Euler gamma function, $\omega_c$ the reservoir cutoff frequency, and $s\geq 0$ the ohmicity parameter. Depending on the value of $s$, the model can exhibit either Markovian or non-Markovian dynamics~\cite{haseli,Haikka2013,Dhar2015,Addis2014,Dakir2025}, with $0\leq s\leq 2$ and  $s > 2$ corresponding to the Markovian and non-Markovian regimes, respectively.
\begin{figure}[H] 
\centering
\includegraphics[width=0.53\columnwidth]{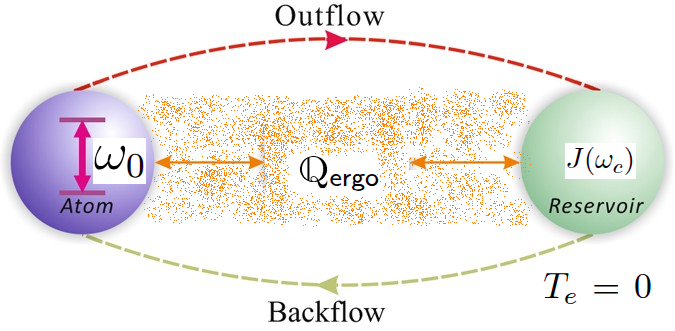}
\caption{(Color online) Qubit of frequency $\omega_0$ interacts with a bosonic reservoir at temperature $T_e = 0$, with spectral density $J(\omega_c)$. When $s \leq 2$, only outflow of information occurs, which corresponds to Markovian behavior. For $s > 2$, both outflow and backflow appear, showing the presence of non-Markovian dynamics. Here, the information flow is interpreted as informational heat, quantified in this example using ergotropy-based heat. Adapted from Ref.\cite{Wang2017}.}
\label{fig:paper3app3a}
\end{figure}
\noindent The solution for a time-independent Hamiltonian, with  
\begin{equation}
 \vec{h}=(0,0,-\omega_0),\\[5pt]   
\end{equation}
and an arbitrary initial state $\vec{r}(0)=(x_0,y_0,z_0)$ is given by
\begin{equation}
 \vec{r}(t)=[x_0\, \Gamma(t),y_0\, \Gamma(t),z_0],\\[5pt]
\end{equation}
where $\Gamma(t)=\exp{(\int^t_0 \gamma(t)dt)}$ \cite{john}. By expressing the initial state in spherical coordinates, $\vec{r}(0)=(r_0\sin{\theta_0}\cos{\phi_0}, \,r_0\sin{\theta_0} \sin{\phi_0},\, r_0\cos{\theta_0})$, we find 
\begin{equation*}
\mathbbm{Q}_\text{ergo}(t, r_0,\theta_0)=-\omega_0\,r_0\,(\sqrt{\cos^2{\theta_0}+\Gamma^2(t)\, \sin^2{\theta_0})}-1)\\[5pt]    
\end{equation*}
for the ergotropy-based heat. Since $\mathbbm{Q}_\text{ergo}$ is monotonically related to the entropy for an arbitrary qubit state, we can use $\mathbbm{Q}_\text{ergo}$ to characterize non-Markovianity for unital maps \cite{haseli,john}.  In this direction, we adopt the generalized approach recently proposed in Ref.~\cite{john} and compute the corresponding non-Markovianity measure 
\begin{equation}
N_{\mathbbm{Q}_\text{ergo}}=\max_{\rho(0)}\, \sum^{}_{i}{\,\big|{\mathbbm{Q}_\text{ergo}(a_i,r_0,\theta_0)-\mathbbm{Q}_\text{ergo}(b_i,r_0,\theta_0)\big|}}, \\[5pt]   
\end{equation}
being $[a_i,b_i]$ the set of time intervals for which  $\gamma(t,s)\leq0$, with $i=1,2,3,...$ labeling the number of such intervals for a given range of $s$. Specifically, for $s\leq 2$, there are no intervals where $\gamma$ becomes negative. For $2<s\leq6$, a single negative interval emerges $(i=1)$, and for $s>6$, the number of such intervals increases $(i=2,3,...)$. Here, we focus on the case of a single interval (see Refs.~\cite{Haikka2013,Addis2014}).
\vspace{0.2cm}\\
\noindent We find that the optimal initial state in the definition of $N_{\mathbbm{Q}_\text{ergo}}$ is a pure state $\vec{r}_\text{max}(0)=(\sin\phi_0,\cos\phi_0,0)$,  with $0\leq\phi_0\leq2\pi$ (see Fig.~\ref{fig:paper3app3b}).

\begin{figure}[H] 
\centering
\includegraphics[width=0.35\columnwidth]{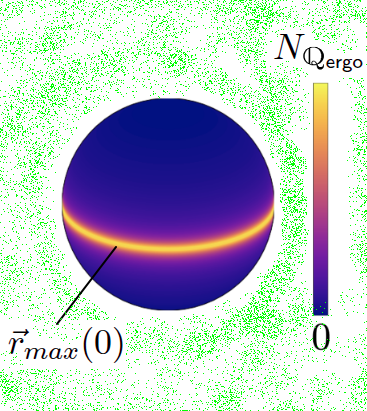}
\caption{(Color online) Schematic representation of the non-Markovian  quantifier $N_{\mathbbm{Q}_{\rm ergo}}$ for a qubit under dephasing with a time-dependent decoherence rate. The color distribution on the surface of the sphere shows the dependence of $N_{\mathbbm{Q}_{\rm ergo}}$ on the initial state, where the curve $\vec{r}_{\max}(0)$ identifies the  pure states that maximizes this measure. Adapted from Ref.\cite{Settimo2022}}
\label{fig:paper3app3b}
\end{figure}

\noindent Consequently,
\begin{equation}\label{nq}
N_{\mathbbm{Q}_\text{ergo}}=\omega_0\sum^{}_{i}{\big|\Gamma(a_i)-\Gamma(b_i)\big|}.\\[5pt]  
\end{equation}
This expression captures the memory effects as the widely employed trace distance-based measure, given by
\begin{equation*}
N_D=\max_{\vec{r}_1(0), \vec{r}_2(0)} \sum_i |D(\vec{r}_1(a_i),\vec{r}_2(a_i))- D(\vec{r}_1(b_i),\vec{r}_2(b_i))|,\\[5pt]    
\end{equation*}
where $D(\vec{r}_1,\vec{r}_2)=|\vec{r_1}-\vec{r_2}|/2$ defines the trace distance between the states $\vec{r}_2$ and $\vec{r}_1$. In fact, the optimal pair of states in $N_{D}$ corresponds to $\vec{r}_2(0)=-\vec{r}_1(0)=(1,0,0)$ \cite{wissmann2012,breuer2016,lorenzo2013}, which leads to
\begin{equation}
 N_{\mathbbm{Q}_\text{ergo}}=\omega_0N_D.   
\end{equation}
Notably, the expression in Eq.~\eqref{nq} also holds for the coherence-based measure \cite{john}. In Fig.~\ref{fig:PDNM}, we compare $N_{\mathbbm{Q}_\text{ergo}}/\omega_0$ with the heat-based alternatives $N_{Q_\text{stand}}/\omega_0$ and $N_{\mathbb{Q}_\text{entro}}/\omega_0$. Note that $N_{Q_\text{stand}}/\omega_0=0$ for all $s$, which means that $N_{Q_\text{stand}}$ is indeed unsuitable as a non-Markovianity measure.
\begin{figure}[H]. 
\centering
\includegraphics[width=0.63\columnwidth]{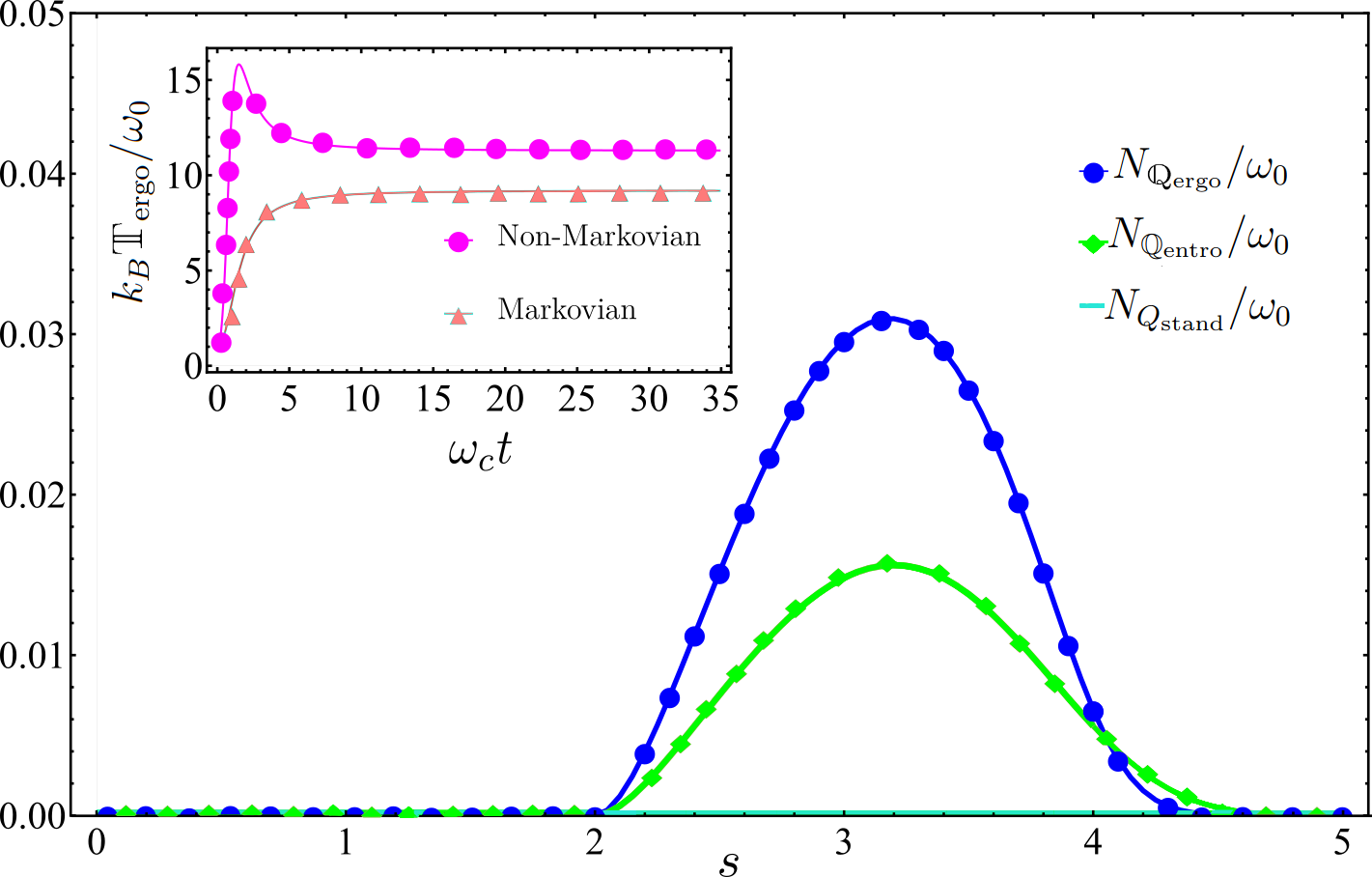}
\caption{(Color online) Dimensionless heat-based non-Markovianity quantifiers $N_{\mathbbm{Q}_\text{ergo}}/\omega_0$, $N_{\mathbb{Q}_\text{entro}}/\omega_0$, and $N_{Q_\text{stand}}/\omega_0$ as functions of the ohmicity parameter $s$. Inset: Dimensionless ergotropy-based temperature $\mathbbm{T}_\text{ergo}$ as a function of $\omega_c t$ for $|\vec{r}(0)|=0.8$.}
\label{fig:PDNM}
\end{figure}

\noindent Concerning $N_{\mathbb{Q}_\text{entro}}/\omega_0$, the quantification provided is, on average, numerically less pronounced than $N_{\mathbbm{Q}_\text{ergo}}/\omega_0$. Specifically, the maximum non-Markovianity is observed  at $s=3.2$, yielding $N_{\mathbb{Q}_\text{entro}}/\omega_0 \approx0.0156$ and $N_{\mathbbm{Q}_\text{ergo}}/\omega_0 \approx 0.0309$.  
Moreover, $N_{\mathbb{Q}_\text{entro}}$ is a more restricted measure, only valid for an energy sign-preserving unital map~\cite{john}. In the inset of Fig.~\ref{fig:PDNM} one can see how the definition of the ergotropy-based temperature, Eq.~\eqref{eq:ergoT}, is applicable as a witness of non-Markovianity through its non-monotonic behavior over time. The figure highlights both Markovian $(s=2)$ and non-Markovian $(s=3.2)$ regimes. For this analysis, the system is initialized in a mixed state $|\vec{r}(0)|=0.8$. Notice that $\mathbbm{T}_\text{ergo}(t)$ successfully captures non-Markovianity within the $xy$-plane of the Bloch sphere, unlike the alternative temperature definitions ${T}_\text{stand}(t)$ and $\mathbb{T}_\text{entro}(t)$, which fail to detect such behavior.

\end{chapter}
\begin{chapter}{Conclusions and Future Perspectives}
\label{conclusion}

Based on the studies and results obtained in the different chapters of this thesis, I extracted relevant conclusions that summarize the achievements reached during my PhD.
\vspace{0.2cm}\\
\noindent The first part, consisting of Chapters 2 and 3, establishes the essential theoretical framework for the development of the works developed in the thesis. First, a foundational theoretical framework for the study of quantum systems was studied, starting with the formal description of quantum states through the density operator, and deepening into its fundamental properties. We analyzed the dynamics of closed systems governed by the von Neumann equation, and later studied two-level systems and their representation on the Bloch sphere. This representation facilitates the understanding of both unitary and non-unitary evolution, helps to understand phenomena of quantum coherence and decoherence, and is particularly relevant because the applications presented in the published works are based on single-qubit systems.
\vspace{0.2cm}\\
\noindent When entering the regime of open systems, we concluded that understanding open-system dynamics requires introducing the concepts of quantum maps and master equations, both Markovian and non-Markovian, which describe rigorously the interaction with thermal environments. To discuss physical processes, the Kraus operator formalism helps to model interactions in a generalized way, leading to the definition of CPTP maps that guarantee the physical consistency of the system’s evolution. In the case of non-Markovian dynamics, the topic is not trivial due to the difficulties in finding the corresponding master equations, and if they exist, how to solve them. For now, we focused on how to identify non-Markovian dynamics using indicators such as the trace-distance-based measure.
\vspace{0.2cm}\\
\noindent With this formal theoretical foundation, we applied it to the thermodynamics of microscopic and finite systems, thus entering the field of quantum thermodynamics. We first reviewed classical thermodynamics and then extended its definitions to the quantum regime, noting that classical quantities like entropy, heat, work, and irreversibility must satisfy the characteristic properties of quantum information and open quantum systems. Thermal states and thermal maps were analyzed to better understand the environment. We then studied the von Neumann entropy and the quantum relative entropy, concluding that these definitions are essential for quantifying irreversibility. It was also necessary to study a concept not commonly found at the microscopic level called ergotropy, which can be roughly defined as the maximum work extractable from a system via cyclic operations. We observed how quantum resources, such as quantum coherence, can influence thermodynamic laws. With this theoretical foundation, we moved to the second part of this thesis, which consists of the works developed during the PhD and presented in Chapters 4, 5, and 6.
\vspace{0.2cm}\\
\noindent The first work, presented in Chapter 4, proposed a characterization of non-Markovianity based on the entropic formulation of quantum thermodynamics, where heat flow becomes a universal witness of quantum memory. We showed that for unital dynamics, the monotonicity of heat flow constitutes a necessary condition for the Markovian regime, and its violation is interpreted as a clear signal of non-Markovianity. This framework allowed linking thermodynamic quantities, such as internal energy and effective work, with measures of quantum coherence, extending the scope of traditional memory quantifiers. In particular, the analysis of dissipative and dephasing channels revealed the compatibility of our indicators with previous entropic measures, showing even proportionality in specific cases. We applied this approach to a bosonic dephasing channel, where heat flow successfully detected non-Markovianity in sub-Ohmic and super-Ohmic spectra. In this way, we established a solid bridge between quantum information theory and thermodynamics, reinforcing the usefulness of the entropic formalism as a fundamental tool to describe memory phenomena in open quantum systems.
\vspace{0.2cm}\\
\noindent The second work, presented in Chapter 5, extended the analysis to the role of ergotropy in the thermodynamics of open systems, identifying how interaction with the environment can generate singular phenomena in its dynamics, such as freezing in non-dissipative channels and sudden death in dissipative processes, in clear analogy with effects already observed in quantum correlations. A crucial result was establishing an explicit relation between the change in ergotropy and the work induced by the environment within the entropic formulation of the first law. We demonstrated that this induced work is upper-bounded by the change in ergotropy, while the excess energy corresponds to the cost of transitioning between passive states. These findings not only consolidate ergotropy as a fundamental resource for work extraction but also place it as a natural framework to analyze irreversibility and non-Markovianity. Additionally, the results provide relevant perspectives for optimizing quantum heat engines and quantum batteries, since controlling the induced work is key for the performance of these devices in contact with realistic environments.
\vspace{0.2cm}\\
Finnally, in the third work presented in Chapter 6, we have introduced an ergotropy-based formulation of quantum thermodynamics. This framework allowed for a direct relationship between heat and von Neumann entropy, which is stronger than the connection found in previous approaches. This is based on the invariance of the ergotropy-based heat under passive state transformations. In this scenario, average heat can then be used as a general measure of non-Markovianity for unital maps. Moreover, by defining the out-of-equilibrium temperature in an ergotropy-based formulation, we can achieve a positive-semidefinite temperature. This means that, even working in an nonequilibrium context, temperature will follow a simple description  typical of equilibrium states, with non-negative values throughout the dynamics. Concerning work, we have obtained that average work is provided by ergotropy variation and an extra passive work contribution, which can be induced by either a controllable parameter of the system or even by the interaction with the environment. As future perspectives, we intend to look at the efficiency of quantum thermal machines in the ergotropy-based scenario, both by theoretical and experimental proposals. In addition, we can also explore the ergotropy-based framework in terms of a resource theory for energy extraction in open quantum systems.
\end{chapter}



\newpage
\phantomsection






\addcontentsline{toc}{chapter}{Bibliography}

\begin{thebibliography}{99}




\bibitem{alexia2022}
A. Auffèves, 
``Quantum technologies need a quantum energy initiative,'' 
\textit{PRX Quantum}, vol.~3, p.~020101, Jun.~2022.

\bibitem{metzler2023}
F. Metzler, J. I. Sandoval, and N. Galvanetto, 
``The emergence of quantum energy science,'' 
\textit{Journal of Physics: Energy}, vol.~5, no.~4, p.~041001, Oct.~2023.

\bibitem{quantuminsider2023}
The Quantum Insider, 
``Quantum computing startups challenging industry leaders,'' 
2023. [Online]. Available: \url{https://thequantuminsider.com}. Accessed: Apr.~3, 2025.

\bibitem{QEI2025}
Second Quantum Energy Initiative (QEI) Workshop, 
Grenoble, France, Jan.~2025. [Online]. Available: \url{https://qei2025.sciencesconf.org/}.

\bibitem{callen1985}
H. B. Callen, 
\textit{Thermodynamics and an Introduction to Thermostatistics}, 
2nd ed., John Wiley \& Sons, New York, 1985.

\bibitem{prasad2016}
R. Prasad, 
\textit{Classical and Quantum Thermal Physics}, 
Cambridge University Press, Cambridge, 2016.

\bibitem{gold20166}
J. Goold, M. Huber, A. Riera, L. del Rio, and P. Skrzypczyk, 
``The role of quantum information in thermodynamics — a topical review,'' 
\textit{Journal of Physics A: Mathematical and Theoretical}, vol.~49, no.~14, p.~143001, Feb.~2016.

\bibitem{binder2019}
F. Binder, L. A. Correa, C. Gogolin, J. Anders, and G. Adesso (Eds.), 
\textit{Thermodynamics in the Quantum Regime: Fundamental Aspects and New Directions}, 
Fundamental Theories of Physics, vol.~195, Springer, 2019. 
doi: \href{https://doi.org/10.1007/978-3-319-99046-0}{10.1007/978-3-319-99046-0}.

\bibitem{Esposito2011} M. Esposito and C. Van den Broeck, ``Second law and Landauer principle far from equilibrium,'' \textit{EPL (Europhysics Letters)} \textbf{95}, 40004 (2011).

\bibitem{Parrondo2015} J. M. R. Parrondo, J. M. Horowitz, and T. Sagawa, ``Thermodynamics of information,'' \textit{Nature Physics} \textbf{11}, 131–139 (2015).

\bibitem{deffner2019}
S. Deffner and S. Campbell, 
\textit{Quantum Thermodynamics: An Introduction to the Thermodynamics of Quantum Information}, 
IOP Concise Physics, Morgan \& Claypool Publishers, 2019.

\bibitem{gemmer} J. Gemmer, M. Michel, and G. Mahler, 
\textit{Quantum Thermodynamics: Emergence of Thermodynamic Behavior within Composite Quantum Systems}, 
Springer, Berlin, 2009.

\bibitem{peter} P. Talkner, E. Lutz, and P. Hanggi, \textit{Fluctuation theorems: Work is not an observable}, Phys. Rev. E \textbf{75}, 050102 (2007).

\bibitem{alicki} R. Alicki, \textit{The quantum open system as a model of the heat engine}, J.Phys A: Math. Gen. \textbf{12},  L103 (1979).

\bibitem{Thomas2018} G. Thomas, N. Siddharth, S. Banerjee, and S. Ghosh, ``Thermodynamics of non-Markovian reservoirs and heat engines,'' \textit{Physical Review E} \textbf{97}, 062108 (2018).

\bibitem{Huang2022} W.-M. Huang and W.-M. Zhang, ``Strong-coupling quantum thermodynamics far from equilibrium: Non-Markovian transient quantum heat and work,'' \textit{Physical Review A} \textbf{106}, 032607 (2022).

\bibitem{Tiwari2024} D. Tiwari, B. Bose, and S. Banerjee, ``Strong coupling non-Markovian quantum thermodynamics of a finite-bath system,'' arXiv preprint arXiv:2404.15915 (2024).

\bibitem{Gorini1976} V. Gorini, A. Kossakowski, and E. C. G. Sudarshan, \textit{ Completely positive dynamical semigroups of N‐level systems }, J. Math. Phys. (N.Y.) \textbf{17}, 821 (1976), [\href{https://doi.org/10.1063/1.522979}{J. Math. Phys. 17, 821–825 }].

\bibitem{Breuer:Book} H.P. Breuer and F. Petruccione, {\it{The Theory of Open Quantum Systems}}, Oxford University Press, USA, 2002.

\bibitem{rivas2014}
Á. Rivas, S. F. Huelga, and M. B. Plenio,  
\textit{Quantum non-Markovianity: characterization, quantification and detection},  
Rep. Prog. Phys. \textbf{77}, 094001 (2014).  

\bibitem{Landi2021} G.~T.~Landi and M.~Paternostro, \textit{Irreversible entropy production: From classical to quantum}, Rev. Mod. Phys. \textbf{93}, 035008 (2021).

\bibitem{binder} F. C. Binder, S. Vinjanampathy, K. Modi and J. Goold, \textit{Quantacell: powerful charging of quantum batteries}.

\bibitem{alipour1} S. Alipour, A. Chenu, A. T. Rezakhani, and A. del Campo, \textit{Entropy-based formulation of thermodynamics in arbitrary quantum evolution} Phys. Rev. A {\bf 105}, L040201 (2022).

\bibitem{alipour2} S. Alipour, A. T. Rezakhani, A. Chenu, A. del Campo, and T. Ala-Nissila, \textit{Shortcuts to Adiabaticity in Driven Open Quantum Systems: Balanced Gain and Loss and Non-Markovian Evolution}, Quantum \textbf{4}, 336 (2020).

\bibitem{Ahmadi2020} B. Ahmadi, S. Salimi, and A. S. Khorashadl, \textit{On the contribution of work or heat in exchanged energy via interaction in open bipartite quantum systems},  Sci Rep \textbf{13}, 160 (2023).

\bibitem{streltsov} A. Streltsov, G. Adesso, and M. B. Plenio, \textit{Colloquium: Quantum coherence as a resource}, Rev. Mod. Phys. \textbf{89}, 041003 (2017).

\bibitem{cramer} T. Baumgratz, M. Cramer, and  M. B. Plenio, \textit{Quantifying Coherence}, Phys. Rev. Lett. \textbf{113}, 140401 (2014).

\bibitem{Binder:18} F. Binder, L. A. Correa, G. Gogolin, J. Anders, and G. Adesso, {\it{Thermodynamics in the Quantum Regime. Fundamental Theories of Physics}}, 
(Berlin:Springer, 2018).

\bibitem{bernardo} B. L. Bernardo, \textit{Unraveling the role of coherence in the first law of quantum thermodynamics}, Phys. Rev. E \textbf{102}, 062152 (2020).

\bibitem{titas} T. Chanda and S. Bhattacharya, \textit{Delineating incoherent non-Markovian dynamics using quantum coherence}, Ann. Phys. (N. Y.) \textbf{366}, 1 (2016).

\bibitem{passo} M. H. M. Passos, P. C. Obando, W. F. Balthazar, F. M. Paula, J. A. O. Huguenin, and  M. S. Sarandy, \textit{Non-Markovianity through quantum coherence in an all-optical setup}, Opt. Lett. \textbf{44}, 2478 (2019).

\bibitem{haseli} S. Haseli, S. Salimi, and A. S. Khorashad, \textit{A measure of non-Markovianity for unital quantum dynamical maps}, Quantum Inf. Process. \textbf{14}, 3581(2015).

\bibitem{kosloff} R. Kosloff, \textit{Quantum Thermodynamics}, Entropy \textbf{15},  2100 (2013).

\bibitem{Allahverdyan:04} A. E. Allahverdyan, R. Balian, and Th. M. Nieuwenhuizen, \textit{Maximal work extraction from finite quantum systems},  Europhys. Lett. {\bf 67}, 565 (2004).

\bibitem{Rossnagel:16} J. Rossnagel, S. T. Dawkins, K. N. Tolazzi, O. Abah, E. Lutz, F. Schmidt-Kaler, and K. Singer, \textit{A single-atom heat engine}, 
Science {\bf 352}, 325 (2016). 

\bibitem{Maslennikov:19} G. Maslennikov, S. Ding, R. Hablützel, J. Gan, A. Roulet, S. Nimmricher, J. Dai, V.Scarani, and D. Matsukevic {\it{et al.}}, \textit{Quantum absorption refrigerator with trapped ions},  Nat. Commun. {\bf 10}, 202 (2019).

\bibitem{Joshi:22} J. Joshi and T. S. Mahesh, \textit{Experimental investigation of a quantum battery using star-topology NMR spin systems},  Phys. Rev. A {\bf 106}, 042601 (2022).

\bibitem{Zhu:23} G. Zhu, Y. Chen, Y. Hasegawa, and P. Xue, \textit{Charging Quantum Batteries via Indefinite Causal Order: Theory and Experiment},  Phys. Rev. Lett. {\bf 131}, 240401 (2023).

\bibitem{vallejo} A. Vallejo, A. Romanelli, and R. Donangelo, \textit{Qubit thermodynamics far from equilibrium: Two perspectives about the nature of heat and work in the quantum regime}, Phys. Rev. E \textbf{103}, 042105 (2021).

\bibitem{maximilian2007}
M. Schlosshauer, \textit{Decoherence and the Quantum-to-Classical Transition}, Springer, Berlin, 2007.

\bibitem{lidar2019} D. A. Lidar, \textit{Lecture Notes on the Theory of Open Quantum Systems}, arXiv:1902.00967 [quant-ph] (2019).

\bibitem{nakahara2008} Mikio Nakahara and Tetsuo Ohmi,
\textit{Quantum Computing: From Linear Algebra to Physical Realizations},2nd edition,CRC Press, Boca Raton, FL, 2008.

\bibitem{manenti} R. Manenti and M. Motta, 
\textit{Quantum Information Science}, 
Oxford University Press, Oxford, (2023).

\bibitem{Band2012}
Yehuda B. Band and Yshai Avishai, 
\textit{Quantum Mechanics with Applications to Nanotechnology and Information Science}, 
Elsevier Inc., (2012).

\bibitem{Nielsen-Book} M. Nielsen and I. Chuang, {\it Quantum Computation and Quantum Information}, Cambridge University Press, Cambridge, (2000).

\bibitem{Preskill2020}
J. Preskill,
\textit{Quantum Computation -- Lecture 4: Channels},
Course slides (Caltech, 2020).
Available at: \url{https://www.preskill.caltech.edu/ph219/Ph-CS-219A-Slides-2020/Ph-CS-219A-Lecture-4-Channels.pdf}

\bibitem{manzano2018} G. Manzano, \textit{Thermodynamics and Synchronization in Open Quantum Systems}, Springer Theses (Springer, Cham, 2018).

\bibitem{DEOLIVEIRA2020}
A.~G. de Oliveira, R.~M. Gomes, V.~C.~C. Brasil, N. Rubiano da Silva, L.~C. Céleri, and P.~H. Souto Ribeiro, 
\textit{Full thermalization of a photonic qubit}, 
Phys. Lett. A \textbf{384}, 126933 (2020).

\bibitem{Srikanth2008} 
R. Srikanth and S. Banerjee, 
\textit{Squeezed generalized amplitude damping channel}, 
Phys. Rev. A \textbf{77}, 012318 (2008).

\bibitem{krantz2019}
P. Krantz, M. Kjaergaard, F. Yan, T. P. Orlando, S. Gustavsson, and W. D. Oliver,
``A Quantum Engineer’s Guide to Superconducting Qubits,''
\textit{Appl. Phys. Rev.}, vol. 6, no. 2, p. 021318, (2019).

\bibitem{Rivas2010} A. Rivas, S. F. Huelga, and M. B. Plenio, \textit{Entanglement and Non-Markovianity of Quantum Evolutions} Phys. Rev. Lett. \textbf{105}, 050403 (2010).

\bibitem{rivas2011}
Á. Rivas and S. F. Huelga,  
\textit{Open Quantum Systems: An Introduction},  
Springer Briefs in Physics (Springer, Heidelberg, 2012).

\bibitem{benenti2004}
G. Benenti, G. Casati, and G. Strini, \textit{Principles of Quantum Computation and Information}, Vol. 1: Basic Concepts, World Scientific, Singapore, (2004).

\bibitem{Lindblad1976}
G. Lindblad, ``On the generators of quantum dynamical semigroups,'' \textit{Commun. Math. Phys.}, vol. 48, no. 2, pp. 119--130, Jun. 1976. 

\bibitem{Breuer2009} H.-P. Breuer, E.-M. Laine, and J. Piilo, \textit{Measure for the Degree of Non-Markovian Behavior of Quantum Processes in Open Systems}, Phys. Rev. Lett. \textbf{103}, 210401 (2009).

\bibitem{Kaya2025}
Ü. Kaya, 
\textit{The dynamics of transitions can be modeled by rate equations}, 
Physical Chemistry (LibreTexts). 
Available at:\url{https://chem.libretexts.org/}

\bibitem{Arie2008}
H. Suchowski, D. Oron, Y. Silberberg, and A. Arie,
\textit{Geometrical representation of sum frequency generation and adiabatic frequency conversion},
Phys. Rev. A \textbf{78}, 063821 (2008).

\bibitem{Renaud2011}
N. Renaud, M. A. Ratner, and V. Mujica,
\textit{A stochastic surrogate Hamiltonian approach of coherent and incoherent exciton transport in the Fenna-Matthews-Olson complex},
J. Chem. Phys. \textbf{135}, 075102 (2011).

\bibitem{Gong2015}
E. Gong, W. Zhou, and S. Schirmer,
\textit{Model discrimination for dephasing two-level systems},
Phys. Lett. A \textbf{379}, 272--278 (2015).

\bibitem{Greiner1997}
W. Greiner, L. Neise, and H. St\"ocker, 
\textit{Thermodynamics and Statistical Mechanics}, 
Classical Theoretical Physics, Springer-Verlag, New York, (1997).

\bibitem{Picatoste2024}
I. A. Picatoste,
\textit{Quantum Thermodynamics of Strongly Coupled Open Systems: Analysis of the Otto Cycle}, Master's thesis, Albert-Ludwigs-Universit{\"a}t Freiburg, (2024).

\bibitem{rivas2019}
Á. Rivas,
\textit{Quantum Thermodynamics in the Refined Weak Coupling Limit},
\textit{Entropy}, vol. 21, no. 8, p. 725, (2019).


\bibitem{Balian} R. Balian, \textit{From Microphysics to Macrophysics: Methods and Applications of Statistical Physics, Volume I}, 2nd ed. (Springer, Berlin, 2007).

\bibitem{vershynina} A. Vershynina,
\textit{Measure of Genuine Coherence Based on Quasi-Relative Entropy},
Quantum Information Processing \textbf{21}, 184 (2022).

\bibitem{salimi2020}
H. Dolatkhah, S. Salimi, A. S. Khorashad, and S. Haseli,
\textit{The entropy production for thermal operations},
Scientific Reports \textbf{10}, 9757 (2020).

\bibitem{nakahara2012}
M. Nakahara and S. Tanaka,
\textit{Lectures on Quantum Computing, Thermodynamics and Statistical Physics},
World Scientific, (2012).

\bibitem{sagawa2022}
T. Sagawa,
``\textit{Entropy, Divergence, and Majorization in Classical and Quantum Thermodynamics},'' Springer Singapore, (2022).

\bibitem{LandiNotes2018}
G. T. Landi,
\textit{Lecture Notes on Quantum Information and Quantum Noise},
Instituto de Física, Universidade de São Paulo (2018).
Available at: \url{http://www.fmt.if.usp.br/~gtlandi/lecture-notes-2.pdf}.

\bibitem{Bertlmann:Book}
R. A. Bertlmann and N. Friis,
{\it Modern Quantum Theory: From Quantum Mechanics to Entanglement and Quantum Information},
Oxford University Press, Oxford, 2023.

\bibitem{Hayashi:Book}
M. Hayashi, S. Ishizaka, A. Kawachi, G. Kimura and T. Ogawa,
{\it Introduction to Quantum Information Science},
Springer, Berlin, 2006.

\bibitem{Pusz78} W. Pusz and S. L. Woronowicz, \textit{Passive states and KMS states for general quantum systems}, Commun. Math. Phys. {\bf 58}, 273 (1978).

\bibitem{Lenard78} A. Lenard, \textit{Thermodynamical proof of the Gibbs formula for elementary quantum systems}, J. Stat. Phys. {\bf 19}, 575 (1978).

\bibitem{Francica:20} G. Francica, F. C. Binder, G. Guarnieri, M.T. Mitchison,  J. Goold,  and F. Plastina, \textit{Quantum Coherence and Ergotropy}, Phys. Rev. Lett. {\bf 125}, 180603 (2020).

\bibitem{xiang2025}
L. Li, S. Zhao, Y.–H. Shi, K. Xu, H. Fan, D. Zheng, and Z. Xiang,
\textit{Experimental extraction of coherent ergotropy and its energetic cost in a superconducting qubit},
Phys. Rev. Research **(accepted)**, (2025).  

\bibitem{Lorenzo2013}
S.~Lorenzo, F.~Plastina, M.~Paternostro,
``Geometrical characterization of non-Markovianity'',
\textit{Phys. Rev. A} \textbf{88}, 020102 (2013).

\bibitem{john} J.~M.~Z.~Choquehuanca, F.~M.~de Paula and M.~S.~Sarandy, \textit{Non-Markovianity through entropy-based quantum thermodynamics}, Phys. Rev. A \textbf{107}, 012220 (2023), [\href{https://link.aps.org/doi/10.1103/PhysRevA.107.012220}{Phys. Rev. A 107, 012220}].

\bibitem{Chruscinski2014} D. Chruscinski, and S. Maniscalco, \textit{Degree of Non-Markovianity of Quantum Evolution}, Phys. Rev. Lett. \textbf{112}, 120404 (2014).

\bibitem{Paula2016} F. M. Paula, P. C. Obando, and M. S. Sarandy, \textit{Non-Markovianity through multipartite correlation measures}, Phys. Rev. A \textbf{93}, 042337 (2016).

\bibitem{Streltsov2018} A. Streltsov, H. Kampermann, S. Wolk, M. Gessner, and D. Bru\ss, \textit{Maximal coherence and the resource theory of purity},  New J. Phys. \textbf{20}, 053058 (2018).

\bibitem{Lombardo2014}
F. C. Lombardo and P. I. Villar,
\textit{Corrections to the Berry phase in a solid-state qubit due to low-frequency noise},
Phys. Rev. A \textbf{89}, 012110 (2014).

\bibitem{Alipour2019arxiv}
S. Alipour, A. T. Rezakhani, A. Chenu, A. del Campo, and T. Ala-Nissila,
\textit{Unambiguous Formulation for Heat and Work in Arbitrary Quantum Evolution},
arXiv:1912.01808v1 (2019).

\bibitem{lang2010}
A. De, A. Lang, D. Zhou, and R. Joynt,
\textit{Suppression of decoherence and disentanglement by the exchange interaction},
Phys. Rev. A \textbf{83}, 042331 (2011).

\bibitem{koch450}
C. Koch,
\textit{PHYS 450 – Theory of Open Quantum Systems, Lecture Notes (Part 2)},
Northwestern University, available at: \url{https://sites.northwestern.edu/koch/teaching/}.

\bibitem{Haikka2013} P. Haikka, T. H. Johnson, and S. Maniscalco, \textit{Non-Markovianity of local dephasing channels and time-invariant discord}, Phys. Rev. A \textbf{87}, 010103 (2013).

\bibitem{Shekhar2015} H. S. Dhar, M. N. Bera, and G. Adesso, \textit{Measuring Quantum Coherence with Entanglement}, Phys. Rev. A \textbf{91}, 032115 (2015).

\bibitem{hadadi2022}
S. Haddadi, M.-L. Hu, Y. Khedif, H. Dolatkhah, M. R. Pourkarimi, and M. Daoud,
\textit{Measurement uncertainty and dense coding in a two-qubit system: Combined effects of bosonic reservoir and dipole–dipole interaction},
Results in Physics \textbf{32}, 105041 (2022)

\bibitem{Choquehuanca2024} J.~M.~Z.~Choquehuanca, P.~A.~C.~Obando, F.~M.~de~Paula, and M.~S.~Sarandy, \textit{Qubit dynamics of ergotropy and environment-induced work}, Phys. Rev. A \textbf{109}, 052219 (2024), \href{https://doi.org/10.1103/PhysRevA.109.052219}{doi:10.1103/PhysRevA.109.052219}.

\bibitem{Utagi:20} S. Utagi, R. Srikanth, and  S. Banerjee, \textit{Temporal self-similarity of quantum dynamical maps as a concept of memorylessness},   Sci Rep {\bf 10}, 15049 (2020). 

\bibitem{TYu:10} T. Yu, and J.H. Eberly, \textit{Entanglement evolution in a non-Markovian environment},  
Optics Communications {\bf 283}, 676 (2010).

\bibitem{Bellomo:07}  B. Bellomo, R. Lo Franco,  and  G. Compagno, \textit{Non-Markovian Effects on the Dynamics of Entanglement}, Phys. Rev. Lett. {\bf 99}, 160502, (2007).

\bibitem{john2025}
J. M. Z. Choquehuanca, P. A. C. Obando, M. S. Sarandy, and F. M. de Paula,  
\textit{Ergotropy-based quantum thermodynamics},  
Phys. Rev. A \textbf{112}, 052220 (2025).  

\bibitem{Shi2022} H.-L.~Shi, S.~Ding, Q.-K.~Wan, X.-H.~Wang, and W.-L.~Yang, \textit{Entanglement, Coherence, and Extractable Work in Quantum Batteries}, Phys. Rev. Lett. \textbf{129}, 130602 (2022).

\bibitem{Niedenzu2018} W.~Niedenzu, V.~Mukherjee, A.~Ghosh, A.~G.~Kofman, and G.~Kurizki, \textit{Quantum engine efficiency bound beyond the second law of thermodynamics}, Nat. Commun. \textbf{9}, 165 (2018).

\bibitem{Alipour2021} S.~Alipour, M.~Afsary, F.~Bakhshinezhad, M.~Ramezani, F.~Benatti, T.~Ala-Nissila, and A.~T.~Rezakhani, \textit{Temperature in nonequilibrium quantum systems}, (2021), arXiv:2105.11915.

\bibitem{Sone2021} A.~Sone and S.~Deffner, \textit{Quantum and Classical Ergotropy from Relative Entropies}, Entropy \textbf{23}, 1107 (2021).

\bibitem{Binder2015} F.~Binder, S.~Vinjanampathy, K.~Modi, and J.~Goold, \textit{Quantum thermodynamics of general quantum processes}, Phys. Rev. E \textbf{91}, 032119 (2015).

\bibitem{Vallejo2020} A.~Vallejo, A.~Romanelli, and R.~Donangelo, \textit{Out-of-equilibrium quantum thermodynamics in the Bloch sphere: Temperature and internal entropy production}, Phys. Rev. E \textbf{101}, 042132 (2020).

\bibitem{Alipour2016} S.~Alipour, F.~Benatti, F.~Bakhshinezhad, M.~Afsary, S.~Marcantoni, and A.~T.~Rezakhani, \textit{Correlations in quantum thermodynamics: Heat, work, and entropy production}, Sci. Rep. \textbf{6}, 35568 (2016).

\bibitem{Yang2024} X.~Yang, X.~Long, R.~Liu, K.~Tang, Y.~Zhai, X.~Nie, T.~Xin, J.~Li, and D.~Lu, \textit{Control-enhanced non-Markovian quantum metrology}, Commun. Phys. \textbf{7}, 282 (2024).

\bibitem{camati2019} P.~A.~Camati, J.~F.~G.~Santos, and R.~M.~Serra, \textit{Coherence effects in the performance of the quantum Otto heat engine}, Phys. Rev. A \textbf{99}, 062103 (2019).

\bibitem{cherian2019} P.~Cherian, S.~Chakraborty, and S.~Ghosh, \textit{On thermalization of two-level quantum systems}, Europhys. Lett. \textbf{126}, 40003 (2019).

\bibitem{zeng2024} J.~Zeng, G.-H.~Xu, W.~Huang, and Y.~Yao, \textit{Classical-quantum correspondence in noise-based dissipative systems}, Phys. Rev. A \textbf{110}, 062219 (2024).

\bibitem{SantosCBPF2024}
M. F. Santos,
\textit{Quantum Jumps in Quantum Optics},
Tutorial presented at the III Brazilian Workshop on Quantum Information and Thermodynamics, CBPF, Rio de Janeiro, Brazil (2024).
Available at: \url{https://sites.google.com/view/iiibrazilianquantinfothermo}

\bibitem{Singh2024} 
G. Singh, R. K. Singh and S. Banerjee,  
\textit{Embedding of a non-Hermitian Hamiltonian to emulate the von Neumann measurement scheme},  
J. Phys. A: Math. Theor. \textbf{57}, 35301 (2024).


\bibitem{Marcantoni2017} S.~Marcantoni, \textit{Thermodynamics of a qubit undergoing dephasing}, J. Phys.: Conf. Ser. \textbf{841}, 012019 (2017).

\bibitem{Garcia2022} J.~J.~García Ripoll, \textit{Quantum Information and Quantum Optics with Superconducting Circuits}, Cambridge University Press (2022).

\bibitem{scopa2019} S.~Scopa, G.~T.~Landi, A.~Hammoumi, and D.~Karevski, \textit{Exact solution of time-dependent Lindblad equations with closed algebras}, Phys. Rev. A \textbf{99}, 022105 (2019).

\bibitem{sears1975} F.~W.~Sears and G.~L.~Salinger, \textit{Thermodynamics, Kinetic Theory, and Statistical Thermodynamics}, Addison-Wesley (1975).

\bibitem{Tahir2020} R.~Tahir-Kheli, \textit{General and Statistical Thermodynamics}, Springer (2020).

\bibitem{cheng2006} Y.~Cheng, \textit{Macroscopic and Statistical Thermodynamics}, World Scientific (2006).

\bibitem{QubitRabi2023}
A. Landa, M. Cattaneo, D. Malz, and A. Nazir,
\textit{Qubit-oscillator relationships in the open quantum Rabi model: the role of dissipation},
Eur. Phys. J. Plus \textbf{138}, 135 (2023).

\bibitem{Urrego2018}
D. F. Urrego, J.-R. Álvarez, O. Calderón-Losada, J. Svozilík, M. Nuñez, and A. Valencia,
\textit{Implementation and characterization of a controllable dephasing channel based on coupling polarization and spatial degrees of freedom of light},
Opt. Express \textbf{26}, 11940–11949 (2018).


\bibitem{Haikka2011} P.~Haikka, S.~McEndoo, G.~De Chiara, G.~M.~Palma, and S.~Maniscalco, \textit{Quantifying, characterizing, and controlling information flow in ultracold atomic gases}, Phys. Rev. A \textbf{84}, 031602 (2011).

\bibitem{Dhar2015} H.~S.~Dhar, M.~N.~Bera, and G.~Adesso, \textit{Characterizing non-Markovianity via quantum interferometric power}, Phys. Rev. A \textbf{91}, 032115 (2015).

\bibitem{Addis2014} C.~Addis, B.~Bylicka, D.~Chruściński, and S.~Maniscalco, \textit{Comparative study of non-Markovianity measures in exactly solvable one- and two-qubit models}, Phys. Rev. A \textbf{90}, 052103 (2014).

\bibitem{Dakir2025} Y.~Dakir, A.~Slaoui, L.~Btissam Drissi, and R.~Ahl Laamara, \textit{Quantifying non-Markovianity via local quantum Fisher information}, (2024), arXiv:2409.10163.

\bibitem{Wang2017}
D. Wang, A.-J. Huang, R. D. Hoehn, F. Ming, W.-Y. Sun, J.-D. Shi, L. Ye, and S. Kais,
\textit{Entropic uncertainty relations for Markovian and non-Markovian processes under a structured bosonic reservoir},
Sci. Rep. \textbf{7}, 1066 (2017).

\bibitem{wissmann2012}
S. Wibmann, A. Karlsson, E.-M. Laine, J. Piilo, and H.-P. Breuer, 
\textit{Optimal state pairs for non-Markovian quantum dynamics}, 
Phys. Rev. A \textbf{86}, 062108 (2012).

\bibitem{Settimo2022}
F. Settimo, H.-P. Breuer, and B. Vacchini,
\textit{Entropic and trace-distance-based measures of non-Markovianity},
Phys. Rev. A \textbf{106}, 042212 (2022).

\bibitem{breuer2016}
H.-P. Breuer, E.-M. Laine, J. Piilo, and B. Vacchini, 
\textit{Colloquium: Non-Markovian dynamics in open quantum systems}, 
Rev. Mod. Phys. \textbf{88}, 021002 (2016). 

\bibitem{lorenzo2013}
S. Lorenzo, F. Plastina, and M. Paternostro, 
\textit{Comparative study of non-Markovianity measures in exactly solvable one- and two-qubit models}, 
Phys. Rev. A \textbf{88}, 020102 (2013).







































































































\end{thebibliography}
\end{document}